\documentclass[prc,showpacs,showkeys,superscriptaddress,twocolumn]{revtex4}
\usepackage{graphics,epsfig}
\usepackage{colordvi,amsmath}
\def\be{\begin{eqnarray}}
\def\ee{\end{eqnarray}}
\def\bc{\begin{center}}
\def\ec{\end{center}}
\def\dsp{\displaystyle}
\def\r{\rho}
\def\ro{\r_0}
\def\om{\omega}

\def\Lb{\Lambda}

\def\mn{m_N}

\def\vec#1{\boldsymbol{ #1}}

\def\Re{{\rm Re\,}}
\def\Im{{\rm Im\,}}
\def\prt{\partial}

\def\lsim{\stackrel{\scriptstyle <}{\phantom{}_{\sim}}}
\def\gsim{\stackrel{\scriptstyle >}{\phantom{}_{\sim}}}

\def\bwt{\begin{widetext}}
\def\ewt{\end{widetext}}
\begin{document}
\title{ Negative Kaons in Dense Baryonic Matter }
\author{Evgeni E.\ Kolomeitsev}
\thanks{
E-mail: E.Kolomeitsev@nbi.dk}
\affiliation{The Niels Bohr Institute, Blegdamsvej 17,
DK-2100 Copenhagen, Denmark\footnote{Current address}}
\affiliation{ECT$^*$, Villa Tambosi, I-38050 Villazzano  (TN),
INFN G.C. Trento, Italy }
\author{Dmitri N.\ Voskresensky}
\affiliation{GSI, Planck Str.\ 1, D-64291 Darmstadt, Germany}
\affiliation{Moscow Engineering Physical Institute, Kashirskoe shosse 31,
RU-115409 Moscow, Russia\footnote{Permanent address}}
\date{\today}
\begin{abstract}
Kaon polarization operator in dense baryonic matter of arbitrary
isotopic composition is calculated including s- and p-wave
kaon-baryon interactions. The regular part of the polarization
operator is extracted from the realistic kaon-nucleon interaction
based on the  chiral and $1/N_c$ expansion. Contributions of the
$\Lambda (1116)$, $\Sigma (1195)$, $\Sigma^* (1385)$ resonances
are taken explicitly  into account in the pole and regular terms
with inclusion of mean-field potentials. The baryon-baryon
correlations are incorporated and fluctuation contributions are
estimated. Results are applied for $K^-$ in neutron star matter.
Within our model a second-order phase transition to the s-wave
$K^-$ condensate state occurs  at $\rho_c \gsim 4 \rho_0$ once the
baryon-baryon correlations are included. We show that the
second-order phase transition to the p-wave $K^-$ condensate state
may occur at densities $\rho_c \sim 3\div 5 \rho_0$ in dependence
on the parameter choice. We demonstrate that a first-order phase
transition to a proton-enriched (approximately isospin-symmetric)
nucleon matter with a p-wave $K^-$ condensate can occur at smaller
densities, $\rho\lsim 2 \rho_0$. The transition is accompanied by
the suppression of hyperon concentrations.

\end{abstract}
\maketitle
\section{Introduction}\label{sec:intro}

Strangeness modes in compressed hadronic matter stay in the focus
of interest over the last decade. Strangeness is considered as a
good probe for dynamics of heavy-ion collisions. The vast amount
of data is accumulated for different collision energy regimes at
GANIL, GSI, CERN and BNL facilities~\cite{hic}. New data are in
advent~\cite{advent}. Understanding of strangeness production
requires a systematic study of the evolution of virtual
strangeness modes in the quark-gluon plasma and in the soup of
virtual hadrons. At break-up stage these virtual modes are
redistributed between real strange particles which can be observed
in experiment.

Other interesting topic is a strangeness  content of neutron
stars. With density increase strangeness can be liberated and face
up in the filling of the hyperon Fermi seas and/or in the creation
of kaon condensates. Both mentioned topics need a better knowledge
of the kaon-baryon interaction in dense baryonic matter. Question
on principal possibility of the kaon condensation in dense nucleon
matter was risen in Refs.~\cite{krive} and~\cite{swave}.
Occurrence of the  kaon condensation  in neutron star interiors
may have interesting observational consequences: (i)~The softening
of the equation of state (EoS), due to the appearance of a kaon
condensate phase, lowers the maximum neutron star mass and could
induce transformation of neutron stars into low-mass black
holes~\cite{bb}. (ii)~Kaon condensation is predicted to be accompanied with
the change of the nucleon isospin composition from the
neutron-enriched star ($N>>Z$) to the 'nuclear' star ($N\sim Z$)
or even to the 'proton' star ($N\lsim Z$), where the electric
charge of protons is compensated by the charge of the condensed
kaons~\cite{kvk95}. (iii)~The enhanced neutrino-emission
processes, occurring on protons in the proton-enriched matter and
on the kaon condensate field, lead to substantially more rapid
cooling of the star~\cite{nscool}.

The $K^-$ condensate is created in neutron stars due to weak
multi-particle processes
\be\label{react}
e^-+ X\to K^-+X'\,,\qquad  n+X\to p + K^-+ X',
\ee
in which electrons are replaced by $K^-$
mesons, and neutrons are converted into protons and
$K^-$~\cite{swave-1}. The symbolic writing (\ref{react})
assumes that surrounding baryons ($X,X'$) assure the momentum
conservation, therefore the critical point is determined by the
energy balance only. These processes become possible, if the
electron chemical potential $\mu_e$ exceeds the minimal $K^-$
energy,
$$\om^{\rm min}(\vec
k_m)
=\min_{\vec k}\{\om^{\rm min}(\vec k)\}\,,$$
where $\om^{\rm min}(|\vec{k}|)$ is the $K^-$ energy at the lowest
(index "$\rm min$")  quasiparticle spectrum branch of $K^-$
excitations in neutron star matter. The
works~\cite{swave,swave-1,swave-2} and subsequent ones have
postulated that there is only one kaon branch, for which
$\om(\vec{k} =0)\rightarrow m_{K}$ ($m_{K}$ is the kaon mass) for
the baryon density $\rho \rightarrow 0$, and the minimum is
achieved at $\vec{k}_{m}=0$. Then the critical point of the s-wave
$K^-$ condensation in a second-order phase transition is
determined by the condition $\om(\vec{k} =0)=\mu_e$. The
first-order phase transition to the kaon condensate state was
investigated in Ref.~\cite{tpl94} applying the  Maxwell
construction principle and in Ref.~\cite{gs} according to Gibbs
criteria. The p-wave kaon-nucleon interaction, which changes the
momentum dependence of the kaon spectrum, was disregarded in those
works. The p-wave $\Lambda (1116)$-nucleon hole and $\Sigma
(1195)$-nucleon hole contributions to the kaon polarization
operator were introduced in Ref.~\cite{pw} in the framework of the
chiral SU(3) symmetry. However then authors focused on the
discussion of the s-wave kaon condensation and considered the
polarization operator at zero momentum.

Ref.~\cite{kvk95} worked out a possibility of the p-wave kaon
condensation. The kaon polarization operator was constructed with
inclusion of the $\Lambda (1116)$--nucleon-hole and $\Sigma
(1195)$--nucleon-hole contributions in the p-wave part of the kaon
polarization operator and the kaon-pion and  kaon-kaon
interactions. The multi-branch spectrum of  $K^-$ mesons was found
and the possibility of the p-wave kaon condensation related to the
population of hyperon--nucleon-hole modes was demonstrated.
Possibilities of first-order  phase transitions in neutron star
interiors to a proton-enriched matter with a p-wave $K^-$
condensate and to a neutron-enriched matter with a p-wave
$\bar{K}^0$ condensate  were suggested.

Ref.~\cite{muto} considered the case of a large hyperon admixture
in the neutron star core. In such a medium both $K^-$ and $K^+$
spectra possess extra branches associated with the particle-hole
excitations $\Xi-\Lambda^{-1}$, $\Xi-\Sigma^{-1}$ for $K^-$ and
$N-\Lambda^{-1}$, $N-\Lambda^{-1}$ for $K^+$ (hole states are
labeled here by tense (-1)). At quite large kaon momenta the
branches of $K^+$ and $K^-$ spectra merge, that signals  an
instability with respect to $K^+K^-$ pair creation.

The analyses~\cite{kvk95,muto} relied heavily on the pole
 approximation for the particle-hole diagrams. The
final width effects were thereby disregarded too. Question on the
presence or absence of the  quasiparticle branches in the $K^-$
spectrum depends on the kaon energy and on the strength of the s-
and p-wave attraction~\cite{kv99}. The short-range baryon-baryon
correlations, which, as a rule, suppress attraction, were
disregarded in~\cite{kvk95,muto} for simplification. A particular
role of the correlations for the p-wave has been taken over in Ref.~\cite{kv98}.
However, the relative strength of s- and p-wave attraction
remained model dependent because no systematic investigation of
the kaon-nucleon interaction including s- and p-waves has been
available that time.

Recently, the $K^-$-nucleon scattering has been studied in the
framework of the relativistic chiral SU(3) Lagrangian imposing
constraints from the $K^+$-nucleon and pion-nucleon
sectors~\cite{lk01}. The covariant coupled-channel Bethe-Salpeter
equation was solved with the interaction kernel truncated to the
third  chiral order including the terms which are  leading in the
large $N_c$ limit of QCD. All SU(3) symmetry-breaking effects are
well under control by combined chiral and large $N_c$ expansions.
This analysis gives an opportunity to extend results of
~\cite{kvk95,kv98} taking into account off-pole (regular
background) contributions to the kaon self-energy. The accurate
fit of experimental data achieved in Ref.~\cite{lk01} fixes
the values of the kaon-nucleon-hyperon coupling constants.
Particularly, the $\Sigma^*(1385)$-pole  contribution to the
kaon-nucleon scattering was proved to be sizable, being not
included in  works Ref.~\cite{kvk95,muto}.

The discussion of the s-wave and, especially, of the p-wave
kaon-baryon interactions in nuclear matter is important for the
kaon production in heavy ion collisions~\cite{kvk96}. The momentum
dependence of kaon yields is experimentally measured~\cite{kaos}.
Also, the multi-branch $K^-$ spectrum can be tested via
$\bar\nu$-scattering on atomic nuclei \cite{kv98}. Peculiarities
of the $K^-$-nucleon interaction near mass shell are of great
importance for the physics of $K^-$ atoms \cite{LF01,Oset}.

In this work we continue the study of the s- and p-wave $K^-$-baryon
interactions in dense baryonic matter of arbitrary isotopic
composition. In particular, we present our results for the
composition typical for the core of neutron stars. Our strategy to
verify possibilities of s- and p-wave condensations is as follows.
For given total baryon density and the composition of the neutron
star matter we find the $K^-$ energy at lowest branch of the
dispersion equation
as a function of the momentum and then find the minimum in respect
to momenta. At density, when $\om^{\rm min}(\vec k_m)$ meets the
electron chemical potential $\mu_e$, the reactions (\ref{react})
become to be possible and the system undergos  a second-order
phase transition developing a classical field with the $K^-$
quantum numbers. If this state is realized for $\vec{k}_m =0$, we
deal with the s-wave kaon condensation and, if $\vec{k}_m \neq 0$,
we deal with the p-wave condensation.

Since solutions depend on many rather
uncertain parameters, we study possibilities of s- and
p-wave condensations separately. For instance, discussing possibility of
the s-wave condensation we put $\vec{k}=0$ in the solution
$\om^{\rm min}(\vec {k})$, assuming that parameters of the p-wave
interaction are such that they do not allow for the p-wave
condensation at the critical density for the s-wave condensation.
Then, trying to find a most appropriate parameter choice we
investigate how the possibility of the s-wave condensation depends on a
parameter variation. Next, in the same manner we study a
possibility of the p-wave condensation. Then we find the energy
and pressure of the system with the kaon condensate at different
spin-isospin compositions and densities. The system selects the
spin-isospin composition corresponding to the minimum of the
energy. We compare the pressure of the system with the kaon
condensate with the pressure of the neutron star matter without
condensate. If at some  density the pressure and the nucleon
chemical potential of the system with the kaon condensate coincide
with those values of the neutron star matter (other density and
isospin composition) without the condensate the system may come to
the condensate state by the first order phase transition. Such a
transition starts at a density described by the Maxwell
construction. It occurs if an effective surface tension parameter
is rather large. Otherwise, if an effective surface tension
parameter is not too large,  a mixed phase is realized starting
from even smaller densities.

In this paper we describe the baryon matter in terms of the
relativistic mean-field model (Sec.~\ref{sec:Baryon}). In
Sec.~\ref{sec:Vacuum} we introduce the kaon-nucleon interaction in
vacuum, as it follows from the partial wave analysis of
Ref.~\cite{lk01}. Then we separate the pole contributions of
$\Lambda(1116)$, $\Sigma(1195)$ and $\Sigma^*(1385)$ hyperons in
p-waves. Sections~\ref{sec:Construction} through \ref{sec:Short}
are devoted to the construction of the kaon polarization operator.
We start in  Sec.~\ref{sec:Construction} with the polarization
operator in the gas approximation but including the mean-field
potentials acting on baryons. Besides the $\Lambda(1116)$,
$\Sigma(1195)$ and $\Sigma^*(1385)$-nucleon-hole contributions it
contents a regular attractive part, weakly dependent on the  kaon
energy. In Sec.~\ref{sec:Decomposition} we separate s- and p-wave
parts of the kaon polarization operator. Occupation of  hyperon
Fermi seas is incorporated in Sec.~\ref{sec:Hyperon}. Repulsive
baryon-baryon correlations are evaluated and included in the
hyperon-nucleon particle-hole channels and in the regular part of
the polarization operator in Sec.~\ref{sec:Short}. In each section
above we  illustrate the strength of  new terms included into the
polarization operator and suggest effective parameterizations. We
relegate the discussion of contributions from kaon fluctuations
(baryon self-energies, multi-loop corrections) to
Appendix~\ref{sec:Fluct}. We argue that these effects do not
modify substantially the kaon polarization operator at zero
temperature in the region of small kaon energies and momenta,
which is of our interest here. In Sect. \ref{sec:Condensation} we
analyze different possibilities of second- and first-order phase
transitions to the s- and p-wave $K^-$ condensate states. In
particular, we argue for the p-wave $K^-$ condensation at $\rho
\lsim 2\rho_0$ ($\rho_0 \simeq 0.17$~fm$^{-3}$ is density of
nuclear saturation)  arising via a first-order phase transition.
In this phase transition all the hyperon Fermi seas are melted and
neutron star matter becomes proton-enriched (with approximately
symmetric isospin composition, $N\simeq Z$). Dependence of the
results on the specifics of the EoS and the corresponding particle
composition is illustrated in Appendix ~\ref{sec:OtherEos}. Some
technical information on the Green's functions and the Lindhard's
function is deferred to Appendices ~\ref{Nonequilibrium} and
\ref{sec:Imag}.

Throughout the paper we use units $\hbar =c=1$.


\section{Baryon Interaction in Relativistic Mean Field Model}\label{sec:Baryon}

\subsection{Lagrangian of the Model}

We consider a dense system consisting of baryons and leptons,
which we describe by the Lagrangian density
containing the baryon and the lepton contributions, respectively:
${\cal L}={\cal L}_B +{\cal L}_l$ .

The baryon matter at densities relevant for neutron star
interiors is convenient to describe by the mean-field solution of
the  Lagrangian~\cite{walecka}:
\begin{eqnarray} \nonumber
&&{\cal L}_B=\sum_{B}\!\bar{B}\!\left(i\, \prt\!\!\!/-g_{\om
B}\om\!\!\!/- g_{\r B}\,\vec{\r}\!\!\!/\cdot \vec{t}_B- m_B
+g_{\sigma B}\sigma\right)\!B
\\ \nonumber
&&+
\frac{\prt_\mu \sigma \prt^\mu\sigma}{2}
-\frac{m_\sigma^2 \sigma^2}{2}
-\frac{m_N b (g_{\sigma N} \sigma)^3}{3}
-\frac{c (g_{\sigma N} \sigma)^3}{4}
\\
&&-\frac{\omega_{\mu\nu}\,\omega^{\mu\nu}}{4}+\frac{
m_\om^2 \om_\mu\,\om^\mu}{2} -
\frac{\rho_{\mu\nu}\,\rho^{\mu\nu}}{4}+\frac{
m_\r^2 \vec{\r}_\mu \vec{\r\,}^\mu}{2},
 \label{lagwal}
\end{eqnarray}
where all states of the baryon ($J^P=\frac12^+$) octet $B=(n,p,
\Lb\,,\Sigma^{\pm,0}\,,\Xi^{-0})$,
interact via exchange  of scalar,
vector and isovector mesons  $\sigma$\,, $\om_\mu$\,, $\vec{\r}_\mu $.
Heavier baryons do not appear at the baryon densities
under consideration ($\rho \leq 6\rho_0$) and are not included, therefore.
In (\ref{lagwal}) $\vec{t}_B$ denotes isospin operator acting on  the baryon
$B$.
The field strength tensors for the
vector mesons  are given by
$\omega_{\mu\nu}=\prt_\mu\,\om_\nu-\prt_\nu\,\om_\mu$ for omega mesons and
$\rho_{\mu\nu}=\prt_\mu\,\vec{\r}_\nu-\prt_\nu\,\vec{\r}_\mu$\, for rho mesons.
The equations of motion for baryons following from (\ref{lagwal})
give
\be\label{baren}
E_B(\vec{p})=\epsilon_B(\vec{p})+V_B\,,\qquad \epsilon_B(\vec{p})=
\sqrt{m_B^{*2}+\vec{p}^2}\,,\ee
 where
$m_B^*=(m_B-g_{\sigma B}\, \sigma)$ is the baryon effective mass,
and $V_B=g_{\om\, B}\,
\om_0+g_{\r\, B}\, \r_{03}\, t_{3\,B},$ with
$\sigma, \om_0, \r_{03}$ being the mean field solutions
of equations of motion for the
meson fields. The composition of the cold
neutron star matter at baryon density $\r$
is determined by the  $\beta$-equilibrium conditions
$E_B(p_{{\rm F}B})=\mu_n-q_B\,\mu_e$.
Here $p_{{\rm F}B}$ is the Fermi momentum,
$q_B$ is the electric charge of the given baryon species $B$, and
$\mu_n$\,, $\mu_e$  are chemical potentials of
neutrons and electrons fixed by the total baryon density $\r$ and the
electroneutrality condition.

The lepton Lagrangian density is the sum of the free electron and $\mu^-$ meson
contributions ${\cal L}_l ={\cal L}_e +{\cal L}_\mu$\,.
The $\beta$-equilibrium requires that  the muon chemical potential
equals to the electron one, $\mu_\mu=\mu_e$, and, therefore,
muons  appear in the system  when $\mu_e$ exceeds the muon mass, $m_\mu$\,.
Electrons can be considered as ultra-relativistic particles.

The energy density of the system is given by
\be\nonumber E_{\rm tot} &=& E_{\rm mes}+\sum_{B} E_{B}^{\rm
kin}(p_{{\rm F} B})
+ \sum_{l=e^-,\mu^-} E_{l}(\mu_e),
\\ \nonumber
E_{\rm mes}&=&\frac13\,b\,\mn\,(g_{\sigma\,N}\,\sigma)^3 +\frac14\,
c\, (g_{\sigma\, N}\,\sigma)^4
\\ \nonumber &+&\frac12\,m_\sigma^2\,\sigma^2+
\frac12 \,m_\om^2\,\om_0^2+\frac12\,m_\r^2\,\r_{03}^2 ,
\\ \nonumber
E_B^{\rm kin}(p_{{\rm F} B}) &=&\frac{1}{\pi^2}\,
\intop_0^{p_{{\rm F}B}}\,dp\, p^2\,\epsilon_B(p) \,,\quad
\\ \label{endens} E_{\rm
lept}(\mu_l)&=& \frac{1}{\pi^2}\,\theta (\mu_l - m_l )
\intop_{m_l}^{\mu_l}\,d\epsilon\, \epsilon^2\,\sqrt{\epsilon^2-
m_l^{2}}, \ee
where $\theta (x)$ is the step function.

\subsection{Coupling Constants}

The coupling constants are  adjusted to reproduce properties of
equilibrium nuclear matter: saturation density $\ro$, binding
energy $E_{\rm bind}$, compressibility modulus $K$, and the effective
nucleon mass  $\mn^*(\ro)$. In the following we take the value
$\ro=0.17$~fm$^{-3}$, and the binding energy $E_{\rm
bind}=-16$~MeV. For the nuclear compressibility  modulus we take
the value $K=$210~MeV, which follows from the variational
calculation~\cite{softeos}. Following Ref.~\cite{walecka} we adopt
the symmetry energy $a_{\rm sym}=36.8$~MeV which lies well in the
interval allowed by the microscopic calculations~\cite{Lombardo}.
We take the effective nucleon mass $\mn^*(\ro)=0.85\, m_N$, cf.
argumentation in Ref.~\cite{MSTV90}. Of course, the values of the
input parameters  can be selected
differently, in order to  fit experimental data. We refer at this point to
the recent work~\cite{Piek} discussing the uncertainties.

The corresponding coupling constants of the Lagrangian
(\ref{lagwal}) are:
\be  \nonumber
&&\frac{\dsp g_{\om N}^2\,m_N^2}{\dsp m_\om^2}=54.6041\,,\quad
\frac{\dsp g_{\sigma N}^2\,m_N^2}{\dsp m_\sigma^2}=164.462\,,
\\ \nonumber
&&\frac{\dsp g_{\rho N}^2\,m_N^2}{\dsp m_\rho^2}=121.690\,,\quad b=0.0202832\,,
\\ \nonumber \label{lagconst}
&&c=0.0471633\,.
\ee
In order to verify the sensitivity of the results to the details of the EoS we
explore in Appendix~\ref{sec:OtherEos} another set of the nuclear matter parameters, which we fit
to reproduce the microscopic calculations of Urbana-Argonne
group~\cite{Akmal}.

Including hyperons, one also has to specify hyperon couplings to
the meson fields. They are expressed via ratios to the nucleon
couplings, $x_{MH}=g_{MH}/g_{MN}$ with $M=(\sigma,\om,\r)$ and
$H=(\Lambda,\Sigma,\Xi )$. Couplings to the vector mesons are
estimated from the quark counting as $x_{\om \Lambda
(\Sigma)}=x_{\r\Sigma}=\frac23$ and $x_{\om \Xi}=x_{\r
\Xi}=\frac13$. Alternatively, relying on SU(3) symmetry~\cite{sbg}
one would find $x_{\r\Sigma}=x_{\r\Xi}=1$ for the $\rho$ meson
couplings. The scalar meson  couplings can be constrained by
making use of hyperon binding  energies in infinite  nuclear
matter at saturation~\cite{glenmos}, extrapolated from
hyper-nucleus data. Denoting the binding energy of a hyperon by
$E_{\rm bind}^H$ we derive the following relation between the
scalar and  vector couplings:
\be\nonumber
E_{\rm bind}^H &=& (g_{\om
N}^2\,\ro/m_\om^2)\, x_{\om H}-(m_N-m_N^*)\, x_{\sigma H}
\\ \nonumber
&=&(80.73\,x_{\om H}-140.70\,x_{\sigma H})~{\rm MeV}.\ee
There is
convincing  evidence from the systematic study of hypernuclei that
for $\Lambda$ particles  $E_{\rm bind}^\Lambda\simeq
-30$~MeV~\cite{lnucl}. For $\Sigma$ hyperons in nuclei, on the
other hand, the data are still controversial, giving the broad
band, $-10~{\rm MeV}<E_{\rm bind}^\Sigma<30~{\rm MeV}$\,, from a
slight attraction to a strong repulsion~\cite{sbind}. Following Ref.~\cite{xibind}
we adopt for $\Xi$ the values $E_{\rm bind}^\Xi\simeq -18$~MeV advocated in
Ref.~\cite{sbg}.

To cover different possibilities we consider four cases:
(I) vector meson couplings are taken according to the quark counting,
$x_{\om \Lambda(\Sigma)}=x_{\r \Sigma}=2\,x_{\om\Xi}=2\,x_{\r
\Xi}=\frac23$
 and $E_{\rm bind}^\Lambda =
-30$~MeV, $E_{\rm bind}^\Xi = -18$~MeV,
$E_{\rm
bind}^{\Sigma}=30$~MeV;
(II) the same as in case~(I)  but $E_{\rm bind}^{\Sigma}=-10$~MeV;
(III) the same as in case (I) but the rho meson
couplings are taken according to SU(3) symmetry, i.e.
 $x_{\r \Sigma,\Xi}=1$;
(IV) the same as in case~(III), but for $E_{\rm bind}^{\Sigma}=-10$~MeV.

\subsection{Particle Concentrations}

\begin{figure*}
\includegraphics[clip=true,width=14cm]{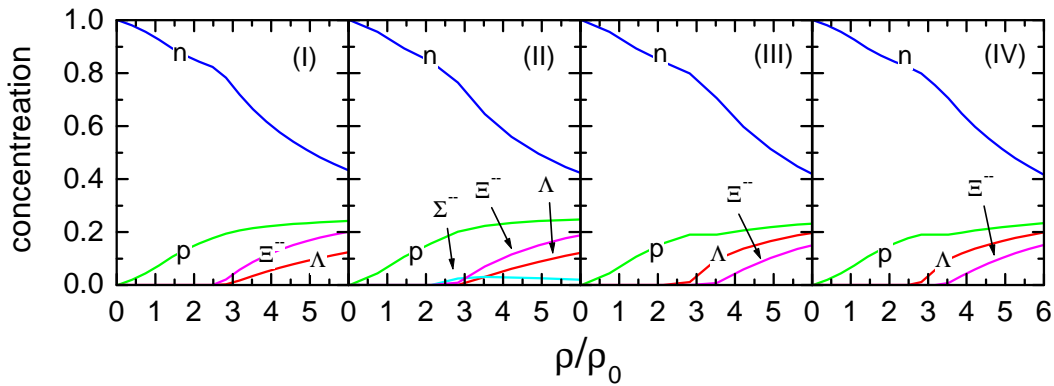}
\caption{
Concentration of baryon species in neutron star matter.
Panels (I-IV) correspond to four choices of the hyperon-nucleon
interaction constants specified in text.}
\label{fig:compos}
\end{figure*}

In Fig.~\ref{fig:compos} we show the resulting concentrations of
different baryon species, as function of the baryon density.
Columns (I-IV) correspond to the four above choices of the
hyperon-meson coupling constants. We see, that hyperons  appear in
neutron star matter in all cases at density $\rho >\rho_{{\rm
c},H}\simeq 2.5\div 3\, \ro$. The latter critical value is rather
insensitive to various choices  of hyperon-nucleon interactions.
However, the order, in which hyperons populate the Fermi seas,
depends crucially on the details of hyperon-nucleon interactions.
The $\Sigma^-$ hyperons do not appear, at least, up to $6\rho_0$
except for case II. However, even in this case their concentration
is very small. The place of $\Sigma^-$ is readily taken by $\Xi^-$
and $\Lambda$ hyperons. This observation is in line with the
results  of work~\cite{gcoupl}. In case IV,  we chose $E_{\rm
bind}^{\Sigma}<0$, nevertheless the $\Sigma$ hyperons do not
appear due to increasing repulsion mediated by $\rho$ meson with
larger coupling constants than in case II. We see that in all
cases I - IV  the proton concentration and the electron chemical
potential saturate with the filling of hyperon Fermi seas,  in
line with the results  of previous works. We also see that all
choices I - IV do not support effects observed in \cite{muto},
where $\Lambda$ hyperons become more abundant than  protons
already at $3\rho_0$. Therefore, the p-wave $K^+K^-$ condensation
discussed in \cite{muto} does not show up in the framework of our
model for all four parameter choices.

Although concentration of hyperons is rather small in all the
cases, their presence may have significant consequences affecting
critical densities of p-wave $K^-$ condensates.

\section{$K^-$-Nucleon Interaction in Vacuum}\label{sec:Vacuum}

The  kaon-nucleon interaction in vacuum results as the
solution of the coupled-channel Bethe-Salpeter equation
\be\label{diag:tij}
\parbox{7.5cm}{
\includegraphics[clip=true,width=7.5cm]{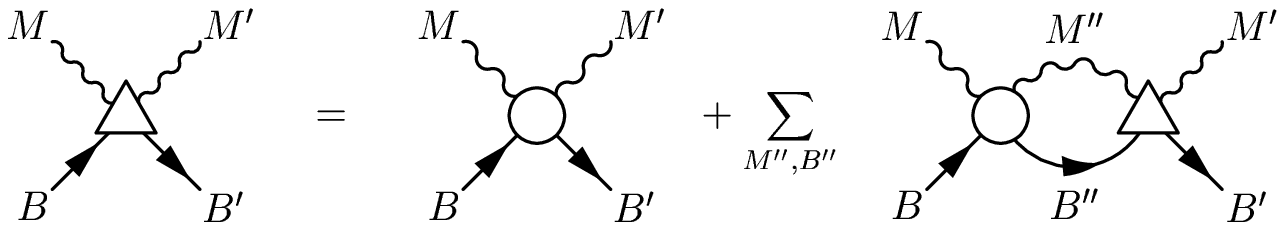}}\,,
\ee
where the triangle is the full meson-baryon scattering amplitude
with strangeness (-1) and the circle stands for the interaction
kernel. This equation involves rescatterings through all
possible meson-baryon intermediate states allowed by the
strangeness conservation. The interaction kernel can be derived
from SU(3) chiral Lagrangians~\cite{kaiser,krippa,or98,om01} or
phenomenologically adjusted to fit the data within the K-matrix
formalism~\cite{kmatrix,keil}.

In view of the general interest to the s-wave kaon condensation
prevailing in the literature so far,  the most attention has been
paid to the kaon-nucleon interaction in the s-wave.
Although far below the threshold
the s-wave kaon-nucleon scattering amplitude is rather smooth
function of the kaon energy, close to the threshold the amplitudes
vary strongly due to the $\Lambda(1405)$ resonance (see
Fig.~\ref{fig:ampl-s}). Therefore, extrapolation of the scattering
amplitude, adjusted to fit the data above the $K^- N$ threshold,
into the subthreshold region crucially depends on the microscopic
model applied. Its in-medium modification is  a matter of debate,
too~\cite{smed,RO00,ske,tolos,KL01}.

Up to recent,  only a scarce information on the p-wave
kaon-nucleon interaction was available. In the isospin-zero
channel the small p-wave amplitudes were not  separated from the
large contribution of the $\Lambda(1405)$ resonance in the s-wave,
which dominates near threshold energies. In the isospin-one
channel determination of the p-wave amplitudes remained  also
uncertain due to a lack of direct  experimental information on the
$K^-n$ scattering at low energies. This gap has been filled by
Ref.~\cite{lk01}, where the relativistic chiral SU(3) Lagrangian
imposed extra constrains from the $K^+$-nucleon and pion-nucleon
sector. This analysis provides reliable estimates for both the s-
and p-wave $K^-N$ scattering amplitudes, which we will use in the
following.

\subsection{Forward Scattering Amplitudes}

The vacuum $\overline{K} N$ forward scattering amplitudes
in the given isospin channel $I$ have contributions from s- and p-partial
waves:
\bwt\be \nonumber
T^{\rm(I)}(s) = T_{\rm S}^{\rm (I)}(s)+T_{\rm P}^{\rm (I)}(s)
&=&\frac{\bar{E}(s,m_N^2,m_K^2)+m_N}{2\,m_N}\,\left[F_{\rm S}^{\rm
(I)}(s)+Q^2(s,m_N^2,m_K^2)\, F_{\rm P}^{\rm (I)}(s)\right]\,,
\\ \label{ampl}
Q^2(s,p^2,k^2) &=& \frac{(s-p^2-k^2)^2-4\, k^2\, p^2}{4\, s}\,,
\ee\ewt
where $s=(p+k)^2$, and $p=(\epsilon_N(\vec{p\,}),\vec{p}\,)$,
$k=(\om,\vec{k\,})$ are 4-momenta of the incoming  nucleon and
kaon, respectively. For nucleons and kaons being on mass-shell,
the quantity  $Q^2(s,m_N^2 ,m_K^2 )$ is nothing
else but the square of the center-of-mass momentum in the
kaon-nucleon channel,
$\bar{E}(s,m_N^2,m_K^2)=\sqrt{m_N^2 +Q^2 (s,m_N^2 ,m_K^2 )}$
is the nucleon energy in the center-of-mass frame,
$$\bar{E}(s,p^2 ,k^2)=\frac{s+p^2-k^2}{2\sqrt s}\,,$$
$m_N$ and  $m_K$
are the free nucleon and kaon masses. We shall neglect the isospin
symmetry breaking within kaon and nucleon isospin multiplets, which are
irrelevant in dense nuclear matter.

\begin{figure*}
\includegraphics[clip=true,width=12cm]{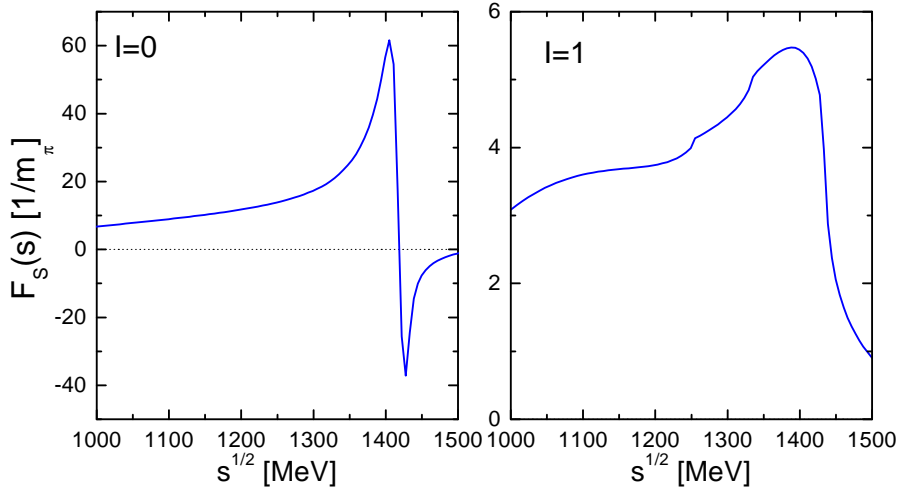}
\caption{Real parts of the s-wave kaon-nucleon scattering
amplitudes in isospin-zero and isospin-one
channels~\protect\cite{lk01}. }\label{fig:ampl-s}
\end{figure*}

\begin{figure*}
\includegraphics[clip=true,width=12cm]{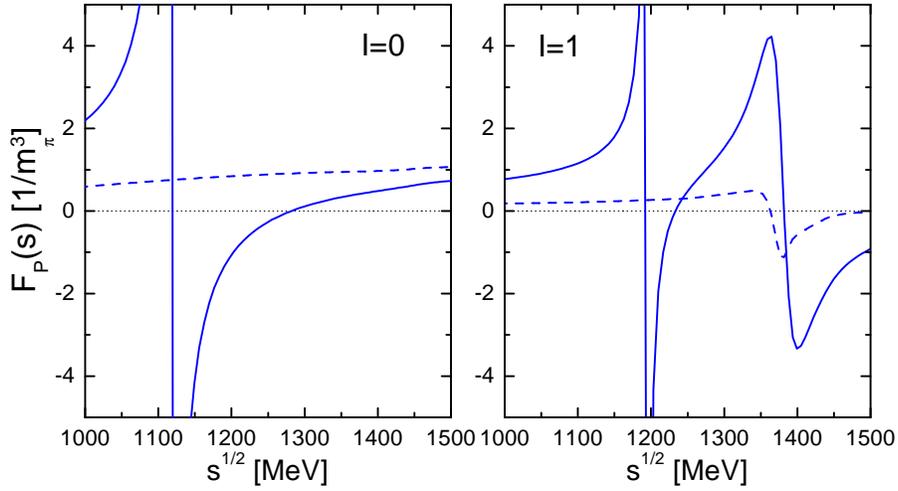}
\caption{Real parts of the p-wave kaon-nucleon scattering
amplitudes $F_{\rm P}$ in isospin zero and isospin one channels
(solid lines)~\protect\cite{lk01}. The dash lines represent
corresponding pole subtracted scattering amplitudes
(\protect\ref{Fsubtr}).} \label{fig:ampl}
\end{figure*}

The invariant partial-wave amplitudes
$F_{\rm S}^{\rm (I)}$ and $F_{\rm P}^{\rm (I)}$
are related to the standard partial wave
amplitudes  (cf.~\cite{Landoldt}, section~3.1 for definitions) as
\be \nonumber
F_{\rm S}^{\rm (I)}(s)&=&\frac{8\,\pi\sqrt{s}}{m_N+\bar{E}(s,m_N^2 ,m_K^2 )}
\,f_{0+}^{\rm (I)}(s),
\\ \nonumber
F_{\rm P}^{\rm (I)}(s)&=&\frac{8\,\pi\sqrt{s}/Q^2(s,m_N^2,m_K^2)}
{m_N+\bar{E}(s,m_N^2,m_K^2)}
\\ \label{F}
&&\times \left(f_{1-}^{\rm(I)}(s) + 2\, f_{1+}^{\rm (I)}(s)\right)\,.
\ee
The partial amplitudes $f_{l\pm}=f_{l,\, J=l\pm\frac12}$ in
(\ref{F}) are the scattering amplitudes
for the given angular momentum $l$ and total momentum $J$. Real
parts of the partial amplitudes $T_{\rm S}^{\rm (I)}$ and
$T_{\rm P}^{\rm (I)}$  are shown in
Fig.~\ref{fig:ampl-s} and Fig.~\ref{fig:ampl} by solid lines.
Pronounced peak structures in the p-wave amplitude are
 due to the $\Lambda (1116)$  pole in the
isospin-zero channel and the  $\Sigma (1195)$ and $\Sigma^*(1385)$ poles
in the isospin-one channel.

\subsection{Separation of the Pole  Terms}

The hyperon s-channel exchanges are responsible for the most
strong variation of the polarization operator at low frequencies
and momenta. Therefore, they deserve a special consideration in view
of the baryon modifications by the mean field-potentials.
To treat them explicitly, we separate
pole contribution from the p-wave amplitudes.
Then we define  the pole and  pole-subtracted
amplitudes
\be\label{Fsubtr} \delta F_{\rm P}^{({\rm I})}(s) &=&
F_{\rm P}^{({\rm I})}-F_{\rm pole}^{({\rm I})}(s)\,,
\\ \label{Fpoles0}
F_{\rm pole}^{(0)}(s)  &=& -2\,
\frac{C_{KN\Lambda}^2\,2\,m_\Lambda\,}{(\bar{E}(s,p^2,k^2)+m_N)^2}
\frac{(m_\Lambda+m_N)^2}{s-m_\Lambda^2+i\,0} ,
\\ \nonumber
F_{\rm pole}^{(1)}(s)   &=& -2\,
\frac{C_{KN\Sigma}^2\,2\,m_\Sigma\,}{(\bar{E}(s,m_N^2,m_K^2)+m_N)^2}\,
\frac{(m_\Sigma+m_N)^2}{s-m_\Sigma^2+i\,0}
\\ \label{Fpoles1}
&&\hspace*{-1.8cm}- \frac43\, \frac{2\,C_{KN\Sigma^*}^2\, (s/m_{\Sigma^*})}{ s
-m_{\Sigma^*}^2+\frac{i}{2}\, \gamma_{\Sigma^*}(s)}\,
\frac{\bar{E}(m_{\Sigma^*}^2 , m^2_N , m^2_K )+m_N}{\bar{E}(s,m_N^2,m_K^2)
+m_N} \,.
\ee
The $\Sigma^*$ width, $\gamma_{\Sigma^*}(s) =(\sqrt
s+m_{\Sigma^*})\, ( \gamma_{\pi\Lambda}(s)+\gamma_{\pi\Sigma}(s)
+\gamma_{KN}(s))$, includes contributions of $\pi\Lambda$\,,
$\pi\Sigma$ and $KN$ channels, $\gamma_{\phi B}(s) =
C_{\phi B\Sigma^*}^2\,(m_B+\bar{E}(s,m_B^2,m_{\phi}^2)) |Q (s,m_B^2,m_G^2)|^3
/(12\,\pi\,
\sqrt{s})\,$, where
$m_B$ is the mass of the corresponding baryon
$B=(n,p,
\Lb\,,\Sigma^{\pm,0}\,,\Xi^{-0})$ and
$m_{\phi}$ is the mass of the corresponding light meson
$\phi =(\vec{\pi},\vec{K})$.
The coupling constants
$C_{\pi\Lambda(\Sigma)\Sigma^*}$ can be extracted from the
partial width of the $\Sigma^*$ hyperon.
The values of the  $KN\Lambda(\Sigma)$ coupling constants
$C_{KN\Lambda}\simeq -0.68/m_\pi$,
$C_{KN\Sigma}\simeq 0.34/m_\pi$ entering
(\ref{Fpoles0},\ref{Fpoles1})  follow directly from the amplitudes
shown in Fig.~\ref{fig:ampl}.
The values of $C_{KN\Lambda}$ and $C_{KN\Sigma}$
are defined as couplings of  $\Lambda$ to $(K^\dagger
N)$ and $\Sigma^\alpha$ to $(K^\dagger \tau^\alpha
N)$ isospin states, where $K$, $N$ stand for isospin-doublets,
$\Sigma^\alpha$ is the isospin-triplet, and  $\tau^\alpha$ are the
isospin Pauli matrices.
Note that the value of
$C_{KN\Lambda}$, obtained on basis of Ref.~\cite{lk01},
is smaller than that fitted by J\"ulich group
\cite{Julich} and then used in Refs.~\cite{pw,kvk95}.

The scattering amplitude nearby the $\Sigma^*$ resonance can be
only approximately described by the  last term in
(\ref{Fpoles1}), since the corresponding amplitude results from a
multi-channel dynamics, which
generates the energy dependent  self-energy of the $\Sigma^*$ resonance to be
included in the denominator. This self-energy, however,  can be neglected
at $\sqrt s$ relevant for calculations below.
Ref.~\cite{lk01} gives the value for the $\Sigma^*$ resonance
coupling  $C_{KN\Sigma^*}\simeq
0.84/m_\pi$ defined by the Lagrangian term
$C_{KN\Sigma^*} \bar{\Sigma}^{*\alpha}_\mu
 \prt^\mu K^\dagger \tau^\alpha N$\,.

The pole subtracted amplitudes $\delta F^{\rm (I)}(s)$ are shown
in Fig.~\ref{fig:ampl} by dash lines. We see that the amplitudes
in both isospin channels are smooth functions of $\sqrt{s}$ in the
subthreshold region $\sqrt s<1300$~MeV of interest.

Nevertheless,
we would like to notice that procedure of the pole subtraction is
not  unambiguous. The forms (\ref{Fpoles0},\ref{Fpoles1}), chosen
here, will later exhibit themselves to be convenient, since they
allow direct comparison with the analysis of
Ref.~\cite{kvk95,kv99,muto}.

Considering further the kaon-nucleon interaction in dense matter,
we also need to take into account the mean-field potentials acting on
baryons according to the Lagrangian~(\ref{lagwal}).

\section{ $K^-$ Polarization Operator from Scattering Amplitudes}
\label{sec:Construction}

\subsection{Gas Approximation and Baryonic Mean-Fields}

Our further aim is to construct
the retarded $K^-$ polarization operator
in baryonic matter $\Pi_R^{\rm tot} (\om,\vec{k})$
related to the propagator
$$D_K^{-1}(\om,\vec{k})=\om^2-\vec{k}^2-m_K^2-\Pi_R^{\rm tot} (\om,\vec{k}).$$

The spectral function is given by $A_K(\om,\vec k) =-2{\rm Im}
D_K(\om , \vec k)$. The quasiparticle branches of the spectrum
appear in some energy-momentum region, if there  the kaon width
$\Gamma_K =-2\Im \Pi_R^{\rm tot} (\om,\vec{k})$ is much smaller
than any other typical energy scale. Then one can put $\Gamma_K
\rightarrow 0$ in the kaon Green's function and the quasiparticle
branches are then  determined by equation $\Re
D_K^{-1}(\om,\vec{k})=0$.

With the help of
the scattering amplitudes in vacuum we may extract
the causal polarization operator $\Pi_C^{(0)} (\om,\vec{k})$
related to the partial wave amplitudes
$F_{\rm S}^{\rm (I)}$ and  $F_{\rm P}^{\rm
(I)}$ in the gas approximation
after integration over the nucleon occupation function:
\bwt\be
\Pi_C^{(0)} (\om,\vec{k})
&=&{\rm I}_{\rm s-wave}(\om,\vec{k})+{\rm I}_{\rm
p-wave}(\om,\vec{k})\,,
\\ \nonumber
{\rm I}_{\rm s-wave}(\om,\vec{k})
&\equiv&
{\rm I}_{{\rm s-wave},p}(\om,\vec{k})+{\rm I}_{{\rm s-wave},n}(\om,\vec{k})
\\  \label{Is}
&=&-\intop
\frac{2\,{\rm d}^3 \vec p}{(2\,\pi)^3}
\frac{\bar{E}(s,m_N^2,k^2)+m_N}{2\,\sqrt{m_N^2+\vec{p}^2}}\,
\left\{\frac12\, \Big( F_{\rm S}^{(0)}(s)+F_{\rm
S}^{(1)}(s)\Big)\, n_{ p}(\vec{p})+ F_{\rm S}^{(1)}(s)\, n_{
n}(\vec{p})\right\} \,,
\\\nonumber
{\rm I}_{\rm p-wave}(\om,\vec{k})&\equiv&
{\rm I}_{{\rm p-wave},p}(\om,\vec{k})+{\rm I}_{{\rm p-wave},n}(\om,\vec{k})
\\  \label{Ip}
&=& -\intop \frac{2\,{\rm d}^3
\vec p}{(2\,\pi)^3}
\frac{\bar{E}(s,m_N^2,k^2)+m_N}{2\,\sqrt{m_N^2+\vec{p}^2}}\,
Q^2(s,m_N^2,k^2)
\left\{\frac12\, \Big( F_{\rm P}^{(0)}(s)+F_{\rm
P}^{(1)}(s)\Big)\, n_{ p}(\vec{p})+ F_{\rm P}^{(1)}(s)\, n_{
n}(\vec{p})\right\} \,,\ee\ewt
where
$n_i (\vec{p} )$ are the nucleon Fermi occupations, $i=(n,p)$,
at zero temperature
$n_p (\vec{p} )=\theta(p_{{\rm F},p}-|\vec{p}|)$ and $n_n
(\vec{p} )=\theta(p_{{\rm F},n}-|\vec{p}|)$,
$s=(\om+\sqrt{m_N^2+\vec{p}^2})^2-(\vec{k}+\vec{p}\, )^2$.

There are simple relations between causal (``-,-'')
and retarded (``R'')
two-point functions,
like Green functions and polarization operators, see eq. (\ref{Fretarded})
in Appendix \ref{Nonequilibrium}. For zero temperature causal and retarded
two-point functions coincide for positive frequencies.
For $T\neq 0$ their real parts continue to
coincide, whereas imaginary parts are different, but can be
interrelated.
Bearing this in mind, we will further suppress for brevity
the subscripts $R$ and $C$, if it does not lead to ambiguity.

Exploiting  decomposition (\ref{Fsubtr}) of the p-wave scattering
amplitude we  present
\bwt\be\label{Ip-dec}
{\rm I}_{\rm p-wave}(\om,\vec{k})=
{\rm I}_{\rm p-wave}^{\rm pole}(\om,\vec{k})+
{\rm I}_{\rm p-wave}^{\rm reg}(\om,\vec{k})\,,
\ee
as the sum of the pole and regular parts.
The pole part is generated by the  hyperon exchanges
\be\nonumber\label{IHp}
{\rm I}_{\rm p-pole}^{\rm pole}(\om,\vec{k})
&=& {\rm I}_\Lambda^{\rm pole}(\om,\vec k)+{\rm I}_\Sigma^{\rm pole}(\om,\vec
k)+{\rm I}_{\Sigma^*}^{\rm pole}(\om,\vec k),
\\ \label{IHpole}
{\rm I}_{\Lambda}^{\rm pole}(\om,\vec k)
&=&
-C_{KN\Lambda}^2\,\intop \frac{2\,{\rm d}^3
\vec p}{(2\,\pi)^3}
\frac{\bar{E}(m_{\Lambda}^2,m_N^2,k^2)-m_N}{2\,\sqrt{m_N^2+\vec{p}^2}}
\frac{(m_\Lambda +m_N)^2}{s^2-m_{\Lambda}^2+i\,0}\,2\,m_\Lambda
\,n_{p}(\vec p),
\\  \label{ISpole}
{\rm I}_{\Sigma}^{\rm pole}(\om,\vec k)
&=&
-C_{KN\Sigma}^2\,\intop \frac{2\,{\rm d}^3
\vec p}{(2\,\pi)^3}
\frac{\bar{E}(m_{\Sigma}^2,m_N^2,k^2)-m_N}{2\,\sqrt{m_N^2+\vec{p}^2}}
\frac{(m_\Sigma +m_N)^2}{s^2-m_{\Sigma}^2+i\,0}\,2\,m_\Sigma
\,(n_{p}(\vec p) +2 n_n (\vec p)),
\\ \label{ISSpole}
{\rm I}_{\Sigma^*}^{\rm pole}(\om,\vec k)
&=& -\frac43\, C_{KN\Sigma^*}^2\,\intop \frac{2\,{\rm d}^3
\vec p}{(2\,\pi)^3}
\frac{\bar{E}(m_{\Sigma^*}^2,m_N^2,k^2)+m_N}{2\,\sqrt{m_N^2+\vec{p}^2}}\,
\frac{s}{m_{\Sigma^*}}\,\frac{Q^2(s,m_N^2,k^2)}{s-m_{\Sigma^*}^2+
\frac{i}{2}\,\gamma_{\Sigma^*}(s)}\, (n_{p}(\vec p) +2 n_n (\vec p))\,.
\ee
The regular part, ${\rm I}_{\rm p-wave}^{\rm reg}$, can be expressed as
\be\label{Ireg}
{\rm I}_{\rm p-wave}^{\rm reg}(\om,\vec k)=
\bar{\rm I}_{\rm p-wave}^{\rm reg}(\om,\vec k)
+\delta{\rm I}^{\rm reg}_{\rm p-wave}(\om,\vec k),
\ee
including the part of integral (\ref{Ip}) evaluated with
$\delta F_{\rm P}^{(\rm I)}$ from (\ref{Fsubtr}),
\be\nonumber
\bar{\rm I}_{\rm p-wave}^{\rm reg}(\om,\vec{k})
&=&-\intop \frac{2\,{\rm d}^3 \vec p}{(2\,\pi)^3}
\frac{\bar{E}(s,m_N^2,k^2)+m_N}{2\,\sqrt{m_N^2+\vec{p}^2}}\,
Q^2(s,m_N^2,k^2)
\nonumber\\ \label{Ipreg}
&&\times
\left\{\frac12\, \Big( \delta F_{\rm P}^{(0)}(s)+\delta F_{\rm
P}^{(1)}(s)\Big)\, n_{ p}(\vec{p})+ \delta F_{\rm P}^{(1)}(s)\, n_{
n}(\vec{p})\right\},
\ee
and the non-pole contributions from the hyperon exchanges
\be \label{deltaI}
&&\delta{\rm I}^{\rm reg}_{\rm p-wave}(\om,\vec k)
=\delta {\rm I}_\Lambda(\om,\vec k)+\delta {\rm I}_\Sigma(\om,\vec k)\,,
\\ \label{dIHpole}
&&\delta{\rm I}_{\Lambda}   (\om,\vec k)
= -C_{KN\Lambda}^2\!\intop \frac{2\,{\rm d}^3
\vec p}{(2\,\pi)^3}
\frac{m_\Lambda\,\sqrt{s} -m_N^2+k^2}{2\,\sqrt{s}\,\sqrt{m_N^2+\vec{p}^2}}
 \frac{(m_\Lambda +m_N)^2}{\sqrt{s}+m_\Lambda}\,n_{p}(\vec p)\,,
\\
&&\delta{\rm I}_{\Sigma}   (\om,\vec k) = -C_{KN\Lambda}^2\!\intop
\frac{2\,{\rm d}^3 \vec p}{(2\,\pi)^3} \frac{m_\Sigma\,\sqrt{s}
-m_N^2+k^2}{2\,\sqrt{s}\,\sqrt{m_N^2+\vec{p}^2}} \frac{(m_\Sigma
+m_N)^2}{\sqrt{s}+m_\Sigma}\,(n_{p}(\vec p)+2n_n (\vec p))\,. \ee
To obtain the last relation we used \be \bar E(s,m_N^2,k^2)-\bar E
(m_H^2,m_N^2,k^2)=\frac{m_H\sqrt s-m_N^2+k^2}{2m_H\sqrt s}
(\sqrt{s}-m_H), \,\,\,H=\{ \Lambda , \Sigma \}. \ee \ewt The above
construction of the  polarization operator, corresponding to the
gas approximation, does not take into account mean-field
potentials acting on baryons, vertex corrections due to the
baryon-baryon correlations, and possible modification of the
scattering amplitudes in the medium. The modification of the
baryon propagator on the mean-field level can  be easily
incorporated in the integrals (\ref{Is},\ref{Ip}) by the
replacement $m_N\to m_N^*$. Effects induced by this modification
in the kinematic prefactors in (\ref{Is},\ref{Ip}) can be easily
traced, as we will demonstrate below. Scaling of nucleon mass in
$s$ is more subtle. Solving the coupled-channel Bethe-Salpeter
equation one sums all the two-particle reducible diagrams for  the
part of the $s$-plane corresponding to $K^-N$ scattering. This
approach is explicitly crossing non-invariant and continuation of
amplitudes far below $K^-N$ threshold can generate artificial
singularities in the scattering amplitude.  In Ref.~\cite{lk01},
from where we borrow the amplitudes, the approximation scheme for
solution of the Bethe-Salpter equation,
 was furnished in such a way that the $K^-N$
and $K^+N$ scattering amplitudes exhibit the
{\it approximate crossing} symmetry, smoothly matching each
other for $\sqrt s\sim m_N$. Therefore, amplitudes depicted in
Fig.~\ref{fig:ampl-s} and Fig.~\ref{fig:ampl} are still physically
well constrained in the corresponding intervals of $\sqrt s$ shown there.
However for somewhat smaller $\sqrt s$ the $K^-N$ s-wave
scattering amplitude gets  unphysical poles.
To cure this problem the complete solution of the  Bethe-Salpeter
equation for $K^-N$ scattering has to be redone with the medium modified
baryon masses.
Fortunately, there are some indications that it would not change the
results drastically. For the integral with the s-wave scattering
amplitude we will demonstrate that the final results can be nicely
modeled  with the polarization operator following from the
leading-order chiral Lagrangian, which has now explicit dependence
on the baryon masses. The loop corrections due to iteration of
the interaction kernel should be suppressed for small kaon frequencies
to keep approximate crossing invariance of the amplitude.
The pole subtracted p-wave amplitude is rather smooth function of
$\sqrt{s}$,  as it is shown in
Fig.~\ref{fig:ampl},
being mainly determined by contact terms of the
chiral Lagrangian and, thus, has a weak baryon effective mass dependence.
Hence, extrapolating the amplitude to somewhat smaller $\sqrt s$,
we do not expect its strong variation.
Opposite, the part of the polarization operator generated by the
hyperon poles, ${\rm I}_{\rm P}^{\rm pole}$,
depends strongly on baryonic mean fields changing
the pole positions in the amplitude. Therefore, this part will  be  treated
explicitly in the course of our consideration.

\subsection{Pole Part of Polarization Operator}

Here, we find contributions to the $K^-$ polarization operator
from the hyperon poles in the $K^-N$ scattering amplitude
determined by (\ref{IHpole},\ref{ISpole},\ref{ISSpole}). Relying
on explicit calculations of Refs~\cite{kvk95,kv98} we can easily
incorporate scalar and vector mean-fields acting on baryons. The
scalar field is taken into account with the help of the
replacement  $m_B\to m_B^*$ and the baryon vector potentials are
included in the pole terms by the shift of the kaon frequency
$\om\to\om+\delta v_{iH}$, with $\delta v_{iH}=V_i-V_H$, $i
=\{n,p\}$, $H=\{\Lambda,\,\Sigma\}\,$, that immediately follows
from the  difference of the baryon energies, see (\ref{baren}).
Please notice that this energy shift is obvious only for the pole
contribution to the polarization operator. Generally, due to the absence
of the gauge invariance for the massive vector fields such a shift
is not motivated  for more complicated diagrams.

Writing down explicitly  all contributions we cast
\be \nonumber
&&\Pi^{({\rm pole}, 0)}(\om,\vec{k}) \equiv {\rm I}
_{\rm p-wave}^{\rm pole}(\om,\vec{k})
\\ \nonumber
&&\quad=\Pi_{p\Lambda}^{({\rm pole}, 0 )}(\om,\vec{k})
+\Pi_{p\Sigma^0}^{({\rm pole}, 0 )}(\om,\vec{k})
+ 2\, \Pi_{n\Sigma^-}^{({\rm pole}, 0 )}(\om,\vec{k})
\\ \label{pi-sum}
&&\quad+ \Pi_{p\Sigma^{*0}}^{({\rm pole},0 )}(\om,\vec{k})
+2\,\Pi_{n\Sigma^{*-}}^{({\rm pole},0 )}(\om,\vec{k})\,, \ee
where
each term is equivalent to the pole  contribution of the hyperon particle -
nucleon hole loop diagram (Schr\"odinger picture)
\be\label{pinh}
\parbox{3cm}{
\includegraphics[clip=true,width=3cm]{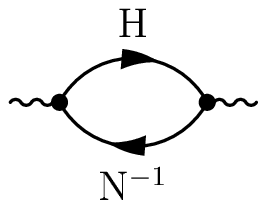}}\,,
\ee
written in terms of the  Lindhard's function~\footnote{For further  convenience
we introduce notation  $i=(n,p)$ and continue to use $N$
when the quantity does not
depend on the nucleon isospin. Please, bear in mind that one uses different
notations for the Lindhard's function in the literature. Our notation
slightly differs
from those in refs \cite{EW88,MSTV90}.} as
\bwt\be\nonumber
&&\Pi_{i H}^{({\rm pole},0 )} (\om,\vec{k})
=C_{KNH}^2\,\left[(m_H^*-m_N^*)^2-(\om+\delta v_{iH})^2+\vec{k}^{\,2}
\right]\, \eta_{NH}^2\,\Phi_{iH}(\om,\vec{k})\,,
\\ \nonumber
&&\Pi_{i\Sigma^*}^{({\rm pole}, 0  )}(\om,\vec{k}) =
C_{KN\Sigma^*}^2\eta_{ i\Sigma^*}^2(\om,\vec k)\,\vec{k}^{\,2}\,
\Phi_{ i\Sigma^*}(\om,\vec{k})\,,
\\ \label{pip}
&&\eta_{NH} = \frac{m_N^{*}+m_H^*}{2\,m_N^*}\,,\quad
\eta_{i\Sigma^*}^2(\om,\vec k) =
\frac{(m_{\Sigma^*}^*+m_N^*)^2-(\om+\delta v_{
iH})^2+\vec{k}^{\,2}} {6\,m_{\Sigma^*}^{*2}}\,.
\ee\ewt
We reserved the
superscript "$(0)$" for each term in (\ref{pi-sum}) indicating
that neither  further self-energy of baryons beyond the  mean
field nor the vertex corrections due to the baryon-baryon
correlations are included. The (retarded)
Lindhard's function $\Phi$ accounting
the relativistic kinematics is defined as
\be\label{li}
\Phi_{ i
H}(\om,\vec{k}) =\intop \, \frac{2\,{\rm
d}^3p}{(2\,\pi)^3\,2\,\epsilon_i(p)}\,\frac{4\,m_N^{*2}}{s-m_H^2+i\,0}
n_i (\vec{p}).
\ee
For the case of zero temperature, on which we will concentrate further
\be\nonumber
\Phi_{ i H}(\om,\vec{k})=\frac{m_N^{*2}}{2\,\pi^2\,| \vec{k}|}\,
\intop_0^{p_{{\rm F}i}} \frac{{\rm
d} p p}{\epsilon_i(p)}
\ln
\left[\frac{\Delta_{ iH}^+(\om,\vec{k},\vec{p})}
{\Delta_{ iH}^-(\om,\vec{k},\vec{p})}\right] ,
\\ \label{lind}
\Delta_{ iH}^\pm(\om,\vec{k},\vec{p}) =
[\om+\delta v_{iH}+\epsilon_i(\vec{p})]^{\,2}-
\epsilon_H^2(|\vec{p}| \mp |\vec{k}|)\,,
\ee
where $p_{{\rm F} i}$ is the Fermi momentum for the given
nucleon species.
Non-relativistic form of the Lindhard's function
used e.g. in Ref.~\cite{MSTV90} (in different normalization),  is recovered
with the help of expansion
$\epsilon_B(p)\approx m_B^*+p^2/(2\, m_B^*)$ in (\ref{lind}).

The imaginary part of the (retarded)
Lindhard's function is obtained as an analytical continuation
  $\ln(x)=\ln|x|+i\,\pi\,
\theta(-x)$ leading to a non-zero contribution for
\be\label{im-inter}
\om_{iH}^-(\vec{k})<\om<\om_{iH}^+(\vec{k}).
\ee
Here $\om_{ iH}^\pm$
are the upper and lower borders of the corresponding particle-hole
continuum
\bwt\be\label{ompl}
\om_{iH}^+(\vec{k})&=&\left\{\begin{array}{lcl}
\sqrt{(m_H^*-m_N^*)^2+\vec{k}^2}+\delta v_{iH} &,& k<p_{{\rm F}
i}\,(\frac{m_H^*}{m_N^*}-1)\\
E_H(|\vec{k}|+p_{{\rm F} i})-E_i (p_{{\rm F} i}) &,& k>p_{{\rm F}
i}\,(\frac{m_H^*}{m_N^*}-1)
\end{array}\right.
\\ \label{ommn}
\om_{iH}^-(\vec{k})&=&E_H(|\vec{k}|-p_{{\rm F} i})-E_i(p_{{\rm F}
i}).
\ee
Baryon energies include vector potentials according to
(\ref{baren}). Properties of ${\rm Im}\, \Phi$ are discussed in
Appendix~\ref{sec:Imag}.

\begin{figure*}
\includegraphics[width=10cm,clip=true]{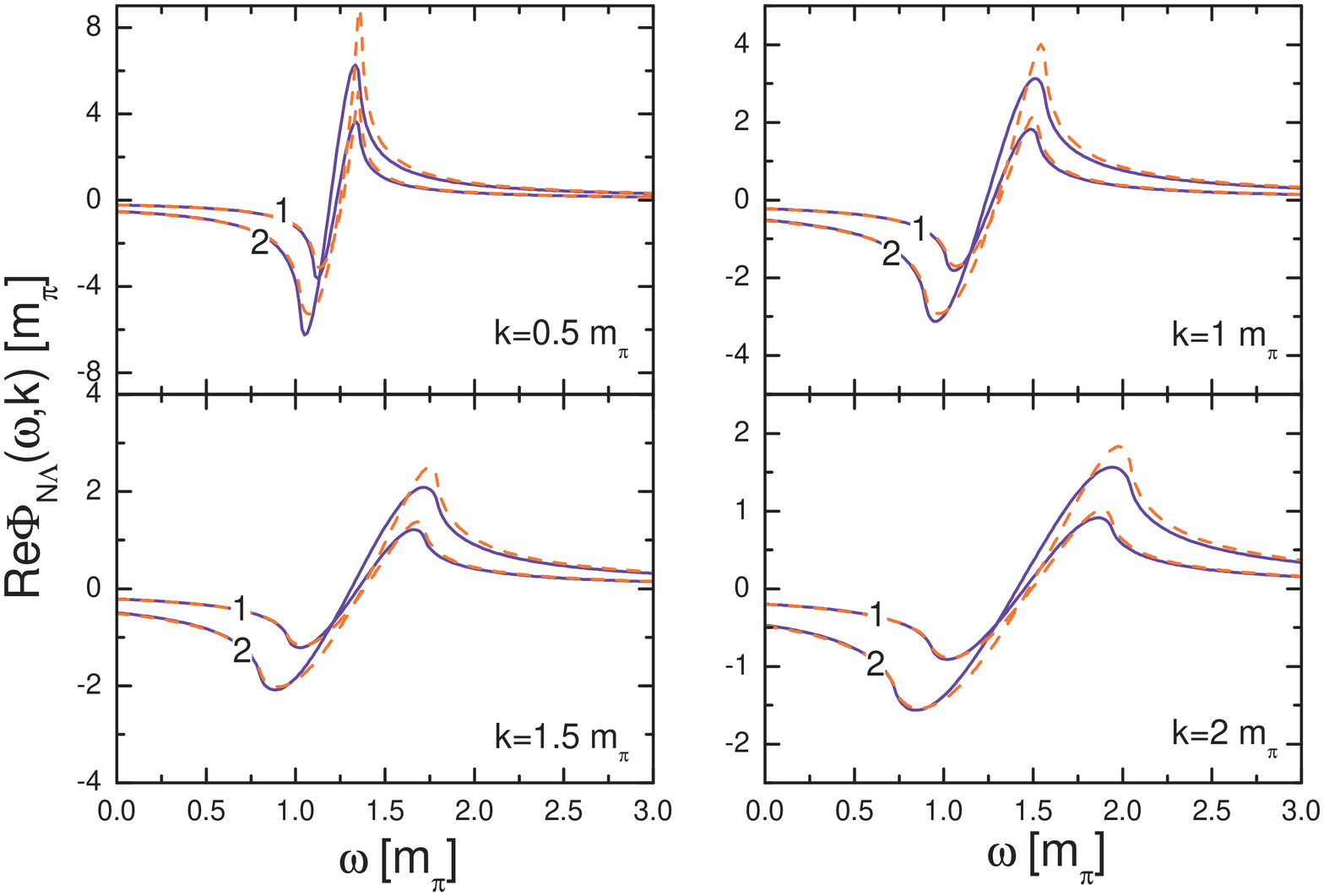}
\caption{The real part of the relativistic  Lindhard's function
(\protect\ref{lind}) is shown by solid lines as a function of
energy for different momenta. Dashed lines correspond to the
approximate relation (\protect\ref{lind-ner}). The curves labeled
with (1) are calculated for the Fermi momentum $p_{\rm F}=300$~MeV
and those labeled with (2) stand for $p_{\rm F}=400$~MeV. }
\label{fig:appacc}
\end{figure*}

An approximate expression  for $\Phi_{ i H}$ renders
\be\nonumber
&&\Phi_{ i H}(\om,\vec{k})
\approx -\frac{m_N^{*2}}{8\,\pi^2\,|\vec{k}|^3\,\epsilon_{{\rm F} i}}
\left[\frac{\widetilde{\Delta}_{ iH}^+ \,\widetilde{\Delta}_{ iH}^-
}{2}\,
\ln\left[\frac{\widetilde{\Delta}_{ iH}^+ }
{\widetilde{\Delta}_{ iH}^- }\right] -\frac{(\widetilde{\Delta}_{ iH}^{+}
)^2
-(\widetilde{\Delta}_{ iH}^{-})^2}{4}\right],
\\\label{lind-ner}
&&\widetilde{\Delta}_{ iH}^{\pm} = \Delta_{ iH}^{\pm}
(\om,\vec{k},p_{{\rm F}i})\,,\quad \epsilon_{{\rm F} i}=
\epsilon_i (p_{{\rm F} i}),
\ee\ewt
being valid for
$\om<(\om_{iH}^++\om_{iH}^-)/2$.
The accuracy of this approximation is illustrated by
Fig.~\ref{fig:appacc}.

For completeness we also quote here the low
momentum limit of (\ref{lind-ner}),
\be\nonumber
\Re \Phi_{ iH}(\om, \vec{k}) \approx
\frac{2\,\r_i\,m_N^{*2}}{\epsilon_{{\rm
F}i}\,\Delta_{ iH}}
\left[1+\frac{\vec{k}^{\,2}}{\Delta_{ iH}} \,
\left(1+\frac45\,\frac{p^2_{{\rm F} i}}{\Delta_{ iH}}\right)\right] ,
\\ \label{phi-exp}
\Delta_{ iH}= \Delta_{ iH}^{\pm}(\om,0,p_{{\rm F} i}), \,\,\,
| \vec{k}| \ll |\Delta_{ iH}(\om,0,p_{{\rm F} i})|/(2\,p_{{\rm F} i}).
\ee
where $\rho_i$ is the density of the given nucleon species $n$ or $p$.

\section{S- and p-wave Parts of the Polarization Operator
}\label{sec:Decomposition}

In our discussion we would like to put a particular
emphasis to in-medium effects,
which modify  the $K^-$ spectrum at finite momenta. For this
purpose we define the momentum independent part,  called  s-wave
part of the polarization operator,  and the momentum dependent part, called
p-wave part  of the
polarization operator,
\be\nonumber \Pi^{(0)}_{C} (\om,\vec{k})  &=& \Pi_{\rm S}^{(0)}(\om)+
\Pi_{\rm P}^{(0)}(\om,\vec{k})
\\ \nonumber
&\equiv&
\Pi_{C}^{(0)} (\om,0)+\Big[ \Pi_{C}^{(0)} (\om,\vec{k})-\Pi_{C}^{(0)}
(\om,0)\Big]
\,.\ee
The term $\Pi_{C}^{(0)} (\om,0)$ does not depend on $\vec{k}$,
whereas the term $\Big[ \Pi_{C}^{(0)} (\om,\vec{k})-\Pi_{C}^{(0)}
(\om,0)\Big]$ depends on $\vec{k}$, vanishing  at $|\vec{k}|=0$.
In order to avoid misunderstanding we point out that the s- and
p-wave scattering amplitudes contribute to both the parts
(\ref{Is}), (\ref{Ip}) of the polarization operator, namely,
\be
\Pi_{\rm S}^{(0)}(\om) &=& {\rm I}_{\rm s-wave}(\om,0)+{\rm I}_{\rm
p-wave}(\om,0),
\\ \nonumber
\Pi_{\rm P}^{(0)}(\om,\vec{k}) &=& [{\rm I}_{\rm s-wave}(\om,\vec{k})-{\rm
I}_{\rm s-wave}(\om,0)]
\\
&+&
[{\rm I}_{\rm p-wave}(\om,\vec{k})-{\rm I}_{\rm
p-wave}(\om,0)]\,.
\ee
Next two subsections are devoted to discussion of the s-and p-wave
parts of the polarization operator.

\subsection{P-wave Part}

Following decomposition (\ref{Ip-dec}) we split the p-wave
part of the polarization operator into the pole and the
regular contributions
\be\label{ppol-split}
\Pi_{\rm P}^{(0)}(\om,\vec{k})=\Pi^{(\rm pole,0)}_{\rm P}(\om,\vec{k})+
\Pi^{(\rm reg, 0)}_{\rm P}(\om,\vec{k}).
\ee
For the pole p-wave part we have
\be\nonumber
&&\Pi^{({\rm pole}, 0)}_{\rm P}(\om,\vec{k})
={\rm I}_{\rm P}^{({\rm pole})}(\om,\vec{k})-
{\rm I}_{{\rm P}}^{({\rm pole})}(\om, 0)
\\ \nonumber
&&\quad={\Pi}^{({\rm P},0)}_{p\Lambda}(\om,\vec{k})
+{\Pi}^{({\rm P},0)}_{p\Sigma^0}(\om,\vec{k})
+2\,{\Pi}^{({\rm P},0)}_{n\Sigma^-}(\om,\vec{k})
\\ \label{pi-sum-P}
&&\quad+
{\Pi}^{({\rm P},0)}_{p\Sigma^{*0}}(\om,\vec{k})
+2\,{\Pi}^{({\rm P},0)}_{n\Sigma^{*-}}(\om,\vec{k}),
\\
&&{\Pi}^{({\rm P},0)}_{ iH}(\om, \vec{k}) =
\Pi_{ iH}^{({\rm pole},0)}(\om,\vec{k})- \Pi_{ iH}^{({\rm pole},0)}(\om,0)\,,
\ee
where we used (\ref{IHp})-(\ref{ISSpole}) and (\ref{pi-sum}).
The contribution of $\Sigma^*$--nucleon-hole term is rather
sensitive to the values of the mean-field potentials acting on
$\Sigma^*$ which are unknown.  Therefore,
we investigate two choices in further. First,
we assume that $\Sigma^*$ couples to the mean field with the same strength as
the $\Sigma$ hyperon ($V_{\Sigma^*}=V_{\Sigma}$,
$g_{\sigma \Sigma^*} =g_{\sigma \Sigma}$).
In the second case, we  detached $\Sigma^*$
from the mean field potentials ($V_{\Sigma^*}=0$, $g_{\sigma \Sigma^*} =0$).

Expanding the p-wave pole part of the polarization operator in
small kaon momenta we have
\bwt\be \nonumber &&\Pi_{iH}^{({\rm
P},0)}(\om,\vec{k})  = C_{KNH}^2 \,\vec{k}^2\,\phi_{iH}^{\rm
P}(\om)+\mathcal{O}(\vec{k}^4)\,,
\\  \label{ppol}
&&\phi_{iH}^{\rm P}(\om) = \eta_{NH}^2\, \Phi_{iH}(\om,0)
+\eta_{NH}^2\big((m_H^*-m_N^*)^2-(\om+\delta v_{iH})^2\big)\,
\frac{\prt
\Phi_{iH}(\om,\vec{k})}{\prt\vec{k}^2}\Big|_{|\vec{k}|=0},
\\ \label{ppols}
&&\Phi_{i\Sigma^*}(\om,\vec k)=
C_{KN\Sigma^*}^2\,\vec{k}^2\,\eta^2_{i\Sigma^*}(\om,0)\,
\Phi_{i\Sigma^*}(\om,0). \ee\ewt

The regular part of the p-wave polarization operator is defined by
\be\nonumber
\Pi_{\rm P}^{({\rm reg},0)}(\om,\vec{k})&=&
[{\rm I}_{\rm s-wave}(\om,\vec{k})-{\rm
I}_{\rm s-wave}(\om,0)]
\\ \nonumber
&+&[{\rm I}_{\rm p-wave}^{\rm reg}(\om,\vec{k})-{\rm I}^{\rm  reg}_{\rm
p-wave}(\om,0)]\,.
\ee

At  small kaon momenta
the real part of $ \Pi_{\rm P}^{(\rm reg ,0)}$ can be
written  as
\be  \nonumber
\Re \Pi_{\rm P}^{({\rm reg},0)}(\om,\vec{k})& = & \vec{k}^{\,2}\, \Big(
b_p(\om)\,\frac{\r_p}{\ro}+ b_n(\om)\frac{\r_n}{\ro}\Big)+\mathcal{O}
(\vec{k}^4)\,,
\\\label{pireg}
 b_i(\om)&=& b_i^{\rm (S)}(\om) + b_i^{\rm (P)}(\om),
\\ \label{bs}
b_i^{\rm (S)} (\om) &=&
\frac{\ro}{\r_i}\,\frac{\prt}{\prt \vec{k}^{\,2}}\, {\rm I}_{{\rm s-wave},i}
(\om,\vec{k})\Bigg|_{\vec{k}=0},
\\ \label{bp}
b_i^{\rm (P)} (\om) &=&
\frac{\ro}{\r_i}\,\frac{\prt}{\prt \vec{k}^{\,2}}\,
{\rm I}_{{\rm p-wave},i}^{\rm reg}(\om,
\vec{k})\Bigg|_{\vec{k}=0}
,\,\,\,
\ee
where we used that the real part of the kaon polarization operator
is even function of the kaon momentum.
\begin{figure*}
\includegraphics[clip=true,width=6cm]{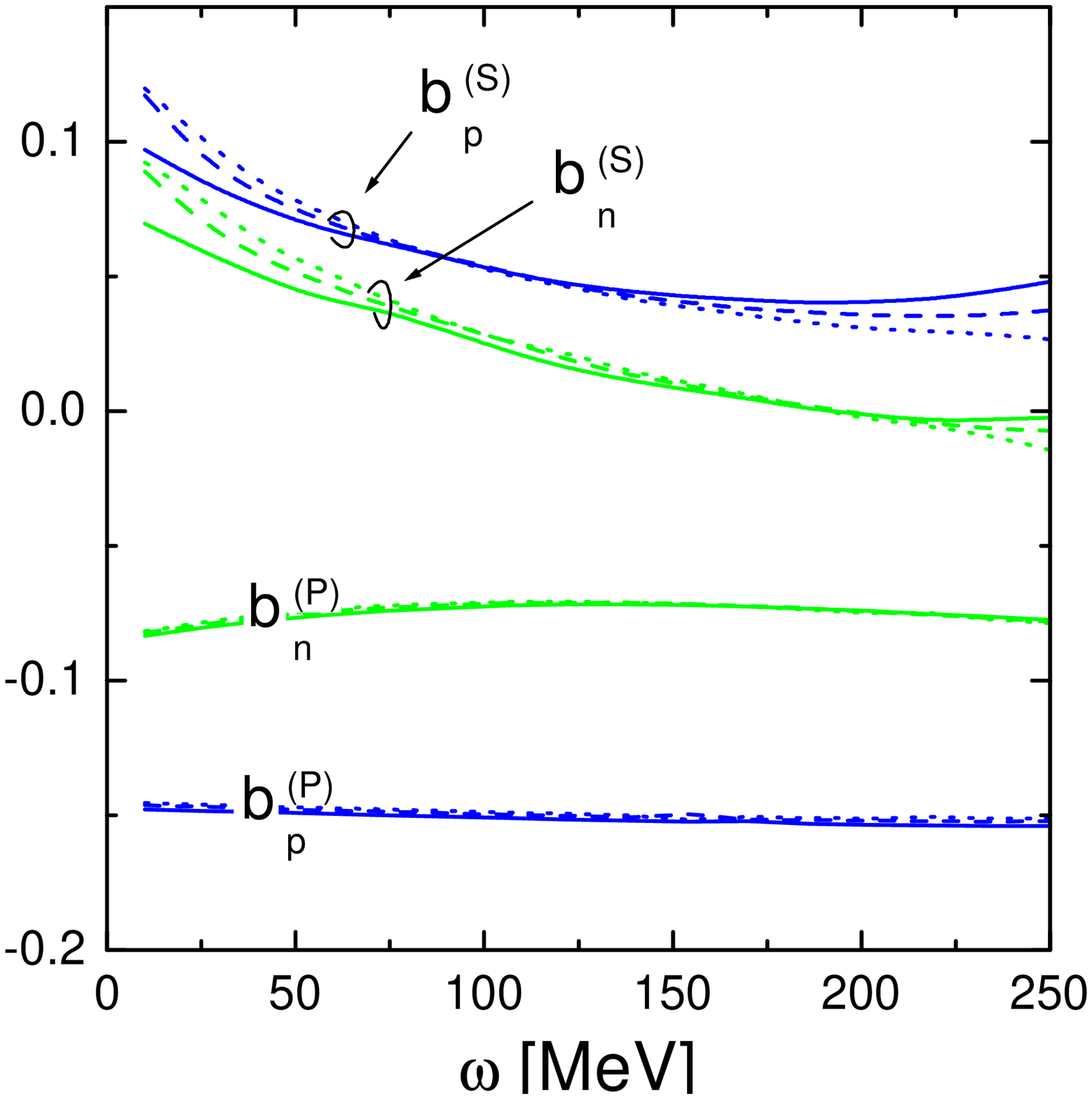}
\quad
\includegraphics[clip=true,width=6cm]{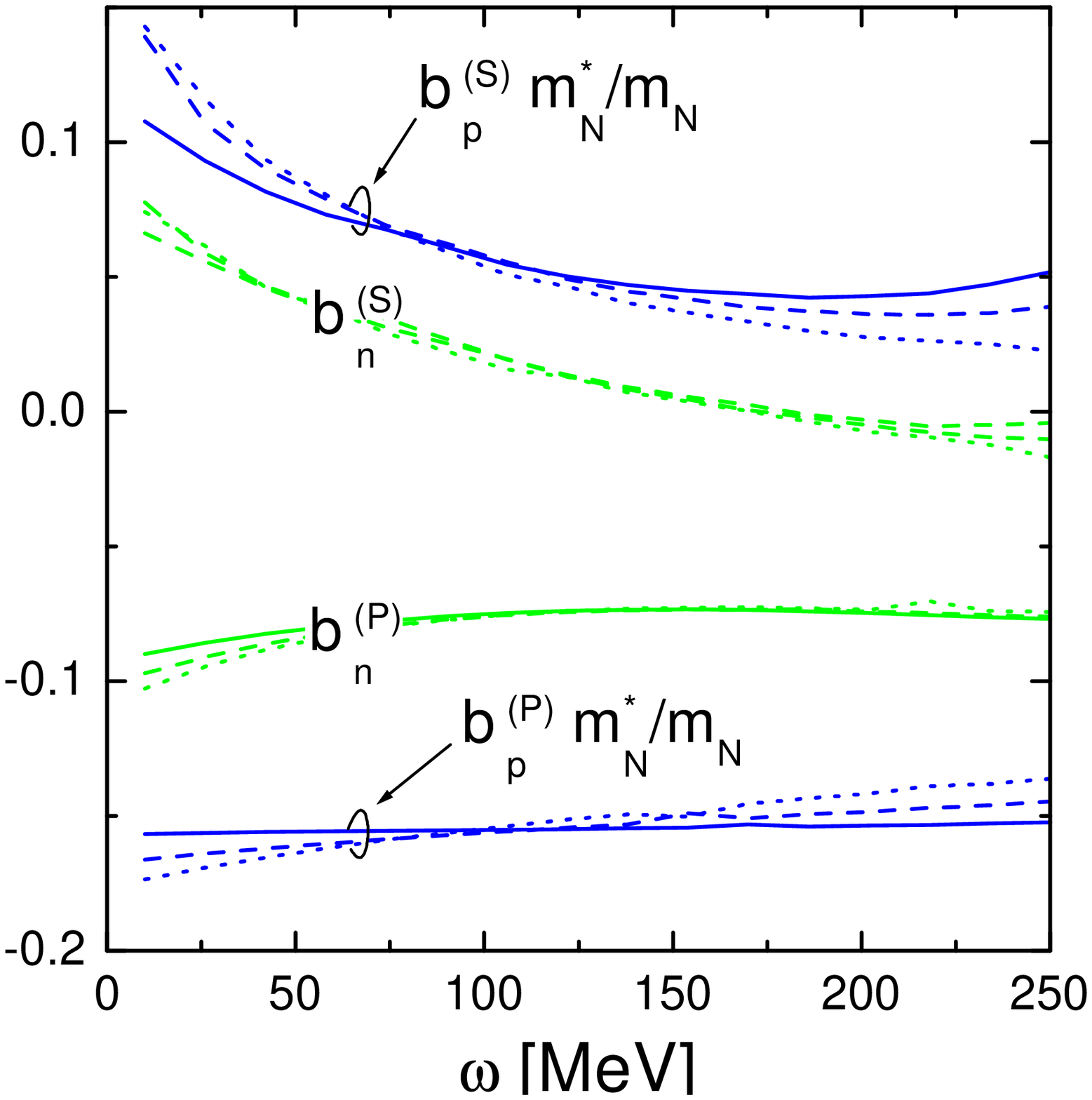}
\caption{Left panel: Coefficients (\protect\ref{bs}) and (\protect\ref{bp}) of
expansion (\protect\ref{pireg}) of the regular
p-wave part of the polarization operator. Solid, dash and dotted lines show
calculations done for nucleon densities $\r_{n,p}= 1, 3$, and $5 \ro$,
respectively.
Right panel: the same as in the left plane but integrals are
evaluated with the effective nucleon mass in kinematical prefactors and
coefficients $b_p^{(S,P)}$ are scaled by $m_N^*/m_N$. }
\label{fig:bnp}
\end{figure*}
The quantities $b_n^{\rm (S,P)}$ and $b_p^{\rm (S,P)}$ are shown
 in Fig.~\ref{fig:bnp} (left plane) for $T=0$ as functions of the
kaon energy for several values of the densities $\rho_i$.
In these calculations integrals
${\rm I}_{\rm S}$ and ${\rm I}_{\rm P}$ have been evaluated with
the free nucleon masses. We see that these coefficients are almost
density independent and only weakly dependent on the kaon energy
in the interval $100~{\rm MeV}<\om<250~{\rm MeV}$ of our interest.
As we have discussed in the beginning of this section, we replace
the baryon masses by the effective masses, as they follow from the
mean field solutions, only in the kinematical prefactors in
(\ref{Is},\ref{Ip}). The results  are shown in the right plane of
Fig.~\ref{fig:bnp}. The coefficients $b_n^{\rm (S,P)}$ depend
moderately on the density, as before, whereas the coefficients
$b_p^{\rm (S,P)}$ exhibit a stronger density dependence, which can
be parameterized by the factor $m_N/m_N^*$, as it is demonstrated in
Fig.~\ref{fig:bnp}. Energy dependence remains to be weak in the
interval $100~{\rm MeV}<\om<250~{\rm MeV}$ of our interest.

Our result (\ref{pireg}) is derived for rather small values of
kaon momenta, $\mid\vec{k}\mid\ll m_{K}$. In order to find the
actual value of the p-wave $K^-$ condensate amplitude in most
general case one needs to deal with momenta up to
$\mid\vec{k}\mid\sim p_{{\rm F},n}\sim m_{K}$. To satisfy the
latter general case we extrapolate our result for the regular part
of the polarization operator to such momenta. Luckily, within our
approach critical points of s- and p-wave condensations deviate
not as much from each other, as we will show it, and in this case
the kaon condensate momentum in the vicinity of the critical
density has extra smallness. Also, the main contribution to the
kaon polarization operator comes from the pole terms, which are
written explicitly for arbitrary momenta (\ref{pi-sum}). Thereby,
ambiguity of mentioned interpolation should not significantly
affect our conclusions.

\subsection{S-wave Part}

The kaon-nucleon interaction determines the following contributions
to the  s-wave part of the $K^-$ meson polarization operator,
\be \nonumber
\Pi_{\rm  S}^{(0)}(\om)
={\rm I}_{\rm s-wave}(\om,0)+
\bar{\rm I}_{\rm p-wave}^{\rm reg}(\om,0)
\\ \label{Ss}
+ \delta \Pi^{({\rm reg},0)}(\om) + \Pi^{({\rm pole},0)}(\om,0)\,,
\ee
where the last two terms correspond
to non-pole and pole parts of the hyperon exchange
terms in the amplitude, respectively.
The term $\Pi^{({\rm pole},0)}$ is given by (\ref{pi-sum}).
Using (\ref{deltaI}) we present
$\delta \Pi^{({\rm reg}, 0)}$ as follows
\bwt
\be \nonumber
&&\delta \Pi^{({\rm reg},0)}(\om)\equiv
\delta {\rm I}^{\rm reg}_{\rm p-wave}(\om,0)
=\delta {\Pi}_{p\Lambda}^{({\rm reg}, 0)}(\om,0)+
\delta{\Pi}_{p\Sigma^0}^{({\rm reg},0)}
(\om,0)+2\,\delta{\Pi}_{n\Sigma^-}^{({\rm reg},0)}(\om,0),
\label{dPi}
\ee
with
\be\label{dPi1}
&&\delta {\Pi}_{iH}^{({\rm reg}, 0)}(\om,0)=
-C_{KNH}^2\,\times \intop \frac{2{\rm d}^3\vec p}{(2\,\pi)^3}
\frac{(m_H^*\,\sqrt{s_0} -m_N^{*\,2}+\om^2)
(m_H^*+m_N^*)^2}
{2\,\epsilon_i(p)\sqrt{s_0}\,(\sqrt{s_0}+m_H^* )}
n_i (\vec{p})\,,
\ee
where $s_0 =(\om +\epsilon_i (\vec{p}))^2 -\vec{p}^2$, and
we also included dependence of effective masses on the mean field.
For $T=0$ and for small kaon energies, the  integral (\ref{dPi1}) can be well
approximated by the following expression
\be\label{dPiapp}
\delta {\Pi}_{iH}^{({\rm reg}, 0)}(\om,0)\approx  -C_{KNH}^2\,
\left(\frac{m_H^{*2}-m_N^{*2}}{2\, m_N^*}\, \r^{\rm scal}_{i}+\om
\,
\r_i-\frac{\om^2}{m_H^*+m_N^*}\, \r_i\right)\,,
\ee
where $\r^{\rm scal}_{i}$ stands for the scalar density of nucleons
defined by
\be
\r^{\rm scal}_{i}
=\intop_0^{p_{{\rm F}, i}}\frac{2\,{\rm d}^3\vec p}{(2\,\pi)^3}
\frac{m_N^*}{\epsilon_i(\vec p)}\,.
\ee
According to (\ref{pip}), (\ref{lind}), the last term in (\ref{Ss}) can be
cast as
\be \label{pols}
\Pi^{({\rm pole}, 0)}(\om,0)
&=&{\Pi}_{p\Lambda}^{({\rm pole}, 0)}(\om,0)+
{\Pi}_{p\Sigma^0}^{({\rm pole}, 0)}(\om,0)+2\,{\Pi}_{n\Sigma^-}^{({\rm pole},
0)}(\om,0),
\\ \nonumber
\Pi^{({\rm pole}, 0)}_{ iH}(\om,0)&=&C_{KNH}^2\,\Big(
(m_H^*-m_N^*)^2-(\om+\delta v_{ iH})^2\Big)\,
\\ \nonumber
&\times&
\frac{(m_N^*+m_H^*)^2}{2\, \pi^2}\, \intop
\frac{{\rm d} p p^2}{\epsilon_{ i}(p)[\Delta_{ iH}^+(\om,0,\vec{p})+i0]}
n_i (\vec{p}),
\ee
that follows from (\ref{lind}) at $| \vec{k}| \rightarrow 0$.
The imaginary part, $\Im\Pi^{({\rm pole},0)}_{iH}(\om,0)$, is given by
\be \nonumber
&&\Im\Pi^{({\rm pole}, 0)}_{ iH}(\om,0)=-i
C_{KNH}^2\,\left((m_H^*-m_N^*)^2-(\om+\delta v_{iH})^2\right)\,
\\\label{ims}
&&\times \frac{(m_N^*+m_H^* )^2}
{8\pi (\om +\delta v_{iH})^2 }
\sqrt{m_H^{*2} - (m_N^* +\om +\delta v_{iH})^2}\,,
\ee\ewt
being non-zero for energies
$\om_{iH}^-(0)=E_H(p_{{\rm F}i})-E_i(p_{{\rm F}i}) <
\om <\om_{iH}^+(0)=E_H(0)-E_i(0) $\,.

\subsection{Energy of the Lowest Branch of the
Dispersion Equation at $\vec k\, =0$}\label{min1}

In this sub-section we illustrate strength of
different terms in (\ref{Ss}) applying it to the problem of the
s-wave kaon condensation.

The neutron star matter becomes unstable with respect to reactions
(\ref{react}) with the production of the zero-momentum $K^-$
meson, when solution ($\om_{\rm S}=\om^{\rm min}(\vec{k})$ at
$\vec{k}=0$) of the dispersion equation \be\label{swde} \om_{\rm
S}^2 -m_K^2-\Re\Pi_{\rm S}(\om_{\rm S})=0\, \ee related to the
lowest branch of the spectrum, meets the electron chemical
potential. Then the s-wave $K^-$ condensation may occur by the
second-order phase transition.

\begin{figure*}
\includegraphics[clip=true,width=13cm]{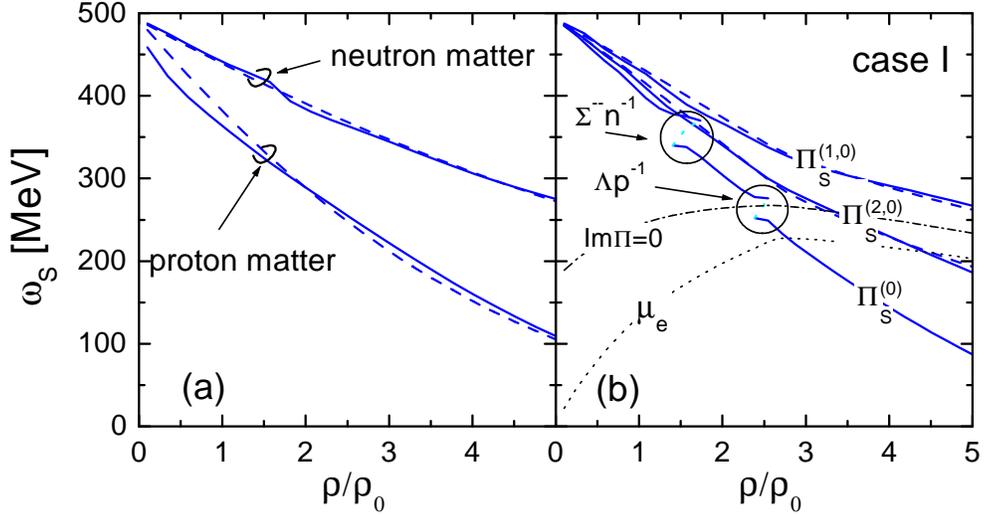}
\caption{Panel (a): The energy  of the lowest $K^-$ branch of the
dispersion equation (\protect\ref{swde})  at $\vec{k}=0$
calculated with $\Pi_{\rm S}^{(1,0)}$ (solid lines) for proton and
neutron matter. Dash lines present approximate spectra given by
(\protect\ref{spec-chipt}). Plane (b): The energy  of the lowest
$K^-$ branch of (\protect\ref{swde})  at $\vec{k}=0$ for real
neutron star matter  (case I of the hyperon-nucleon interaction)
for the s-wave polarization operators $\Pi_{\rm S}^{(1,0)},
\Pi_{\rm S}^{(2,0)}, \Pi_{\rm S}^{(0)} $ given by
(\protect\ref{tps1},\protect\ref{tps2},\protect\ref{tps3}) (solid
lines). Dash lines are solutions obtained with approximate
relations (\protect\ref{dPiapp},\protect\ref{spol-chipt}). Dotted
line shows the electron chemical potential. Dash-dotted line
depicts the border of the imaginary part of the $K^-$ polarization
operator (border of hyperonization). Short-dashes show the
boarders of the regions of non-physical solutions of
(\protect\ref{swde}) characterizing by negative sign residues. }
\label{fig:swave}
\end{figure*}

In Fig.~\ref{fig:swave} we present the  energy at the lowest $K^-$
branch
of the dispersion
equation~(\ref{swde}),
as function of the density
at the momentum $\vec{k}=0$ \footnote{Note that it is not
necessarily one and the same branch at different densities.}.
The hyperon interaction is
taken according to case I.
In order to  illustrate the strength of different
contributions to the s-wave part of the polarization  operator (\ref{Ss})
we consider several test polarization operators
\be\label{tps1}
\Pi_{\rm S}^{(1,0)}(\om) &=& {\rm I}_{\rm s-wave}(\om,0)+
\bar{\rm I}_{\rm p-wave}^{\rm reg}(\om,0),
\\ \label{tps2}
\Pi_{\rm S}^{(2,0)}(\om) &=&\Pi_{\rm S}^{(1,0)}(\om)+
\delta \Pi^{(\rm reg,0)}(\om),
\\ \label{tps3}
\Pi_{\rm S}^{(0)} (\om)&=&\Pi_{\rm S}^{(2,0)}(\om)
+ \Pi^{({\rm pole},0)}(\om,0).
\ee
In panel (a) of Fig.~\ref{fig:swave} solid lines show the  energy
of the lowest
branch
solution of (\ref{swde})
with $\Pi_{\rm S}^{(1,0)}$ for the cases of pure proton and neutron matter.
The contribution of $\bar{{\rm I}}_{\rm p-wave}^{\rm reg}(\om,0)$ is found to be
very small, at the level of few percent.
It is instructive to compare this our
result with that given by the frequently
used parameterization of the $K^-$ spectrum motivated by the
leading-order chiral perturbation theory ($\chi{\rm PT}$)
expansion of $K^- N$ interaction, cf. ~\cite{swave-1},
\be\nonumber
&&\om_{\rm S}^{\chi{\rm PT}}(\r_n,\r_p)
=\sqrt{m_K^2-S_K+V_K^2}-V_K\,,
\\ \nonumber
&&S_K=\frac{1}{f^2}\left(\Sigma_{KN}
\,(\r^{\rm scal}_{p}+\r^{\rm scal}_{n})+C\,
(\r^{\rm scal}_{p}-\r^{\rm scal}_{n})\right)\,,\,\,\,\,
\\ \label{spec-chipt}
&&V_K=\frac{(2\,\r_p+\r_n)}{4\,f^2}\,,
\ee
where $f\simeq 90$~MeV is the pion decay constant in the chiral
limit~\cite{lk01}, and $\Sigma_{KN}$ and $C$ stand for the isoscalar and
isovector
kaon-nucleon $\Sigma$-terms related to the explicit chiral symmetry breaking.
The SU(3) symmetry predicts $C=m_K^2\,(2\,m_\Xi-3\, m_\Sigma+m_\Lambda)/
(16\, (m_K^2-m_\pi^2))\approx 66$~MeV, cf.~\cite{swave}.
The model polarization operator leading to the dispersion
relation~(\ref{spec-chipt}) can be written as, cf. \cite{swave-1},
\be\nonumber
\Pi_{\rm S}^{(\chi{\rm PT},0)}(\om)&=&-\frac{\Sigma_{KN}}{f^2}
\,(\r^{\rm scal}_{p}+\r^{\rm scal}_{n})-
\frac{C}{f^2}\, (\r^{\rm scal}_{p}-\r^{\rm scal}_{n})
\\ \label{spol-chipt} &-&\frac{2\,
\r_p+\r_n}{2\, f^2}\, \om\,.
\ee
Spectrum (\ref{spec-chipt}), calculated using
$\Sigma_{KN}=150$~MeV, is shown in Fig.~\ref{fig:swave} (panel
(a)) by the dash lines. We observe a good agreement of the model
spectrum with that follows from numerical evaluation of the
integrals $\Big[{\rm I}_{\rm s-wave}^{}(\om,0)+\bar{{\rm I}}_{\rm
p-wave}^{\rm reg}(\om,0)\Big]$. Note, that obtained value of the
effective kaon-nucleon $\Sigma$-term is 2--3 times smaller than
that used as an ad-hoc parameter in Ref.~\cite{swave-1}, where the
same  parameterization (\ref{spol-chipt}) has been exploited.

The results for the realistic  composition of neutron star matter,
shown in Fig.~\ref{fig:compos}, are presented in panel (b) of
Fig.~\ref{fig:swave} for case I of the hyperon-nucleon interaction
as a representative example. The solid lines depict the
energy at the lowest
branch of the dispersion equation calculated for
$\vec{k}=0$ with
$\Pi^{(1,0)}_{\rm S}$, $\Pi^{(2,0)}_{\rm S}$ and  $\Pi_{\rm
S}^{(0)}$. Dash lines show solutions obtained with the approximate
expressions (\ref{spol-chipt}) in $\Pi_{\rm S}^{(1,0)}$ and
(\ref{dPiapp}) for $\delta \Pi^{(\rm reg,0)}$ in $\Pi_{\rm
S}^{(2,0)}$. Excellent coincidence of the curves  justifies the
accuracy of (\ref{spol-chipt}) and (\ref{dPiapp}).

Crossing point of the solid and dotted lines corresponds to the
critical density of the s-wave condensation for the case of a
realistic neutron star composition. We observe that the lines
corresponding to $\Pi_{\rm S}^{(1,0)}$ do not meet the chemical
potential (dotted line). Therefore, the s-wave kaon-nucleon
interaction, following from Ref.~\cite{lk01},  would  not support
a second-order phase transition into the s-wave $K^-$ condensate
state due to the  small value of the kaon-nucleon sigma term
following from the analysis ~\cite{lk01}. However, an additional
attraction comes from the term $\delta{\rm I}_{\rm p-wave}^{\rm
(reg)}$ included in $\Pi_{\rm S}^{(2,0)}$. It makes the reactions
(\ref{react})  possible at  density $\geq 4.5\ro$. Another
attractive piece is the pole term ${\rm I}_{\rm p-wave}^{\rm
pole}(\om,0)$ taken into account in $\Pi_{\rm S}^{(0)}$. The
significance of these  terms, originating both from the hyperon
exchange diagram in $\overline{K}N$ interaction, was pointed in
Ref.~\cite{kvk95}. Both mentioned contributions were, however,
disregarded in works~\cite{swave-1,swave-2,bb,nscool,tpl94,gs,pw}
discussing s-wave $K^-$ condensation.

The curves  $\omega_{\rm S}$  calculated  with the full s-wave
polarization operator $\Pi_{\rm S}^{(0)}$ have cuts. In the region
between the cuts equation (\ref{swde}) has no solutions with
positive residues, cf. \cite{kvk95}. The dash-dotted line depicts
the border of the imaginary part of the $K^-$ polarization
operator. We see that, fortunately, within our approach the curve
calculated with $\Pi_{\rm S}^{(0)}$ and  $\mu_e$ meet at energy
below  the region of the imaginary part. Thus, with the full
polarization operator (\ref{Ss}), (\ref{tps3}) we recovered
statement of previous
works~\cite{swave-1,swave-2,bb,nscool,tpl94,gs,pw} (where,
however, the $2\div 3$ times larger $\Sigma$ term was used) on the
possibility of the $K^-$ condensate production in reaction
(\ref{react}) at rather moderate densities, $\rho_{\rm c,S}\simeq
2.7\rho_0$ in our case. The reader should bear in mind that
baryon-baryon correlations are  still not included in the above
analysis. They will increase $\rho_{\rm c, S}$. This issue will be
addressed in section~\ref{sec:Short}.

\section{Contributions of the Hyperon Fermi-seas to the
Polarization Operator}\label{sec:Hyperon}

When nucleon density exceeds the critical density of hyperonization
$\rho_{{\rm c},H}$, the  Fermi sea of the hyperon $H$ begins to grow, and the
$K^-$  polarization operator receives new contributions.

\subsection{Pole Terms}

New contributions  to the $K^-$ polarization operator relate to
the  diagrams (in Schr\"odinger picture)
\be
\parbox{3cm}{
\includegraphics[clip=true,width=3cm]{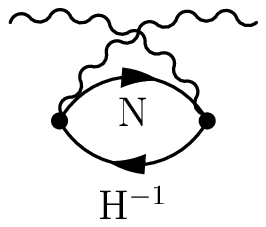}
}\,,
\ee
where the hyperon and the nucleon interchange their  roles compared to the
above discussed the hyperon--nucleon-hole  $NH$ terms.
At finite temperatures these diagrams contribute for all densities. But this
contribution
is exponentially suppressed  for $\rho < \rho_{{\rm c},H}$
and small temperatures.

Account of the hyperon contribution to the pole part of the
polarization operator $\Pi^{({\rm pole}, 0)}$
is simply done with the help of the replacement in (\ref{pip})
\be\label{hyp}
\eta^2_{iH}\Phi_{iH} (\om , \vec{k})\rightarrow
\eta^2_{NH}\Phi_{iH} (\om , \vec{k})
+\eta^2_{HN}\Phi_{Hi} (-\om , -\vec{k})\,,
\ee
where the last term implies interchange of all indices $i\leftrightarrow H$ in
(\ref{li}), (\ref{lind}), (\ref{lind-ner}).
This replacement results in appearance of an extra term
\be\nonumber
\delta\Pi_{\rm hyp}^{(\rm pole,0)} (\om , \vec{k})&=&
\Pi_{\Lambda p}^{(\rm pole,0)}(-\om , -\vec{k})
+\Pi_{\Sigma^0 p}^{(\rm pole,0)}(-\om , -\vec{k})
\\ \label{pip-h}
&+&2\,\Pi_{\Sigma^-  n}^{(\rm pole,0)}(-\om , -\vec{k}),
\ee
to be added to the total polarization operator.
This term depends on  the hyperon density and contributes
only at densities $\r>\r_{{\rm c},H}$.

\subsection{Regular Terms}

There are no any
experimental constraints  on the hyperon contribution
to the regular part of the polarization operator so far.
As a rough estimation we may suggest an extension of the model
polarization operator (\ref{spol-chipt}) to the hyperon sector according
to the leading order terms of the chiral Lagrangian
\be\nonumber
\delta \Pi_{{\rm S,hyp}}^{(\chi{\rm PT}, 0)}(\om)&=&
-\frac{\Sigma_{KN}}{f^2}\,(\r^{\rm scal}_{\Lambda}+\r^{\rm scal}_{\Sigma^{-}}
+\r^{\rm scal}_{\Xi^-})
\\ \nonumber
&-&\frac{C}{f^2}\, (\frac13\, \r^{\rm scal}_{\Lambda}-\r^{\rm scal}_{\Sigma^-}
+\r^{\rm scal}_{\Xi^-})
\\ \label{spolh}
&+&\frac{C_\Lambda}{f^2}\,\r^{\rm scal}_{\Lambda}+
\frac{\r_{\Sigma^-}+2\, \r_{\Xi^-}}{2\,f^2}\,\om . \ee In this
expression we utilize the value of $\Sigma_{KN}$ from the fit with
the formula (\ref{spec-chipt}) to the numerical results in
Fig.~\ref{fig:swave} ($\Sigma_{KN}\simeq 150$~MeV), whereas the
values of coefficients $C\approx 66$~MeV  and
$C_\Lambda=m_K^2\,(m_\Xi-m_\Lambda)/(12\, (m_K^2-m_\pi^2))\approx
34$~MeV and  $f\simeq 90$~MeV are predicted by the chiral SU(3)
symmetry. To estimate the non-pole contribution from the nucleon
u-channel exchange (analogous to $\delta\Pi^{(\rm reg, 0) }$) we
use approximate relations (\ref{dPiapp}) \be \nonumber &&\delta
\Pi_{\rm S,hyp}^{({\rm reg},0)}(\om)=-C_{KN\Lambda}^2\,
\\ \nonumber
&&\times
\left[\frac{m_N^{*2}-m_\Lambda^{*2}}{2\, m_\Lambda^*}\, \r^{\rm scal}_{\Lambda}
-\om\,\r_\Lambda-\frac{\om^2}{m_\Lambda^*+m_N^*}\, \r_\Lambda\right]
\\\nonumber
 &&\qquad\qquad\qquad\,\,-2\,C_{KN\Sigma}^2
\\ \label{dPisH}
&&\times \left[\frac{m_N^{*2}-m_\Sigma^{*2}}{2\, m_\Sigma^*}\,
\r^{\rm scal}_{\Sigma^-}
-\om\,\r_{\Sigma^-}-\frac{\om^2}{m_\Sigma^*+m_N^*}\,
\r_{\Sigma^-}\right] \,. \ee With this estimation we do not take
into account contributions to the regular p-wave part of the
polarization operator $\propto \vec{k}^2\, \rho_{H}$.

\subsection{Energy of the Lowest Branch of the Dispersion Equation at $\vec{k}=0$ }
\label{min2}

\begin{figure}
\includegraphics[clip=true,width=7cm]{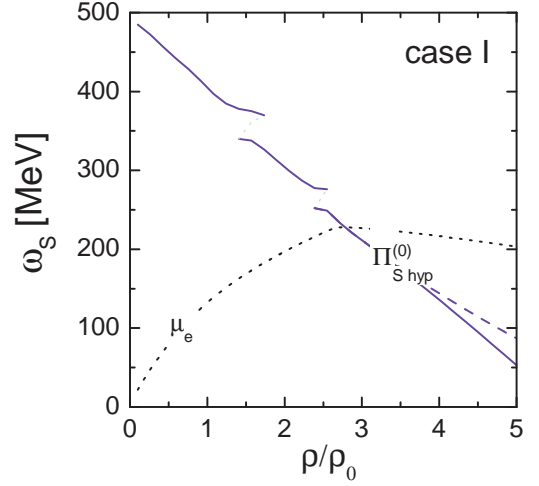}
\caption{The  energy of the lowest branch of the dispersion
equation at $\vec{k}=0$ calculated with the polarization operator
$\Pi_{\rm S, hyp}^{(0)}$ given by (\protect\ref{tps3h}) is
depicted by solid lines. Dashed line shows the  corresponding
solution with the polarization operator (\protect\ref{tps3}). The
dotted line depicts the electron chemical potential.
}\label{fig:swave-h}
\end{figure}

Solutions  related to the
lowest branch of the dispersion equation (\ref{swde})
for $\vec{k}=0$ calculated
with the
polarization operator
\be\nonumber
\Pi_{{\rm S},{\rm hyp}}^{(0)} (\om) &=&
\Pi_{\rm S}^{(0)}(\om) + \delta \Pi_{{\rm S},{\rm hyp}}^{(0)} (\om),
\\ \nonumber
\delta \Pi_{{\rm S},{\rm hyp}}^{(0)} (\om) &=&
 \delta\Pi_{{\rm S},{\rm hyp}}^{(\chi{\rm PT},0)}(\om)
+\delta\Pi^{({\rm reg}, 0)}_{\rm  hyp }(\om)
\\ \label{tps3h}
&+&\delta\Pi^{({\rm
pole}, 0)}_{\rm  hyp}(\om , 0)\,
\ee
are shown in
Fig.~\ref{fig:swave-h} by solid lines in comparison with
corresponding solutions obtained  without inclusion of hyperons
(dash lines). Calculations are done for hyperon coupling constants
corresponding to case I. We see that presence of hyperons produces
an additional small attraction only at rather high densities
($>4\rho_0$). The reasons are partial cancellation of the
attractive $\Pi_{{\rm S},{\rm hyp}}^{(\chi{\rm PT},0)}$ term and
the repulsive $\delta\Pi^{(\rm reg, 0)}_{\rm hyp}$ term and that
in  the framework of our model for description of the neutron star
matter hyperon concentrations are much smaller than the neutron
concentration and even smaller than the proton one. Thus,
population of the hyperon Fermi seas only slightly affects the
s-wave part of the polarization operator.

This allows us not to care much about the hyperon Fermi sea occupations
considering $\vec{k}=0$ case.

\section{Baryon-Baryon  Correlations}\label{sec:Short}

Operating with the polarization operator constructed by
integrating the meson-nucleon scattering amplitude over the
nucleon Fermi sea, e.g., as in (\ref{Is},\ref{Ip}), one assumes
that all multiple meson-nucleon interactions are independent from each other and
have the same probability proportional to the nucleon local
density $\rho(\vec{r\,})$. However the successive meson-nucleon
scatterings in dense nuclear matter are not independent because of
the core of nucleon-nucleon interactions and the Pauli exclusion
principle \cite{Bruk,jb78}. The probability to find two nucleons $i$
and $ i'$ at the positions $\vec{r}_1$ and $\vec{r}_2$,
respectively, is proportional  to the two-particle density
$$\r_{ii'}(\vec{r}_1,\vec{r}_2)=
[1+C_{ii'}(|\vec{r}_1-\vec{r}_2\,|)]\,\r_i(\vec{r}_1\,)\,\r_{i'}(\vec{r}_2\,)$$
with the correlation function $C_{ii'}(r)<0$ and is, therefore,
reduced in comparison  to the product of two single particle
densities. The correlation function can be approximately written as
\be\label{c-c}
C_{ii'}(r) \approx C^{\rm core}(r)+\delta_{ii'}\,C^{\rm Pauli}_i (r)\,
[1+C^{\rm
core}(r)]
\ee%
with  contributions from the hard core, $C^{\rm core}$, and the
Pauli exclusion principle, $C^{\rm Pauli}$, assuming that both
correlations contribute multiplicatively. The former can be taken
from  the description of nuclear matter with the realistic
nucleon-nucleon interaction. The convenient parameterization  was
suggested in Ref.~\cite{bbow}, $C^{\rm core}(r)\approx -j_0(m_0\,
r)$ with $m_0\approx 5.6 m_\pi$, where $j_l(x)$ stands for the
spherical Bessel function. For the Pauli correlation we use
expression for the ideal fermion gas~\cite{FW},
$C^{\rm Pauli}_i(r)=-9\, j_1^2(p_{{\rm F} i}r)/(2\,p_{{\rm F} i}^2 r^2)$\,.

\subsection{Correction of s-wave and Regular p-wave Terms}

General derivation of the corrections to the meson propagation in
dense nuclear matter due to the nucleon-nucleon correlations
(so-called Ericson-Ericson-Lorentz-Lorenz corrections) can be
found in Ref.~\cite{ee66,f73} for pions.
In Ref.~\cite{corr-pand,wrw97} it was extended for kaons.

In  diagrams, correlation processes can be  presented by symbolic equation
\be \label{reg-cor}
\parbox{6cm}{
\includegraphics[width=6cm,clip=true]{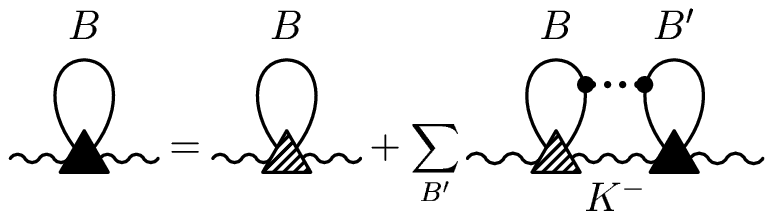}}\quad ,
\ee where the wavy line relates to the kaon, the sum goes over the
baryon species. The line without arrow means that both particles
and holes are treated on equal footing (the conservation of
charges, e.g. strangeness, baryonic number etc., in each vertex is
implied). The hatched triangle is the bare scattering amplitude
(scattering on the particle or on the hole) and the full triangle
stands for the amplitude including   baryon-baryon correlations.
The dotted line symbolically depicts the two-baryon correlation
function $C_{BB'}$ due to the  $BB'$ correlations through the core
and the Pauli principle. There are nor experimental information
neither theoretical estimations for the hyperon-nucleon and
hyperon-hyperon correlations. Since the latter ones are less
relevant for our discussion below we will neglect them. Thus we
actually include only minimal correlations given by $C_{ii'}$
nucleon-nucleon correlation functions.

We consider first the correlation corrections to the s-wave part
of the kaon polarization operator,
$\Pi_{\rm S}^{(0)}$ given by (\ref{Ss}), and the regular p-wave part ,
$\Pi_{\rm P}^{({\rm reg}, 0)}$ from
(\ref{pireg}).
We separate contributions induced by the
scattering on a nucleon of a given isospin species $i=\{n,p\}$.
\be \nonumber
\Pi_{\rm S }^{(0)}(\om)&=&
\Pi_{{\rm S},n}^{(0)}(\om)+\Pi_{{\rm S},p}^{(0)}(\om)\,,
\\ \nonumber
\Pi_{{\rm S},i}^{(0)}(\om) &=& {\rm I}_{{\rm s-wave}
,i}(\om,0)+{\rm I}_{{\rm p-wave},i}(\om,0), \ee cf.
(\ref{Is},\ref{Ip}). Then, adopting results of
Refs.~\cite{w77,wrw97} we may present the polarization operator
terms corrected by baryon-baryon correlations as
\bwt\be\label{chicorS}
&&\Pi_{\rm S}(\om)= \frac{\widetilde{\Pi}_{{\rm S},n}(\om) +
\widetilde{\Pi}_{{\rm S},p}(\om) + 2\, \widetilde{\Pi}_{{\rm
S},n}(\om)\, \widetilde{\Pi}_{{\rm S},p}(\om)\,
\xi_{pn}^{\rm(S)}(\om) } {1-\widetilde\Pi_{{\rm S},n}(\om)\,
\widetilde \Pi_{{\rm S},p}(\om) \,(\xi_{pn}^{\rm (S)}(\om))^2 }\,,
\qquad
\widetilde{{\Pi}}_{{\rm S},i}(\om)=\frac{{\Pi}_{{\rm S},i}^{(0)}(\om)}
{1-\xi_{ii}^{\rm(S)}(\om)\,{\Pi}_{{\rm S},i}^{(0)}(\om)}\,,
\\  \label{chicorP}
&&\Pi_{\rm P}^{\rm reg}(\om,\vec k)=
\vec{k}^2\,\frac{\widetilde{b}_p(\om)
+\widetilde{b}_n(\om)\, +2\,
\widetilde{b}_p(\om)\,\widetilde{b}_n(\om)
 \xi^{\rm (P)}_{np}(\om)} {1-
 \widetilde{b}_p (\om)\,\widetilde{b}_n(\om)
(\xi_{np}^{\rm (P)}(\om))^2}\,, \qquad \widetilde{b}_i(\om) =
\frac{b_i(\om) \,(\r_i/\r_0)}{1-b_i(\om)\,\r_i\, \xi_{ii}^{\rm
(P)}(\om)/\r_0}\,. \ee\ewt Functions, $\xi_{ii'}^{\rm (S)}$ and
$\xi_{ii'}^{\rm (P)}$ are defined as, cf.~Ref.~\cite{wrw97},
\be\label{xiab} \xi_{ii'}^{\rm (S)}(\om) &=& \intop {\rm d}^3 r
 C_{ii'}(r)\, D_K^0 (\om,r)
\,,
\\ \nonumber
\xi_{ii'}^{\rm (P)}(\om) &=& \frac13\, \intop {\rm d}^3 r
 C_{ii'}(r)\, \nabla^2\, D_K^0 (\om,r)
 \\ \label{xiabP}
 &=& -\frac13\, \Big( C_{ii'}(0)-(m_K^2-\om^2)\, \xi_{ii'}^{\rm (
 S)}(\om)\Big)
\,,\ee
containing Pauli and core contributions following  (\ref{c-c}),
$i,i' =\{n, p \}$, and
$D_K^0 (\om,r)=-\exp(-\sqrt{m_K^2-\om^2}\, r)/(4\,\pi\, r)$, as the free
kaon propagator in the "mixed" representation.

\begin{figure*}
\includegraphics[width=7cm,clip=true]{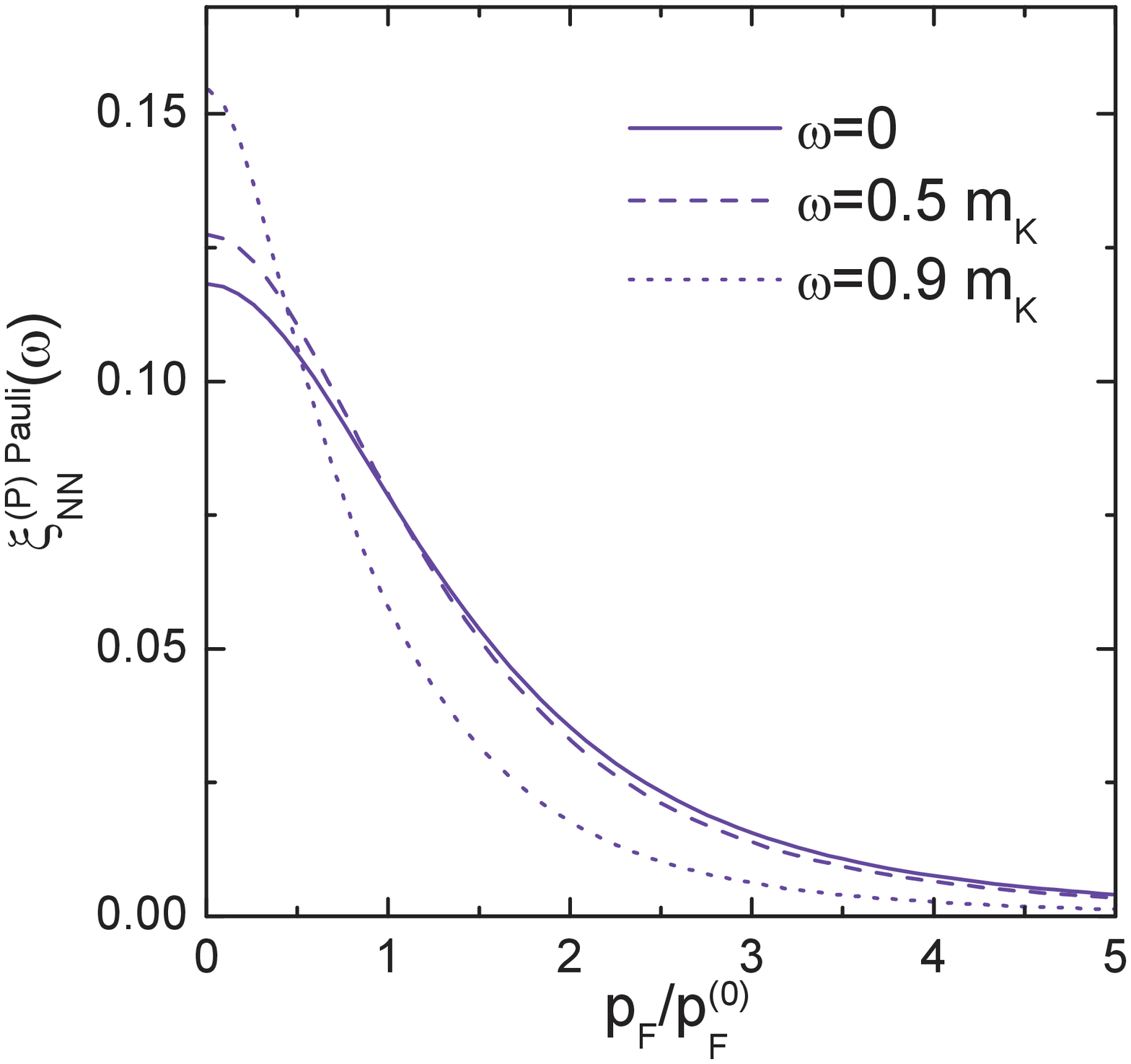}\qquad
\includegraphics[width=7cm,clip=true]{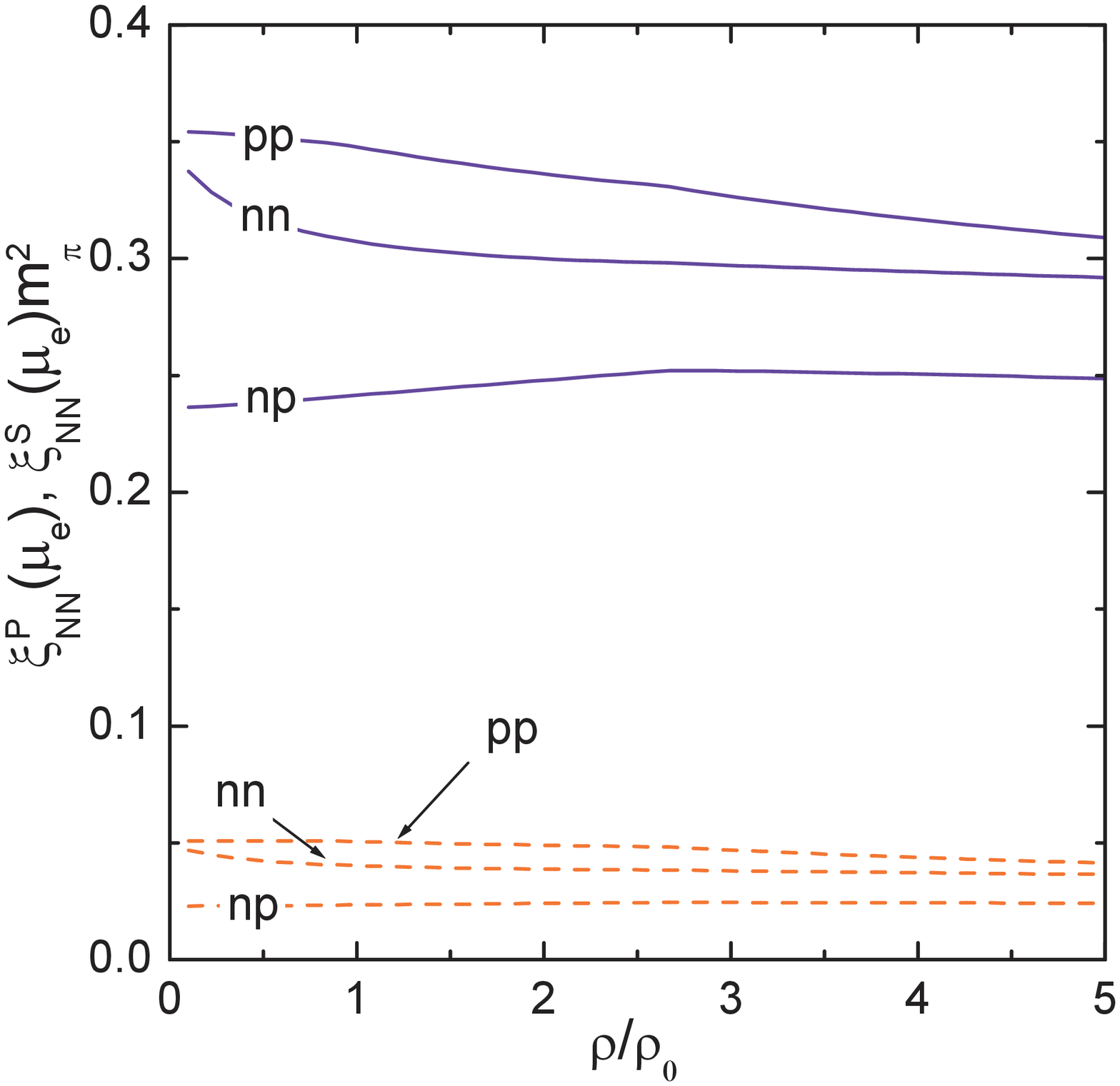}
\caption{Left panel: contribution
of the Pauli spin-correlations  to the
p-wave correlation
function (\protect\ref{xiabP}) vs. the Fermi momentum for different
kaon energies, $p_{\rm F}^{(0)}$ is the Fermi momentum  for $\rho=\rho_0$.
Right Panel: the s-wave (dash lines)  and p-wave (solid lines)
correlation functions
(\protect\ref{xiab},\protect\ref{xiabP})
evaluated for $\om=\mu_e$ in neutron star matter (case I) as function
of the total
baryon density.}
\label{fig:xipauli}
\end{figure*}

Using (\ref{reg-cor}) one  finds that the repulsive core
contributes with \be\label{kpc} \xi_{ii'}^{\rm (P,\, core)}
=\xi_K^{\rm (P,\,
core)}(\om)=\frac13\,\frac{m_0^2}{(m_0^2+m_K^2-\om^2)} \ee to the
p-wave $\xi$'s and with \be \xi_{ii'}^{\rm (S,\, core)}
=\xi_K^{\rm (S,\, core)}(\om)=\frac{1}{(m_0^2+m_K^2-\om^2)} \ee to
the s-wave ones. The contribution from Pauli spin-correlation
(second term in $C_{ii'}$) to the p-wave correlation function
$\xi^{\rm (P)}_{ii}$ is shown in Fig.~\ref{fig:xipauli} (left
panel) as a function of the Fermi momentum for different kaon
energies. We see that the correlation parameter decreases with
density since the baryon-baryon correlations hold baryons  apart
from each other and suppress, thereby, effect of the Pauli
exclusion principle. The right panel of Fig.~\ref{fig:xipauli}
presents the values of the correlation function (\ref{xiab},
\ref{xiabP}) calculated for $\om=\mu_e$ in the neutron star matter
with the hyperon coupling constants according to case~I.

We leave the contributions from the hyperon Fermi seas to the
regular part of the polarization operator, $\delta \Pi_{\rm S,
hyp}^{(0)}$, without corrections due to the baryon-baryon
correlations in view of their small contributions.

\subsection{Correction of p-wave Pole Terms}

We turn now to consideration of effect of correlations in the
particle-hole channel.

If we approximate the free $\overline{K}N$ scattering amplitude
(hatched triangle) in (\ref{reg-cor}) by the hyperon-exchange
diagram, the same one, which produces the  particle-hole diagrams
(\ref{pinh}),  we can see that the account of correlations via
(\ref{reg-cor}) is equivalent  to the replacement~\footnote{The
replacement (\protect\ref{loopmod},\protect\ref{verteq}) can be
explicitly proven in the non-relativistic limit. Working with
relativistic kinematics, we apply it only to the pole part of the
diagram (\ref{pinh}) written in terms of the Lindhard's function
(\ref{li}). Thereby, we preserve the correct transition to the
non-relativistic limit. } \be\label{loopmod}
\parbox{6cm}{
\includegraphics[width=6cm,clip=true]{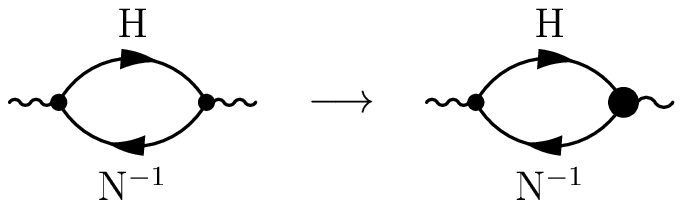}}
\ee
with a modified vertex (fat point) obeying  equation
\be\label{verteq}
\parbox{7.7cm}{
\includegraphics[clip=true,width=7.7cm]{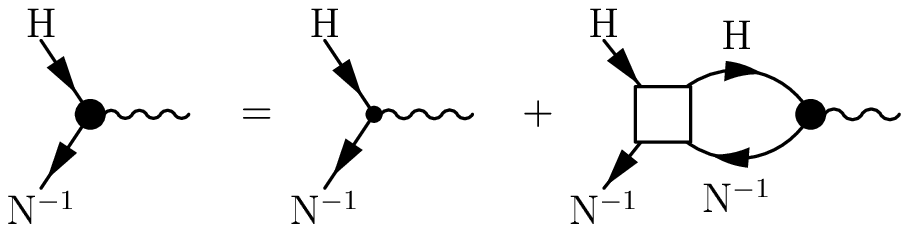}}\,,
\ee
where the particle-hole irreducible box ${T}_{ HN^{\prime}}^{\rm
loc}$ (light square) can be expressed in terms of $\xi^{\rm P}$ and
kaon-nucleon-hyperon coupling constants.

Below we would like to put the discussion on a more
phenomenological level. According to the argumentation of the
Fermi-liquid theory~\cite{tkfs}, the particle-hole irreducible box
${T}_{ HN^{\prime}}^{\rm loc}$ has a weak dependence on incoming
energies and momenta  and can be, therefore, parameterized in
terms of phenomenological Landau-Migdal parameters:
\bwt\be\nonumber
{T}_{\Lambda { N}}^{\rm loc} &=&
\parbox{2cm}{
\includegraphics[width=2cm,clip=true]{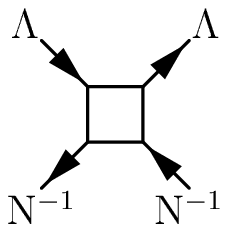}}
= C_0\,f'_\Lambda\,s_{12}\,,
\qquad\qquad
{T}_{\Sigma { N}}^{\rm loc} =
\parbox{2cm}{
\includegraphics[width=2cm,clip=true]{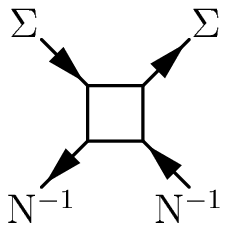}}
= C_0\,g'_\Sigma\,s_{12}\,{
P}_{\Sigma {\rm N}}^{(1/2)}\,,
\\  \label{tloc}
{T}_{\Lambda \Sigma^\alpha}^{\rm loc} &=&
\parbox{2cm}{
\includegraphics[width=2cm,clip=true]{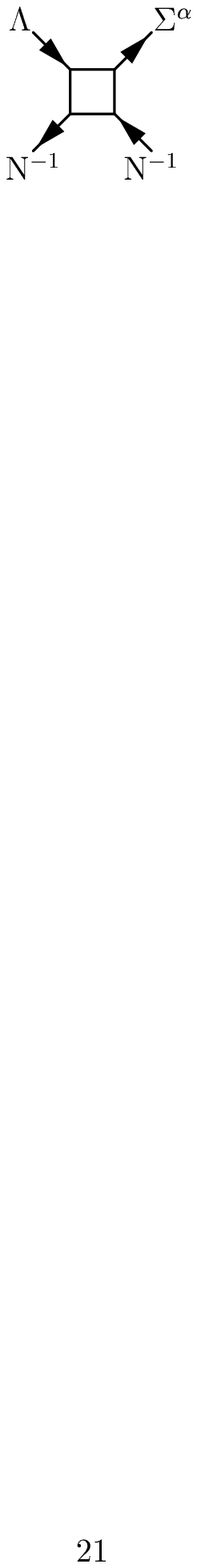}}
= C_0\,f'_{\Lambda\Sigma}\,s_{12}\,\tau^{\alpha}
 \,, \qquad
{T}_{\Sigma^{*\alpha}\rm N}^{\rm loc} =
\parbox{2cm}{
\includegraphics[width=2cm,clip=true]{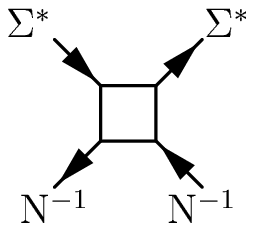}}
=C_0\, g'_{\Sigma^{*}}\, S_{12}\,{P}_{\Sigma^* {\rm N}}^{(1/2)}
\,,
\ee\ewt
with $\alpha =1,2,3$. The amplitudes are normalized with
 $C_0=300\, {\rm MeV}\cdot {\rm fm}^3$, that allows to compare the
 values for the hyperon-nucleon correlation parameters with those
 for the nucleon-nucleon correlations introduced in Ref.~\cite{MSTV90}.
In (\ref{tloc})
${P}_{\Sigma N}^{(1/2)} = (1-\vec{t}_\Sigma\cdot\vec{\tau})/3$
is the projection operator onto the given $\Sigma N$ state with
isospin $1/2$, $\vec{t}_\Sigma$ are the isospin-1 matrices and
$\vec{\tau}$ are the  Pauli matrices of the  nucleon isospin, as
above. The projector ${P}_{\Sigma^* {\rm N}}^{(1/2)}$ is defined analogously.
The spin-spin operators in the particle-hole channel are
given by  $s_{12}=(\vec{\sigma}_1 \vec{\sigma}_2)$, and
$S_{12}=(\vec{S}_1\vec{S_2}^\dagger)$, with $\vec{S}$
standing for  spin-operator, which  couples spin 1/2 and 3/2
states.

The inclusion of correlations according to (\ref{verteq}) brings
the pole polarization operator (\ref{pi-sum}) into the form
\bwt\be\nonumber {\Pi}_{\rm }^{\rm pole}(\om,\vec k)&=& \frac{
\widetilde{\Pi}^{\rm }_{p\Lambda}(\om,\vec k) +
\widetilde{\Pi}^{\rm }_{\Sigma}(\om,\vec k) + 2\,
c\,f'_{\Lambda\Sigma}\,\widetilde{\Pi}^{\rm }_{p\Lambda}(\om,\vec
k)\, \widetilde{\Pi}^{\rm }_{\Sigma}(\om,\vec k)/3}
{1-c^2\,f^{'2}_{\Lambda\Sigma}\,\widetilde{\Pi}^{\rm
}_{p\Lambda}(\om,\vec k)\, \widetilde{\Pi}^{\rm
}_{\Sigma}(\om,\vec k)/3}+{\Pi}^{\rm }_{\Sigma^*}(\om,\vec k)\,,
\\ \nonumber
\widetilde{\Pi}^{\rm }_{p\Lambda}(\om,\vec k) &=&
\frac{\Pi^{\rm (0)}_{p\Lambda}(\om,\vec k)}{1-f'_\Lambda\,
C_0\,\Phi_{p\Lambda}(\om,\vec k)}\,,\qquad
c=C_0/(C_{KN\Lambda}\, C_{KN\Sigma}),
\\ \nonumber
\widetilde{\Pi}^{\rm }_{\Sigma}(\om,\vec k) &=&
\frac{\Pi^{\rm (0)}_{p\Sigma^0}(\om,\vec k)
+2\,\Pi^{\rm (0)}_{n\Sigma^-}(\om,\vec k)}
{1-g'_\Sigma\, C_0\,(\Phi_{p\Sigma^0}(\om,\vec k)
+2\,\Phi_{n\Sigma^-}(\om,\vec k))/3}\,,
\\ \label{pi-pole-cor}
{\Pi}_{\Sigma^*}(\om,\vec k) &=&
\frac{\Pi^{\rm (0)}_{p\Sigma^{*0}}(\om,\vec k)
+2\,\Pi^{\rm (0)}_{n\Sigma^{*-}}(\om,\vec k)}
{1-g'_{\Sigma^*}\,
C_0\,(\Phi_{p\Sigma^{*0}}(\om,\vec k)+2\,\Phi_{n\Sigma^{*-}}(\om,\vec
k))/3}\,.
\ee\ewt

\subsection{Correlation Parameters}

To our best knowledge there is no direct experimental information
on the values of the Landau-Migdal parameters for the
hyperon-nucleon interactions $f'_\Lambda$,
$g'_{\Sigma(\Sigma^*)}$,  and $f'_{\Lambda\Sigma}$. In principle,
this information could be extracted from the data on multi-strange
hyper-nuclei, which, however, are rather poor, if not absent. In
the work \cite{kv98} the Landau-Migdal parameter
$f^{\prime}_{\Lambda}$ was estimated in line with
Refs.~\cite{correst}, where Landau-Migdal parameters of the
nucleon-nucleon interaction were calculated within
Ericson-Ericson-Lorentz-Lorenz approach. We will follow this
approach, estimating these parameters. Further corrections
can be computed, as in Ref.~\cite{bbow}.

Following~\cite{correst}  we assume that  the squared block in (\ref{tloc})
is determined by exchanges of the kaon and
the heavy strange vector meson $K^*$ with the mass
$m_{K^*}\simeq 892$~MeV. This can be shown  in diagrams as
\be\label{dia-tloc}
\parbox{7cm}{
\includegraphics[width=7cm,clip=true]{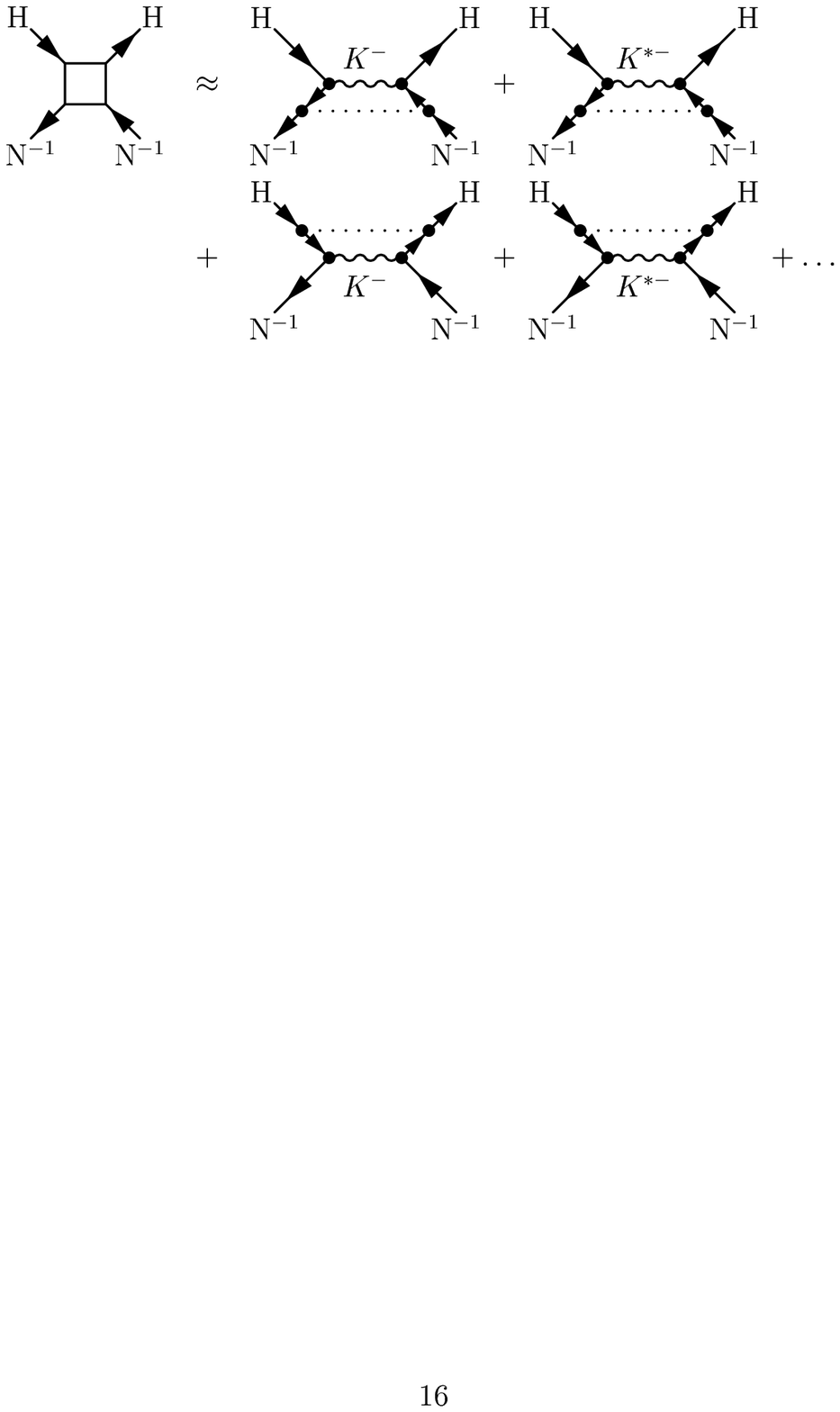}}\,\,.
\ee
Intermediate states in the processes (\ref{dia-tloc}),
(\ref{reg-cor}) involve large momenta suppressing medium effects.
In this approximation, correlation parameters are equal for
$HN^{-1}$ and $NH^{-1}$, for the nucleon and the hyperon of given
species. Then, including hyperon Fermi seas, the pole term of the
polarization operator is corrected with the help of the
replacement (\ref{hyp}) in (\ref{pi-pole-cor}). The block
(\ref{dia-tloc}), being evaluated at zero momentum and energy
transfer, contributes to the local interaction in (\ref{tloc}) as
\be \nonumber
T_{HN}^{\rm loc} &\approx&
C_{KNH}\, C_{KNH}\,\xi^{\rm (P,\, core)}_K(0)
\\ \label{correst}
&+& C_{K^* NH}\, C_{K^* NH}\, \xi^{\rm (P,\,core)}_{K^*}(0)\,,
\\ \nonumber
&& \qquad H=\Lambda, \Sigma, \Sigma^*\,,
\\ \nonumber
T_{\Lambda\Sigma}^{\rm loc}(\om) &\approx&
 C_{KN\Lambda}\, C_{KN'\Sigma}\,\xi^{\rm (P,\, core)}_K(0)
\\ \label{correst1}
&+& C_{K^* N\Lambda}\, C_{K^* N'\Sigma}\, \xi^{\rm (P,\,core)}_{K^*}(0)\,
\,.\ee

For shortness we do not write here explicitly the spin and isospin
operators which are exactly the same as in (\ref{tloc}).
The vector meson  coupling constants
$C_{K^*NH}$ in (\ref{correst}) correspond to the non-relativistic vertex
$\propto [\vec{\sigma}\times \vec{k}]$.
Coupling constants can be, e.g., taken  from the analysis of the
J\"ulich model
of the hyperon-nucleon interaction
via the meson exchange~\cite{Julich}:
$C_{K^*N\Lambda}=-\frac{1.3}{m_\pi}$\,,
$C_{K^*N\Sigma}=\frac{0.07}{ m_\pi}$\,, and
$C_{K^*N\Sigma^*}=\frac{0.7}{m_\pi}$\,.
These values account for the form-factors used in~\cite{Julich}.
In particular, the
form-factor related to a very soft energy range
is responsible for a strong suppression
of the $C_{K^*N\Sigma}$ vertex.

Thus, the Landau-Migdal  correlation parameters (\ref{tloc}) can be cast
as
\be\nonumber
{C_0}f'_\Lambda &=&{C_{KN\Lambda}^2}\,\left[\xi_K^{\rm (P, \, core)
}(0)
+R_{\Lambda\Lambda}\,\xi_{K^*}^{\rm (P, \, core)}(0)\right]\,,
\\ \nonumber
{C_0}g'_\Sigma &=& 3\, {C_{KN\Sigma}^2}\,\left[\xi_K^{\rm (P,\,
core)}(0)
+ R_{\Sigma\Sigma}\,\xi_{K^*}^{\rm (P, core)}(0)\right]\,,
\\ \nonumber
{C_0}f'_{\Lambda\Sigma} &=& {C_{KN\Lambda}\,C_{KN\Sigma}}\,\left[
\xi_K^{\rm(P,\, core)}(0)+R_{\Lambda\Sigma}\,\xi_{K^*}^{\rm (P,\,
core)}(0)\right]\,,
\\ \nonumber
{C_0}g'_{\Sigma^*} &=& 3\, {C_{KN\Sigma^*}^2}\,\left[\xi_K^{\rm (P),\,
core}(0)
+ R_{\Sigma^*\Sigma^*}\,\xi_{K^*}^{\rm (P),\, core}(0)\right)]\,,
\ee
where the first term was introduced in (\ref{kpc}),
$\xi_K^{\rm  (P,\,core)}(0)\simeq 0.24$,
the second one is equal to $\xi_{K^*}^{\rm (P,\, core)}(0)
= \frac23\,m_0^2/(m_0^2+m_{K^*}^{2})\simeq 0.28$, cf.~\cite{correst},
 $R_{HH'}= C_{K^*NH}\, C_{K^*NH'}/( C_{KNH}\, C_{KNH'})$ with
$R_{\Lambda\Lambda}\simeq 3.7$, $R_{\Sigma\Sigma}\simeq
0.04$, $R_{\Lambda\Sigma}\simeq 0.39$, and
$R_{\Sigma^*\Sigma^*}\simeq 0.69$.
Additional factor 2 in $\xi^{\rm (P,\, core)}_{K^*}$ compared to
$\xi^{\rm (P,\, core)}_{K}$,
originates from the  reduction $[\vec{\sigma}_2\times
\vec{k}]\,[\vec{\sigma}_1\times
\vec{k}]\to s_{12}$.
Finally, we estimate the following values of the  correlation parameters in
(\ref{tloc}) as
\be\label{corparH}
f'_\Lambda\simeq 0.9\,,\quad g'_\Sigma \simeq 0.1\,,\quad
f'_{\Lambda\Sigma}
\simeq -0.1\,,\quad g'_{\Sigma^*} \simeq  1.2 \,.
\ee
Compared to \cite{kv98}
we obtained a  smaller value of the corresponding parameter
$C_0 f'_\Lambda\simeq 0.6/m_\pi^2$,
since we included here the form-factors mentioned above
\footnote{Note that
different normalizations of Landau-Migdal parameters are used here and in
Ref.~\cite{kv98}.}.

\subsection{Energy of the Lowest Branch of the Dispersion Relation at
$\vec{k}\,=0$.}\label{min3}

\begin{figure}
\includegraphics[clip=true,width=7cm]{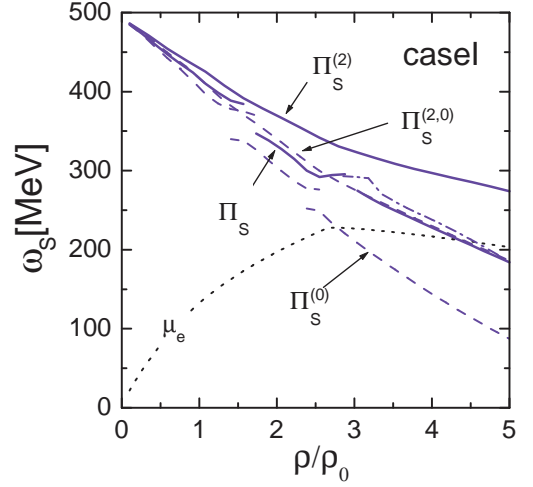}
\caption{ Solid lines present the energy of the lowest branch of
the dispersion equation at $\vec{k}=0$ calculated with the
polarization operators ${\Pi}_{{\rm S}}^{(2)}$ and ${\Pi}_{{\rm
S}}$ including effects of  baryon-baryon correlations. Dash-dotted
continuations of solid curves demonstrate effect of the filling of
the hyperon Fermi seas on the s-wave polarization operator.
Correlation parameters are taken according to
Fig.~\protect\ref{fig:xipauli} (right panel), (\protect\ref{xiab})
and (\protect\ref{corparH}). Dashed lines show  solutions
${\Pi}_{\rm S}^{(2,0)}$ and ${\Pi}_{\rm S}^{(0)}$ without
correlation effects, as in Fig~\protect\ref{fig:swave}, plane (b).
The dotted line depicts the electron chemical potential.
}\label{fig:swave-c}
\end{figure}

Fig.~\ref{fig:swave-c} illustrates how much baryon-baryon
correlations affect the terms of the bare polarization operator.
For case I we show the energy of the lowest  branch  of the
dispersion equation at $\vec{k}=0$ calculated with the
polarization operator ${\Pi}_{\rm S}^{(2)}$ constructed from
${\Pi}_{\rm S}^{(2,0)}$ according to (\ref{chicorS}) and the
polarization operator ${\Pi}_{\rm S}(\om)={{\Pi}}_{\rm
S}^{(2)}(\om)+ {\Pi}^{\rm pole}(\om,0)$, where ${\Pi}^{\rm
pole}(\om,0)$ follows from (\ref{pi-pole-cor}) with parameters
(\ref{corparH}).

At $\rho >\rho_{c,H}$ we have to include correlations in the term
$\delta\Pi_{\rm S, hyp}^{(0)}$ in (\ref{tps3h}). The pole term
$\delta{\Pi}^{(\rm pole ,0)}_{\rm hyp} (\om,0)$ is included in
(\ref{pi-pole-cor}) with the help of the replacement (\ref{hyp}).
The other terms $\delta \Pi_{\rm S, hyp}^{(\chi{\rm PT},0)}(\om)$
and $\delta{\Pi}_{{\rm S},h}^{(0)}$ can be corrected in the same
manner as the ${\Pi}_{{\rm S}}^{(0)}$ term. However these terms
are rather small as it is demonstrated by Fig.~\ref{fig:swave-h}.
Therefore we omit correlations in them  in further.

From Fig.~\ref{fig:swave-c} we see that baryon-baryon correlations
suppress the s-wave part of the polarization operator, in
agreement with the statements~\cite{corr-pand,wrw97}. It results
in an increase of the critical density of the s-wave condensation,
from $2.7\ro$ to $4.3\ro$ for case I chosen by us for an
illustration.

\section{$K^-$ Condensation in Neutron Stars}\label{sec:Condensation}

In sections \ref{sec:Construction}-\ref{sec:Short}  we have
constructed the  $K^-$ polarization operator.
Now we use it to study a possible instability of the system
with respect to a phase transition into a state with $K^-$
condensate.

For this aim we first investigate solutions of the $K^-$
dispersion relation \be\label{dys-ful} \om^2-\vec
k^2-m_K^2-\Re\Pi^{\rm tot}(\om,\vec k) =0, \ee where the complete
polarization operator is given by \be\Pi^{\rm tot}(\om,\vec k) &=&
\Pi_{{\rm S}}(\om)+ \Pi^{\rm reg}_{{\rm P}}(\om,\vec k)+\Pi^{\rm
pole}_{\rm P}(\om,\vec k)
\nonumber\\
&+& \delta \Pi_{\rm hyp}^{(\chi{\rm PT},0)}(\om)+
\delta \Pi_{\rm hyp}^{\rm reg,0}(\om)
\label{poltot}\,.\ee
It contains the s-wave part, regular p-wave and pole parts of the
polarization operator given by (\ref{chicorS},\ref{chicorP}) and
(\ref{pi-pole-cor}), respectively, and the terms determined by the
hyperon populations (\ref{spolh}) and (\ref{dPisH}). The
correlation parameters are taken according to
(\ref{xiab},\ref{xiabP}) and (\ref{corparH}).

There are two different possibilities. The $K^-$ condensation may
occur in the neutron star matter via a second-order phase
transition or a first-order phase transition. The dynamics of both
phase transitions are quite different. Therefore, both
possibilities might be realized at different physical conditions
related to the different stages of neutron star evolution.

In the case of a second-order phase transition, at the moment,
when the density in the neutron star center achieves the critical
density $\rho_c^{\rm II}$, the reactions (\ref{react})  come into
the game.  At this second order phase transition the isospin
composition and the density may change only soothly. For typical
times $\tau \propto \tau_{\rm react}\sqrt{\rho_c^{\rm
II}}/\sqrt{\rho -\rho_c^{\rm II}}$ the system creates an
energetically favorable condensate state, $\tau_{\rm react}$ is
the typical  time of weak processes (\ref{react}). The condensate
appears within the region where $\rho
>\rho_c^{\rm II}$. If it happens during the supernova
explosion, the typical size of the condensate region might become
of the order of the neutron star radius.  Due to the energy
conservation the positive energy is released in such a transition.
When the condensate region is heated up to the temperatures $T\geq
T_{\rm opac} \sim (1\div 2)$~MeV neutrinos are trapped. At
this stage the cooling time is determined by the neutrino heat
transport from the condensate interior of the neutron star to its
exterior \cite{MSTV90}. When the star cools down to smaller
temperatures, $T<T_{\rm opac}$, the neutron star begins to be
transparent for neutrinos. They can be  directly radiated away. A part
of the energy is radiated by photons from the star surface. In
binary long lived systems the critical density in the neutron star
center  can be achieved by the accretion process. Then the
transition is characterized by the typical large time of the
accretion.

In the case of a first-order phase transition the final state
might significantly differ from the initial one by its isospin
composition and the density. Thus this new state can't be prepared
in microscopic processes. Too small droplets of the new phase are
not energetically favorable due to a positive surface energy
contribution.  When the density in the star center begins to
exceed the value $\rho_c^{\rm I}<\rho_c^{\rm II}$ the system
arrives at a metastable state. When in a fluctuation a droplet of
the new phase (with density $\rho_c^{\rm fin}> \rho_c^{\rm I}$) of
rather large (overcritical) size   is prepared it starts to grow.
At zero temperature, the probability of the creation of such a
droplet via quantum fluctuations is very small but increases
greatly with the temperature \cite{MSTV90}. Thus, the first-order
phase transition occurs most likely at an initial stage of the
neutron star formation or cooling when the temperature is rather
large. If the density in the center of a star exceeds the value
$\rho_c^{\rm II}$, a second-order phase transition may also occur.
In binary stellar systems, where the neutron star slowly accretes
the mass from the star-companion and the temperature is  small,
the second-order phase transition might be a more probable one
(depending on the accretion rate).

\subsection{II Order Phase Transition to the s-wave Condensate State}

Let us first analyze possibility of the s-wave $K^-$ condensation.

\begin{figure*}
\includegraphics[width=14cm,clip=true]{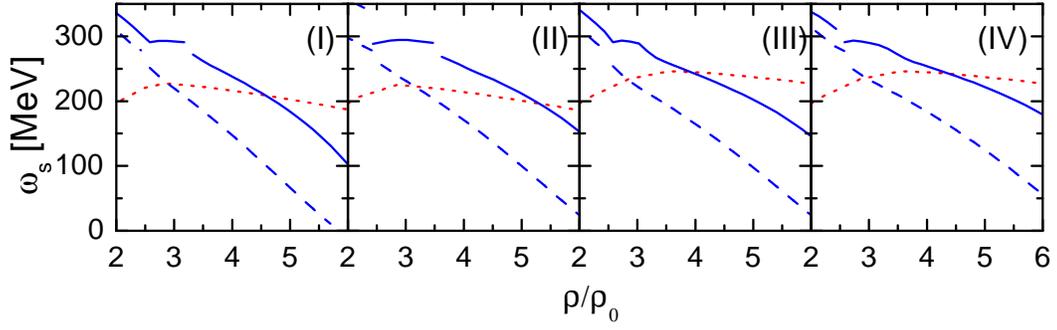}
\caption{The  energy of the lowest $K^-$ branch of the dispersion
relation (\protect\ref{dys-ful}) for $\vec{k} =0$ is depicted by
solid lines. The dotted lines show the electron chemical
potentials. The dashed lines are calculated without baryon-baryon
correlations. Different panels correspond to the interaction given
by the cases I-IV. } \label{fig:swave-col}
\end{figure*}

In Fig.~\ref{fig:swave-col}, summarizing the results of sections
\ref{min1}, \ref{min2}, \ref{min3},
we show the energy of the lowest  $K^-$
branch
of the kaon  dispersion relation  (\ref{dys-ful}) for $\vec{k}
=0$ together with the electron chemical potential. For the given
parameter choice (see different panels),
the crossing points of the lines indicate the
critical density of the s-wave condensation via the second order
phase transition.

We see that for all  the models the neutron star matter is
unstable with respect to reactions (\ref{react})  in the density
interval $(3\div 5.2)\rho_0$ in dependence on the choice of the
correlation parameters and the parameters of the hyperon-nucleon
interactions. Recall, that in the framework of our model the
attraction due to the s-wave kaon-nucleon interaction corresponds
to the effective kaon-nucleon sigma term $\Sigma_{KN}$ equal to
150~MeV, which is much smaller than the value $\Sigma_{KN}\simeq
(300\div 400)$~MeV used in the previous works studying the s-wave
$K^-$ condensation. Additional  suppression comes from the
nucleon-nucleon correlations. Nevertheless, we found also an extra
attraction associated with $\delta \Pi^{\rm reg}$, $\Pi^{\rm
pole}$ and $\delta \Pi_h$ terms. Thus, until correlations are not
included, we support a conjecture of previous works on the
possibility of the s-wave condensation at $\rho_{\rm c, S}^{\rm
(II)}\sim 3\rho_0$. However, the baryon-baryon correlations
additionally shift the condensation critical density to larger
values than those ones discussed in
Refs.~\cite{swave-1,swave-2,bb,nscool,tpl94,gs,pw}.

Note that here we checked the
necessary condition of the $s$-wave condensation but we did not yet
minimize  the energy of the lowest branch  of the dispersion equation,
$\om^{\rm min} (\vec{k})$,
over  $|\vec{k}|$.
Thus we still can't conclude whether we deal with the s-wave or the
p-wave condensation.

\subsection{II Order Phase Transition to the p-wave Condensate State}

In this section we are going to study the principal possibility of
the p-wave  $K^-$ condensation in neutron star matter at the
assumption of the second order phase transition. We would like to
investigate whether the p-wave condensation second order phase
transition can occur at densities smaller than that for the s-wave
condensation, and how much such a situation is sensitive to the
parameter choices for hyperon-nucleon interactions and
correlations.

Let us first assume that the s-wave
$K^-$ condensation is indeed possible at some critical density $\rho_{\rm
c,S}$. The  lowest energy branch of the $K^-$ spectrum
at small momenta is given by
\be \label{kprop}
\om\approx\om_{{\rm S}}+\alpha (\om_{{\rm S}})\,Z_{\rm S}(\om_{\rm S})\,
\vec{k}^{\,2}\,, \ee
where $\om_{\rm S}$ is, as before,
$\om (\vec{k}=0)$ for
the lowest energy branch of the dispersion law given by the solution of the
equation
$\om_{\rm
S}^2=m_K^2+\Re\Pi_{\rm S} (\om_{\rm S },\vec{k}=0 )$,
and
\be\nonumber
Z^{-1}_{\rm S}(\om) &=& \left( 2\,\om_{\rm S}-\frac{\partial
\Re\Pi_{\rm S}(\om ,\vec{k})}{\partial \om }\Big|_{\vec k =0}  \right)>0 \,,
\\ \nonumber
\alpha (\om) &=&
1+\frac{\prt\Re\Pi_{\rm P}(\om ,\vec{k}) }{
\prt \vec{k}^{\,2}}\Big|_{\vec k = 0}\,.\ee
If $\alpha (\om_{\rm S})<0$ at $\rho_{\rm c,S}$, then
instead of the s-wave condensation  we, actually, have the p-wave
condensation at a somewhat smaller density. The aim of this section is to
find  the value $\alpha(\om_{\rm S})$ in
(\ref{kprop})  at the critical point of  the s-wave condensation
i.e., when $\om_{\rm S} = \mu_e$.

Taking into account the p-wave kaon-baryon interaction which we
have determined in Sect.~\ref{sec:Decomposition}, we find
\be\label{alpha} \alpha(\om) =1+\alpha_{\rm pole}(\om)+\alpha_{\rm
reg}(\om)\,. \ee Without  baryon-baryon correlations, the
contribution of the pole part is \be \alpha_{\rm pole}^{(0)} =
\frac{\prt}{\prt \vec{k}^2}\,\Re\Pi^{(\rm{ pole}, 0)}(\om,\vec
k)\Big|_{\vec k =0}= \alpha_{\Lambda p}^{(0)} +
\alpha^{(0)}_\Sigma+\alpha^{(0)}_{\Sigma^*}, \ee with
$\alpha^{(0)}_\Sigma=\alpha^{(0)}_{\Sigma^0
p}+2\,\alpha^{(0)}_{\Sigma^- n}$ and $\alpha_{\Sigma^*}^{(0)}
=\alpha^{(0)}_{p\Sigma^{*0} }+2\,\alpha^{(0)}_{n\Sigma^{*-} }$\,.
From (\ref{ppol},\ref{ppols}) we have \be\nonumber
\alpha_{iH}^{(0)}(\om) &=&
C_{KNH}^2\,\Big(\eta_{NH}^2\,\phi_{iH}^{\rm P}(\om) +
\eta_{HN}^2\,\phi_{Hi}^{\rm P}(-\om)\Big)\,,
\\ \nonumber
\alpha_{i\Sigma^*}^{(0)}(\om) &=&
C_{KN\Sigma^*}^2\,\eta_{N\Sigma^*}^2\,\Phi_{i\Sigma^*}(\om,0)
\,,\quad H=\{\Lambda, \Sigma\},\,. \ee The term
$\eta_{HN}^2\,\phi_{iN}^{\rm P}(-\om)$ accounts for the
contribution of the hyperon Fermi sea. Once the baryon-baryon
correlations are included in the pole part of the polarization
operator according to (\ref{pi-pole-cor}), $\alpha^{(0)}_{\rm
pole}$ is to be replaced by \be\nonumber \alpha_{\rm
pole}(\om)=\frac{\prt}{\prt \vec{k}^2}\,\Re\Pi^{\rm pole}(\om,\vec
k)\Big|_{\vec k =0} .\ee The regular part follows from
(\ref{pireg}), $\alpha^{(0)}_{\rm reg}(\om)=b_p(\om)\,\r_p /\r_0
+b_n(\om)\,\r_n/\r_0 $, with the coefficients defined in
(\ref{bs},\ref{bp}). Suppression of the regular part $\alpha_{\rm
reg}$ due to the baryon-baryon correlations can be taken into
account according to (\ref{chicorP}) \be \nonumber
 \alpha_{\rm
reg}(\om)&=&\frac{\widetilde{b}_p(\om) +\widetilde{b}_n(\om) +2\,
\widetilde{b}_p(\om) \, \widetilde{b}_n(\om) \xi_{np}^{\rm
(P)}(\om)} {1- \widetilde{b}_p(\om)\,\widetilde{b}_n(\om)
(\xi_{np}^{\rm (P)})^2(\om)},
\\ \nonumber
\widetilde{b}_p(\om) &=& \frac{b_p(\om)\r_p /\r_0}{1-b_p(\om)\,\r_p\,
\xi_{pp}^{\rm (P)}(\om)/\r_0}\,,
\\ \label{regcor}
\widetilde{b}_n(\om)  &=& \frac{b_n(\om)\,\r_n
/\r_0}{1-b_n(\om)\,\r_n\, \xi_{nn}^{\rm (P)}(\om)/\r_0}.
\ee
Although the $\xi_{ii'}^{\rm (P)}(\om)$-functions can be evaluated as in
(\ref{xiabP}),
we will treat them here as free energy-independent parameters $\xi_{ii'}$, in
order to investigate the sensitivity
of the results to them.

\begin{figure*}
\includegraphics[clip=true,width=14cm]{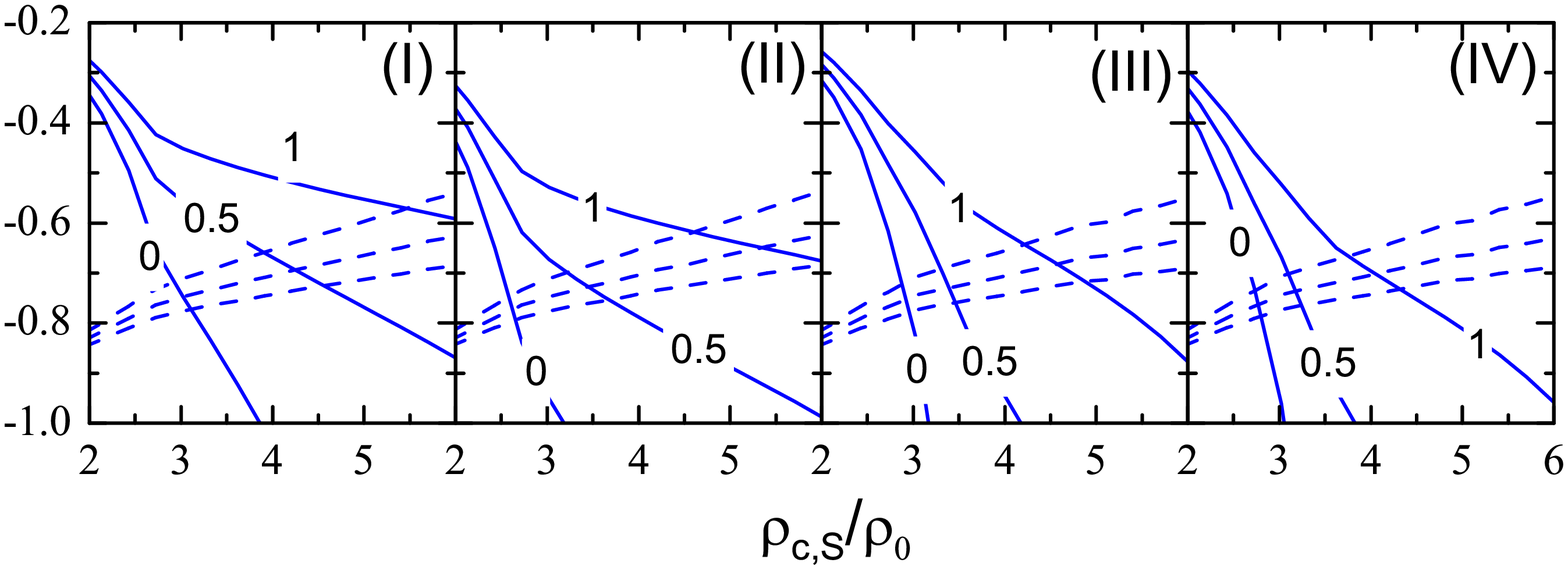} \\
\includegraphics[clip=true,width=14cm]{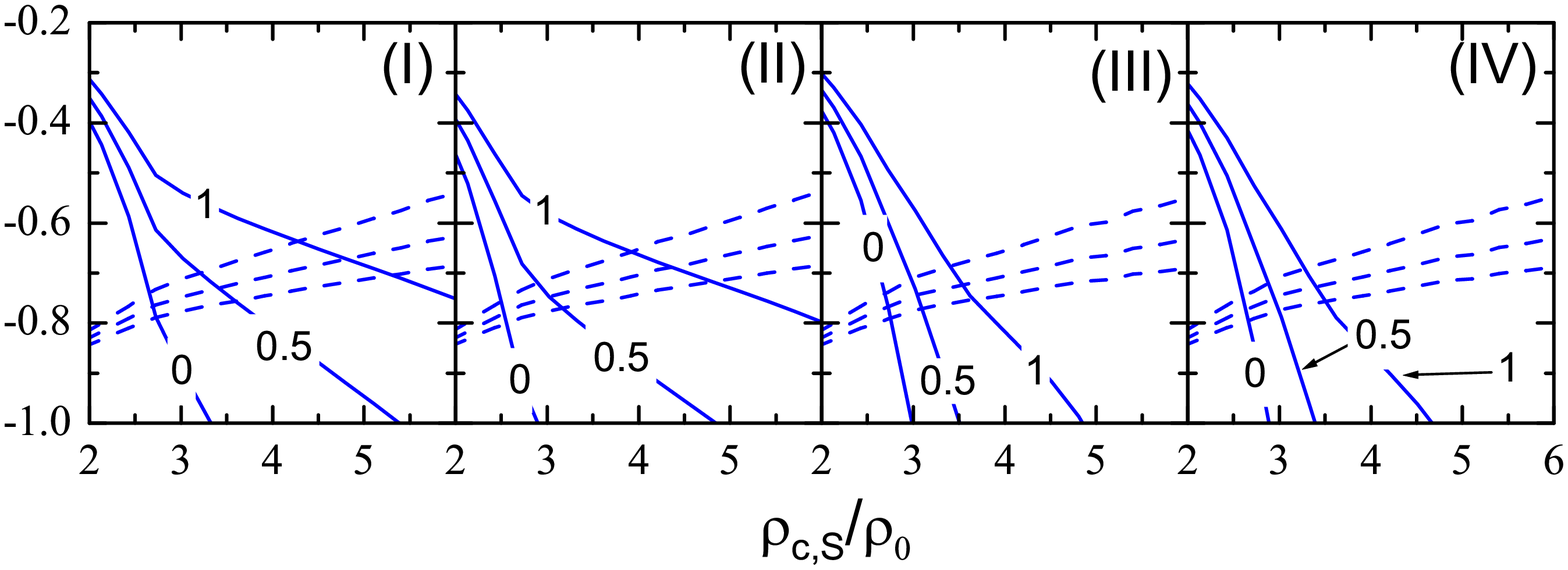}
\caption{The quantities ${\alpha}_{\rm pole}(\mu_e)$,
$\om_{\rm S}=\mu_e$, (solid lines)
and $-1-\alpha_{\rm reg}$ (dash lines) are shown
vs. critical density of a s-wave kaon condensation for various correlation
parameter choices. Labels
show values of $f'_\Lambda=g'_\Sigma$.
Solid lines are drown for
$f'_{\Lambda\Sigma}=0$, and for $f'_\Lambda=g'_\Sigma =0,0.5,1$.
Dashed lines are drown for three choices of the
correlation parameters used in calculation of $\alpha_{\rm reg}$,
$\xi_{nn}=\xi_{pp}=\xi_{np}=0,0.5,1$ (from
the upper line to the lower one).
Cases (I-IV)  correspond to those in Fig.~\ref{fig:compos}.
In the upper plot the
$\Sigma^*$ contributions  are taken with $g'_{\Sigma^*}=g_{\Sigma}$,
$V_{\Sigma^*}=V_{\Sigma}$ and $g_{\sigma\Sigma^*}=g_{\sigma\Sigma}$.
In the lower plot the $\Sigma^*$ is detached from the mean-field potentials
$V_{\Sigma^*}=0$ and $g_{\sigma\Sigma^*}=0$.
Hyperon populations are included in mean fields but not
in the polarization operator.
  }
\label{fig:pc-hs}
\end{figure*}

\begin{figure*}
\includegraphics[clip=true,width=14cm]{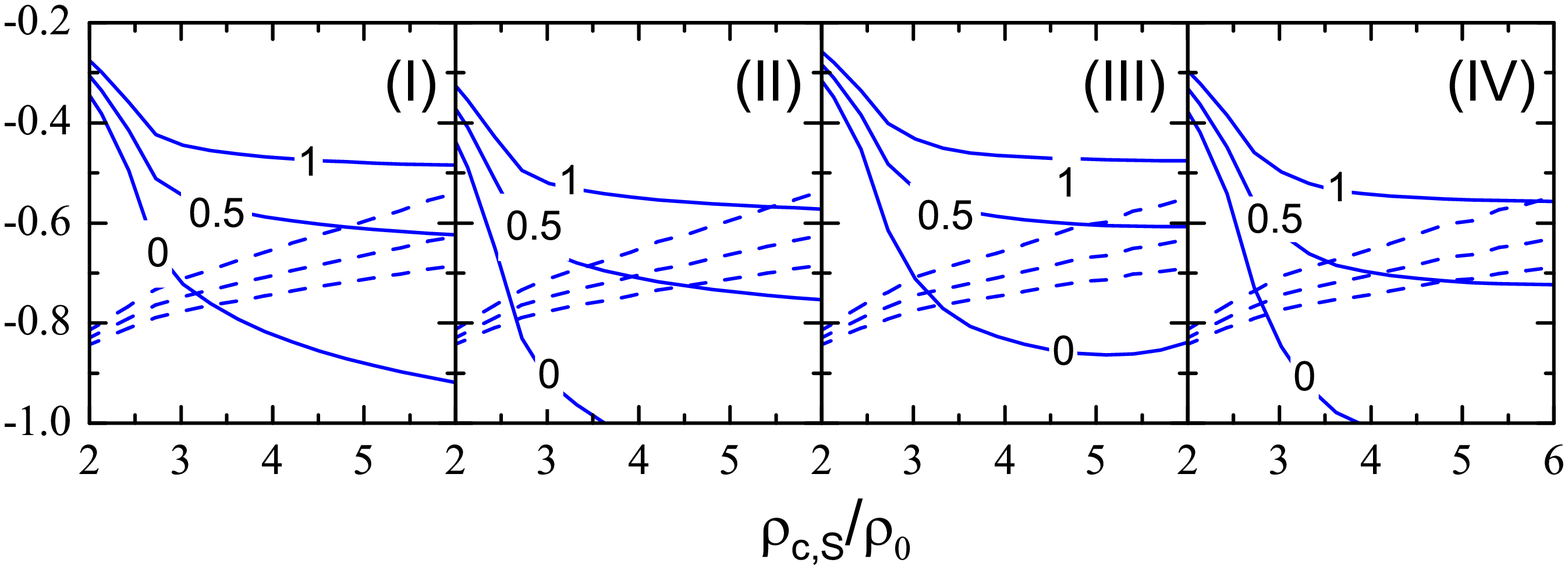}\\
\includegraphics[clip=true,width=14cm]{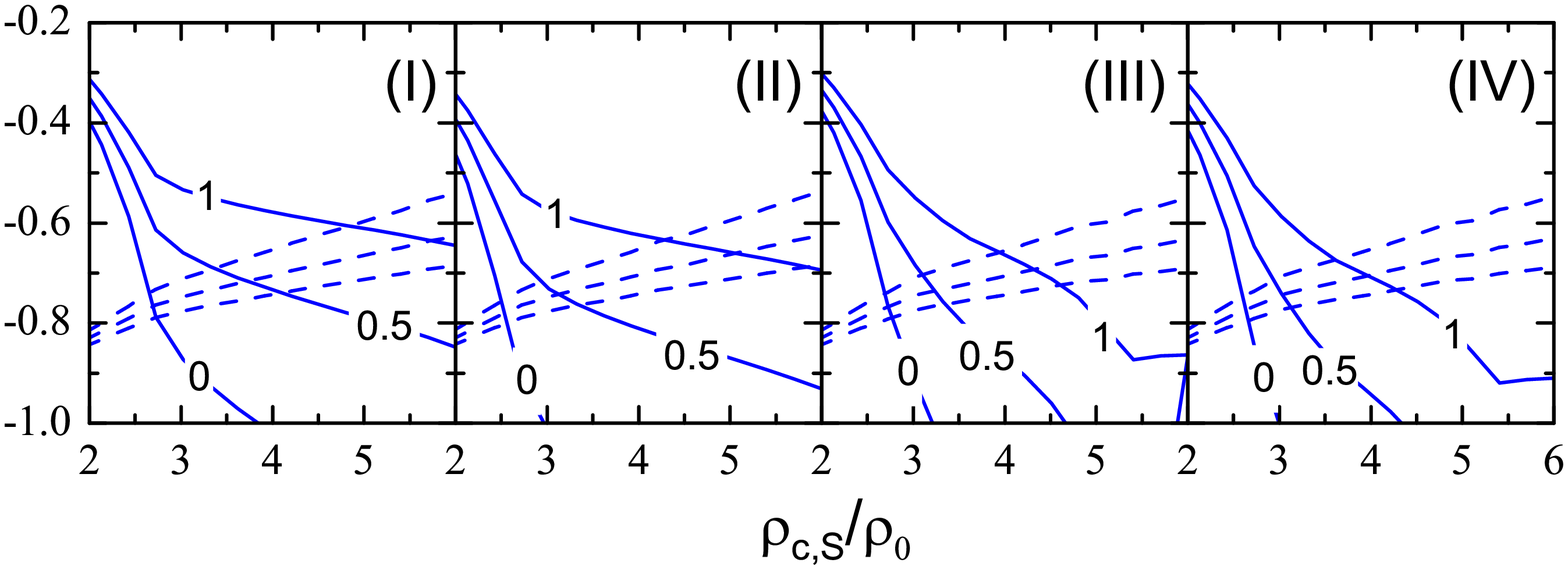}
\caption{The same as in Fig.~\protect\ref{fig:pc-hs},
but with  account for the  hyperon population in the polarization operator.
 }
\label{fig:pc-hsh}
\end{figure*}

In Figs.~\ref{fig:pc-hs}--\ref{fig:pc-hsh} we show values of
$\alpha_{\rm pole}(\mu_e)$ (solid lines) and $-1-\alpha_{\rm reg}$
(dash lines), calculated as a function of  $\r_{c,{\rm S}}$
 for various baryon-baryon correlation
parameters and for cases I--IV determining the hyperon couplings
in (\ref{lagwal}). The $\Sigma$ hyperon contribution is proved to
be very small and most part of the strength is due to the $\Lambda
p^{-1}$ and  $\Sigma^* n^{-1}$ contributions. Filling of the
hyperon Fermi seas is not incorporated  (we artificially suppress
terms $\propto\Phi_{Hi}$ in $\Pi^{\rm pole}$). As we mentioned,
embedding $\Sigma^*$ into the mean-field model (\ref{lagwal}) is
quite uncertain due to the absence  of any empirical constrain on
the coupling constants. Therefore we consider two different cases.
In the upper plot in Fig.~\ref{fig:pc-hs} we  assume that
$\Sigma^*$ couples to the mean-field with the same strength as the
$\Sigma$-hyperon ($V_{\Sigma^{*}}=V_{\Sigma}$,
$g_{\sigma\Sigma^*}= g_{\sigma\Sigma}$), whereas in the lower plot
in Fig.~\ref{fig:pc-hs} we detach $\Sigma^*$ from the mean field
potentials ($V_{\Sigma^{*}}=0$, $g_{\sigma\Sigma^*}=0$).

At the crossing point of the solid line with the corresponding
dash line we have $\alpha_{\rm pole}=-1-\alpha_{\rm reg}$ and
therefore $\alpha =0$, that means that given density is the
critical density for both the s- and the p-wave condensations.
When at the given value of $\rho_{\rm c,S}$ the solid line is
below the corresponding dash line we have $\alpha_{\rm
pole}<-1-\alpha_{\rm reg}$. This means that $\alpha <0$ and for
such a parameter set $K^-$ condensation occurs in the p-wave state
at somewhat smaller density than the assumed s-wave condensation.

The contributions from the hyperon Fermi seas are included in
Figs.~\ref{fig:pc-hsh}.  From Figs.~\ref{fig:pc-hs}  and
\ref{fig:pc-hsh} we observe that $\Sigma^*$  contributes
essentially to the $K^-$ polarization operator increasing
attraction in the p-wave. The most favorable case for the p-wave
condensation is realized if $\Sigma^*$ is detached from the mean
field potentials. Comparing Figs.~\ref{fig:pc-hs}  and
\ref{fig:pc-hsh}  we also see that the effects from the hyperon
population on the $p$-wave terms are large.

\begin{figure*}
\includegraphics[clip=true,width=14cm]{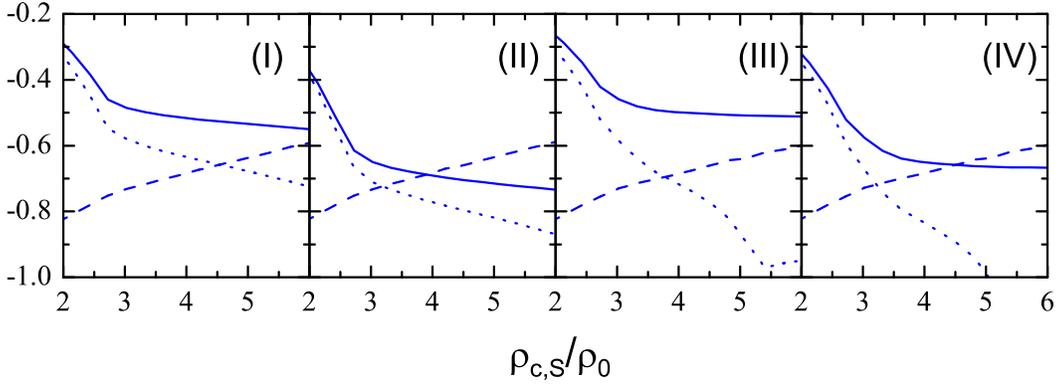}
\caption{Values $\alpha_{\rm pole}$ (solid lines) and
$-1-\alpha_{\rm reg}$ (dash lines), $\om_{\rm S}=\mu_e$,
calculated for the correlation
parameters (\ref{corparH}). Dotted lines present calculations
with $\Sigma^*$ detached from  mean field potentials.  Planes
correspond to the cases I-IV of hyperon couplings. Possibility of
the hyperon population is also
incorporated in the polarization operator.}\label{fin}
\end{figure*}

In Fig.~\ref{fin}  we show the results of our calculations for the
values of the correlation parameters, which we have evaluated in
Sect.~\ref{sec:Short}. For the  $\Sigma^*$ detached from the
mean-field potentials our model predicts a preference of the
p-wave condensation. The p-wave condensation may occur already at
$\rho \leq 3\rho_0$ for the cases II, IV, and for $\rho \simeq
4\rho_0$ and $\rho \simeq 4.5\rho_0$ for cases III and  I provided
the s-wave softening of the spectrum is also rather high. For
$\Sigma^*$ coupled with the same strength as $\Sigma$ the s-wave
condensation might be preferable compared to the p-wave
condensation. Since a variation of other parameters, besides
$\Sigma^*$ strength, is certainly also allowed the results are
more uncertain in the latter case, cf. Figs ~\ref{fig:pc-hs} and
\ref{fig:pc-hsh}.

\subsection{I Order Phase Transition to the $K^-$ Condensate State }

In the previous two sub-sections we have  studied the possibility
of the  $K^-$ condensation assuming that it occurs by a
second-order phase transition. Now we will investigate properties
of $K^-$ excitations in the baryonic matter of different particle
compositions, to figure out whether an abrupt ( of the first
order)  phase transition into a state with a new particle
composition and another baryon density is energetically favorable.

We shall consider: (i)~nucleon-hyperon matter (NHM)
composition which we have discussed above, see
Fig.~\ref{fig:compos}; (ii)~neutron-proton matter (NPM)
composition, the matter consisting only of protons and neutrons
assumed to be in $\beta$-equilibrium; (iii)~isospin-symmetrical
nuclear matter (ISM) composition consisting of protons and
neutrons with $\rho_p=\rho_n$ and leptons,  compensating the
electric charge of the protons; and (iv)~the proton-enriched
matter (PEM) consisting of protons and neutrons with the
concentration $Y_p=\rho_p/\rho=0.7$ and the charge compensated by
the leptons. Thus in cases (ii)-(iv) we switch off the hyperons in
our mean-field model (\ref{lagwal}) and in cases (iii), (iv) we
also freeze the value of the proton concentration.  For each case
we shall calculate the total energy of the system with and without
$K^-$ condensation. The idea behind this procedure is as follows.
We consider the given configuration with the corresponding proton
concentration and the condensate field as the most energetically
favorable configuration until we did not prove  the opposite.
The full reaction balance is assumed to be recovered, thereby. Below we
show that the first order phase transition occurs from NHM
starting at the density $\rho^{\rm I}_c <\rho^{\rm II}_c$ to
approximately ISM at $\rho^{\rm I}_{\rm fin}>\rho^{\rm I}_c$.

\begin{figure*}
\begin{center}
\parbox{6cm}{
\includegraphics[clip=true,width=6cm]{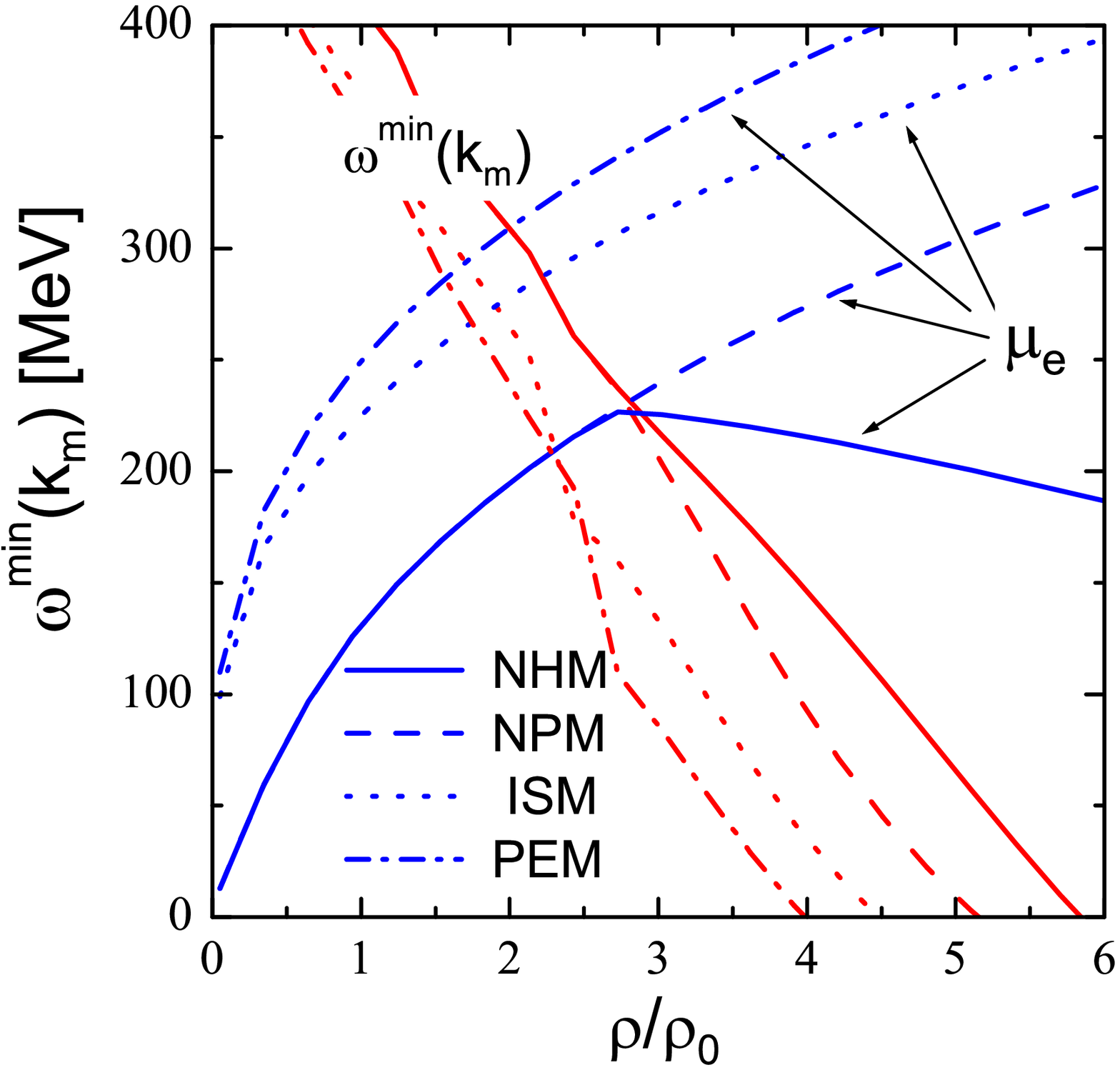}}\qquad
\parbox{6cm}{\includegraphics[clip=true,width=6cm]{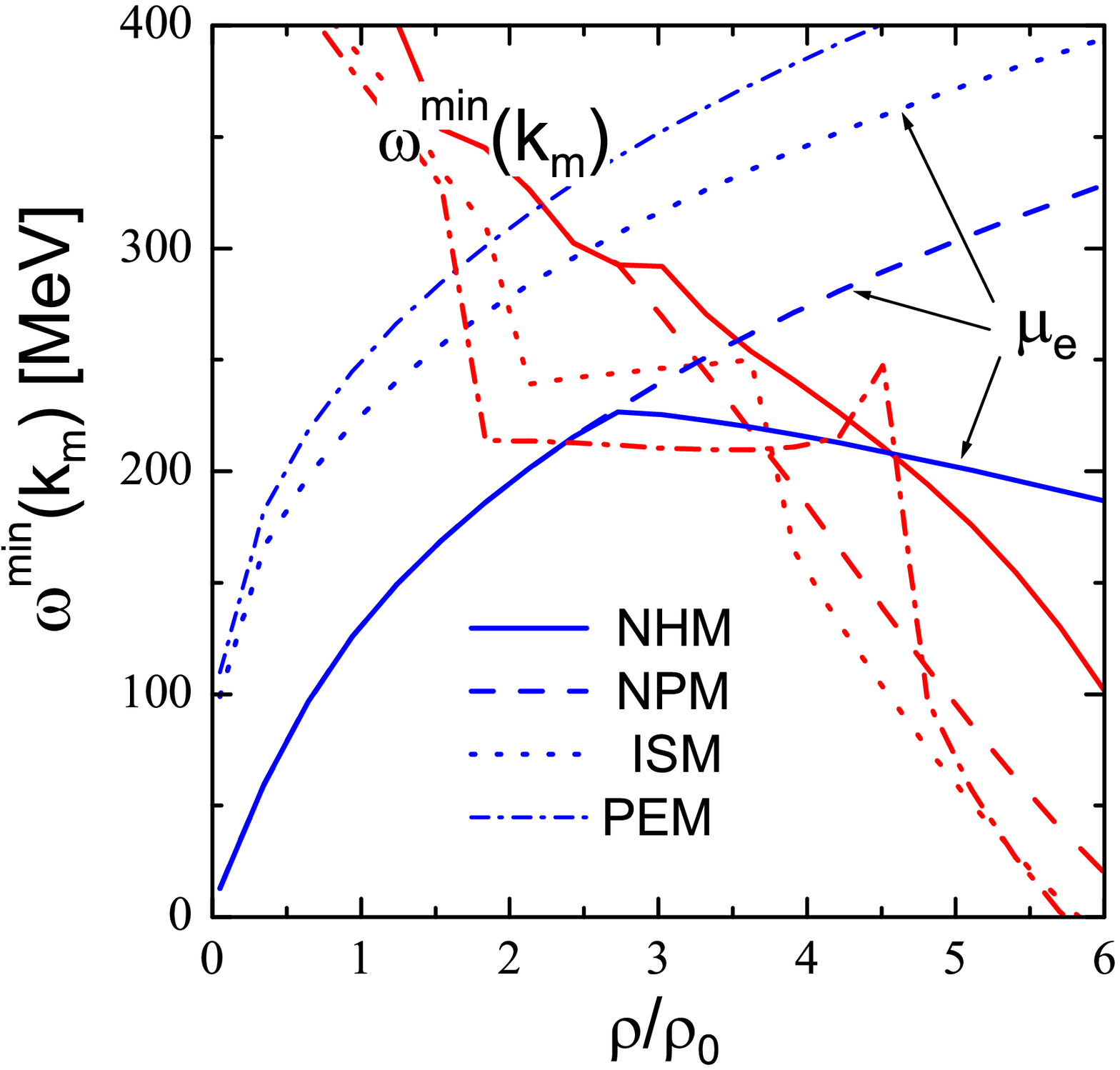}}
\end{center}
\caption{ The energy of the lowest $K^-$ branch $\om^{\rm min}
(\vec{k}_m , Y_p , \rho )$ minimized in $|\vec{k}|$, and the
electron chemical potential, calculated for the NHM, the NPM,  ISM
and PEM. In the right and left panels the results are presented
with and without baryon-baryon correlation effects, respectively.
In the NHM case interaction forces of case I are used and the
$\Sigma^*$ contributions  are taken with
$g'_{\Sigma^*}=g_{\Sigma}$, $V_{\Sigma^*}=V_{\Sigma}$ and
$g_{\sigma\Sigma^*}=g_{\sigma\Sigma}$. } \label{fig:sc-difm}
\end{figure*}

In Fig.~\ref{fig:sc-difm} we show the energy of the lowest $K^-$
branch of the dispersion equation (\ref{dys-ful}) $\om^{\rm min}
(\vec{k}_m , Y_p ,\rho)$ minimized with respect to the momentum,
as a function of the baryon density for different proton
concentrations. We see that the more protons exist in the matter
the smaller is the value of the density at which the minimal  kaon
energy $\om^{\rm min} (\vec{k}_m , Y_p ,\rho)$ meets the electron
chemical potential. Therefore if the energy gain due to the
condensation were large enough to compensate an energy loss in the
fermion kinetic energies the system would undergo a first-order
phase transition from the NHM state to the state with a proton
enriched composition and a different density.

Let us first compare the energies of NHM, NPM, ISM and PEM taking
into account a possibility of the $K^-$ condensation in each case.
For densities, when $\om^{\rm min} (\vec{k}_m , Y_p ,\rho)>\mu_e$,
there is no condensation and the energy density of the system is
given by (\ref{endens}). When,  at given $Y_p$ and $\rho$, we have
$\om^{\rm min} (\vec{k}_m , Y_p ,\rho)<\mu_e$, the $K^-$
excitations appear, partially replacing the leptons. These
excitations occupy single state and form a $K^-$ condensate. The
electron chemical potential is fixed now as $\mu_e=\om^{\rm min}
(\vec{k}_m , Y_p ,\rho)$. The energy density of the system with
the $K^-$ condensate  reads:
\be \nonumber
E_{\rm tot}^{(K)} &=& E_{\rm mes}+\sum_B E_B^{\rm kin}
\\ &+& \sum_{l=\mu\,e} E_{l}(\om^{\rm min}
(\vec{k}_m , Y_p ,\rho) ) + E_{\rm cond}^{(K)} \,,
\ee
where $E_{\rm cond}^{(K)}$ is the energy density of the condensate field
related to the mean field Lagrangian
\be {\cal L}_K&=& \left( \om^2 -m^2_K
- \vec{k}^2 -\Pi (\om ,\vec{k})\right)|\phi_{\vec{k\,}}|^2 .
\ee
The $\phi_{\vec{k}}$ is the $K^-$ condensate mean field with the
wave vector $\vec{k}$. This field component   should be found from
the minimization of the appropriate thermodynamical potential. For
simplicity we use here the linear model neglecting the higher
order terms, as $\propto |\phi_{\vec{k\,}}|^4$, which incorporate
an effective kaon-kaon interaction in dense baryonic matter. The
effective kaon-kaon interaction depends on the structure of the
mean field. In absence of the non-linear effective kaon-kaon
interaction, within the variational procedure we obtain the field
of the running plane wave type, $\phi_{\vec{k}}=
\mbox{exp}(-i\om^{\rm min} (\vec{k}_m , Y_p ,\rho)t +i\vec{k}_m
\vec{r})$. Using that the dispersion relation (\ref{dys-ful}) is
fulfilled for  $\vec{k}=\vec{k}_m$ and $\om=\om^{\rm min}
(\vec{k}_m , Y_p ,\rho)$, and that the density of the charged kaon
condensate $\rho_K$ is fixed by the electro-neutrality condition
$$\rho_K = \sum_{B}\,q_B\, \rho_B-\rho_e -
\rho_{\mu^-}\,,$$
we find the energy density of the kaon condensate equal to
\be\nonumber
E_{\rm cond}^{(K)}=\om^{\rm min} (\vec{k}_m , Y_p ,\rho)\,
 \Big(\sum_{B}\,q_B\, \rho_B-\rho_e -
\rho_{\mu^-}\Big).
\ee

\begin{figure*}
\begin{center}
\parbox{6cm}{
\includegraphics[clip=true,width=6cm]{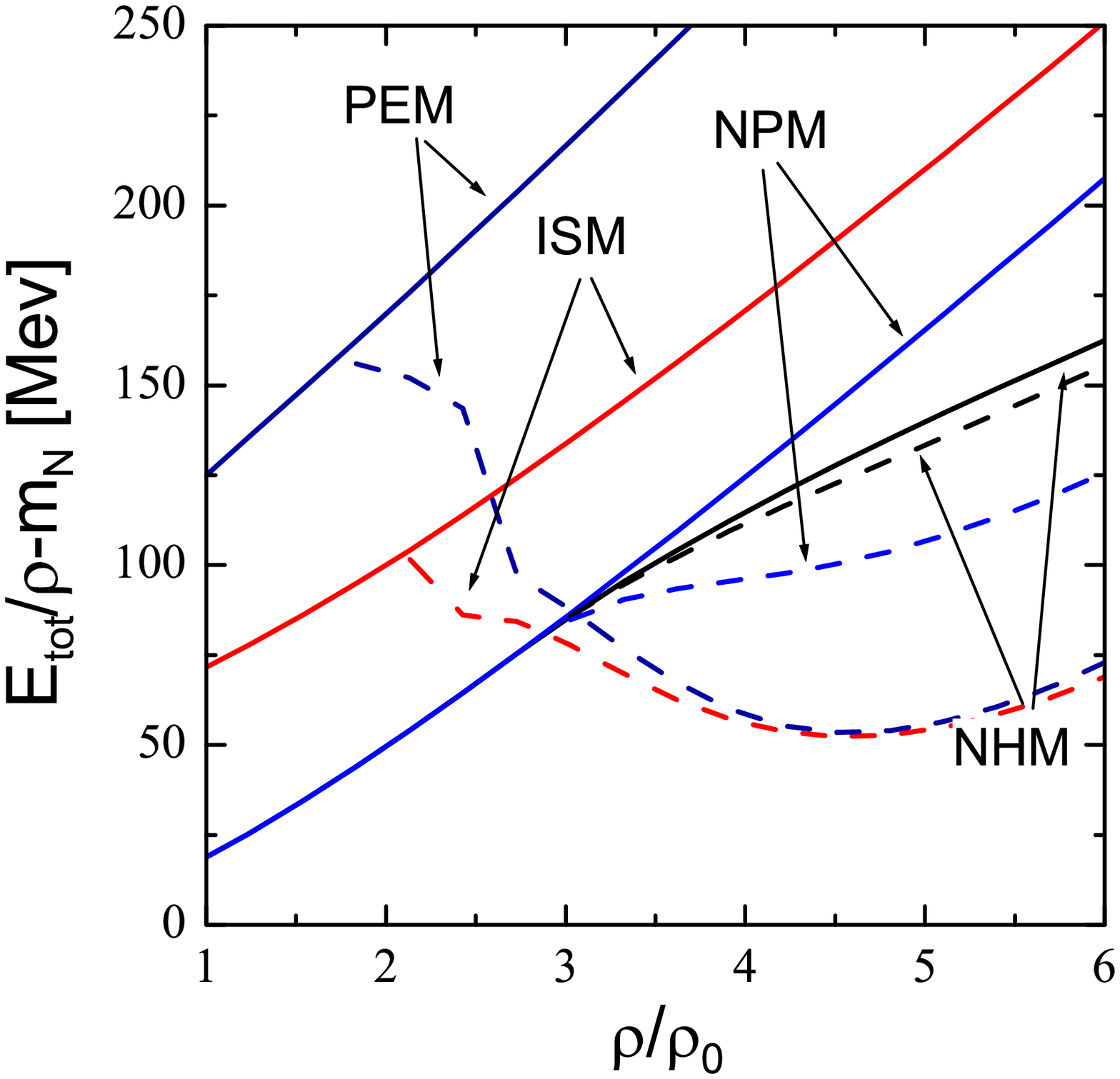}}\qquad
\parbox{6cm}{
\includegraphics[clip=true,width=6cm]{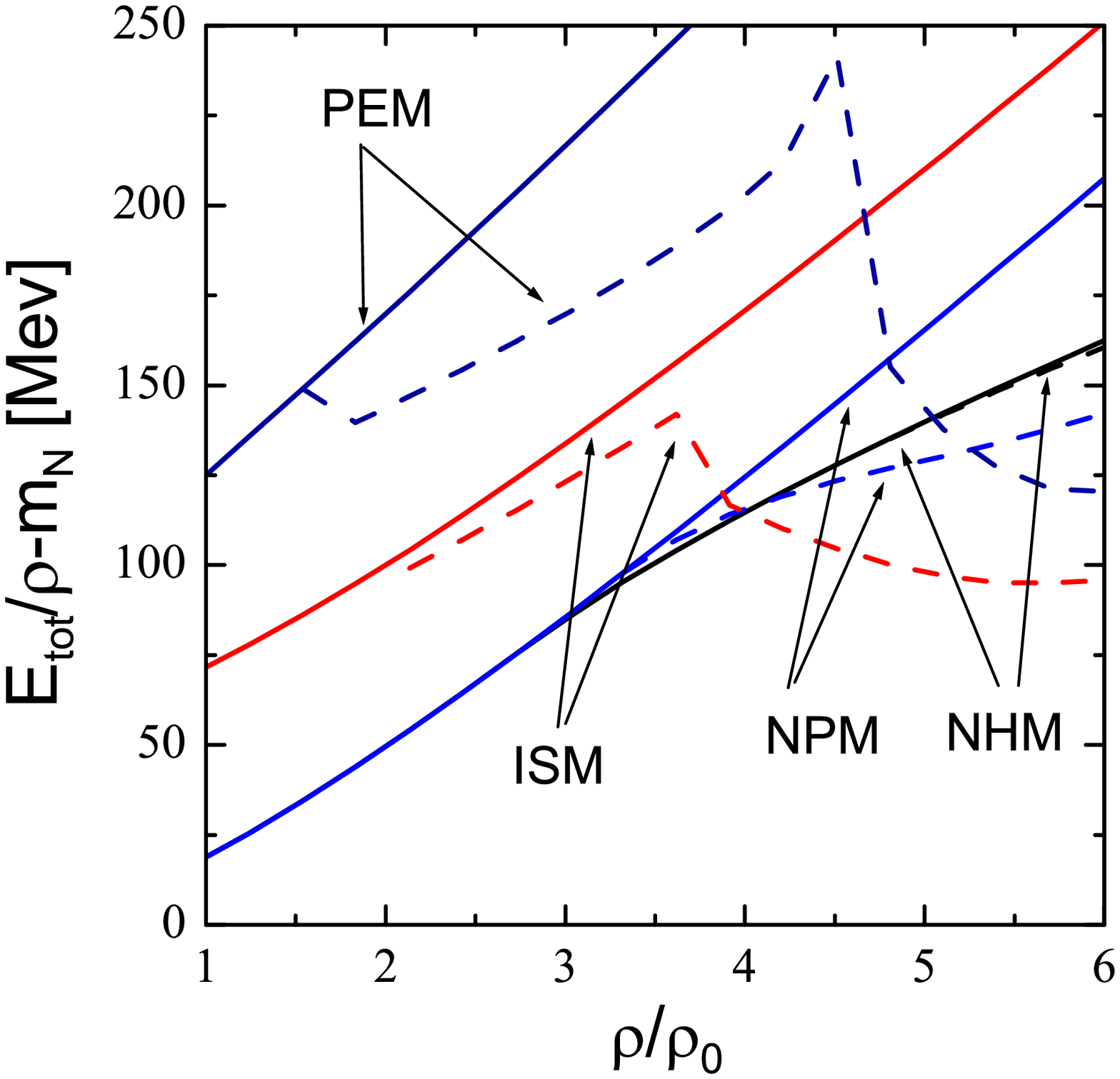}}
\end{center}
\caption{Total energy per baryon of the matter with different particle
compositions: without a $K^-$ condensate (solid lines) and
with the condensate (dashed lines). Abbreviations are the same, as
in Fig.~\protect\ref{fig:sc-difm}. The curves in the right and the left
panels are drawn with and
without inclusion of correlations. In the NHM case
interaction forces of case I  are used and the
$\Sigma^*$ contributions  are taken with $g'_{\Sigma^*}=g_{\Sigma}$,
$V_{\Sigma^*}=V_{\Sigma}$ and $g_{\sigma\Sigma^*}=g_{\sigma\Sigma}$.}
\label{fig:energ-s}
\end{figure*}

In Fig.~\ref{fig:energ-s} we show the energy per baryon in the
various baryonic systems, NHM, NPM, ISM, and PEM ($Y_p \simeq
0.7$), with and without the kaon condensate (dashed and solid
lines, respectively). We see that for the density $\rho_{\rm
c}\simeq 3\, \rho_0$ without baryon-baryon correlations and
$\simeq 4\, \rho_0$ including correlations, the condensate state
of ISM becomes to be energetically favorable compared to NHM. The
transition  to the new more symmetric isospin configuration would
increase the Fermi energies of the leptons (the latter ones  are
needed to compensate a larger charge of protons), but reduce  the
symmetry energy of the nuclear matter and the total Fermi energy
of nucleons.  Without the $K^-$ condensate, the energy loss is
larger than the gain and the system chooses the composition shown
in Fig.~\ref{fig:compos} for the given EoS, cf. that in
Fig.~\ref{fig:sc-difm} the solid lines labeled ISM are lying above
the NPM lines. With account for a possibility of the $K^-$
condensation one additionally gains on the kinetic energy of the
leptons, which are replaced by kaons. The latter energy gain is
large enough, leading to the preference of the ISM. The hyperon
Fermi seas are not filled. It is more energetically preferable to
create extra $K^-$ condensate mesons instead of a filling of  a
Fermi sea of the hyperons. From Fig.~\ref{fig:energ-s} we see that
PEM ($Y_p \simeq 0.7$) has a larger energy than ISM. As it is also
seen from Fig.~\ref{fig:energ-s} the resulting isospin composition
can be only slightly above $Y_p=1/2$. Further on we neglect this
difference considering ISM as the final configuration.

\begin{figure*}
\parbox{6cm}{
\includegraphics[clip=true,width=6cm]{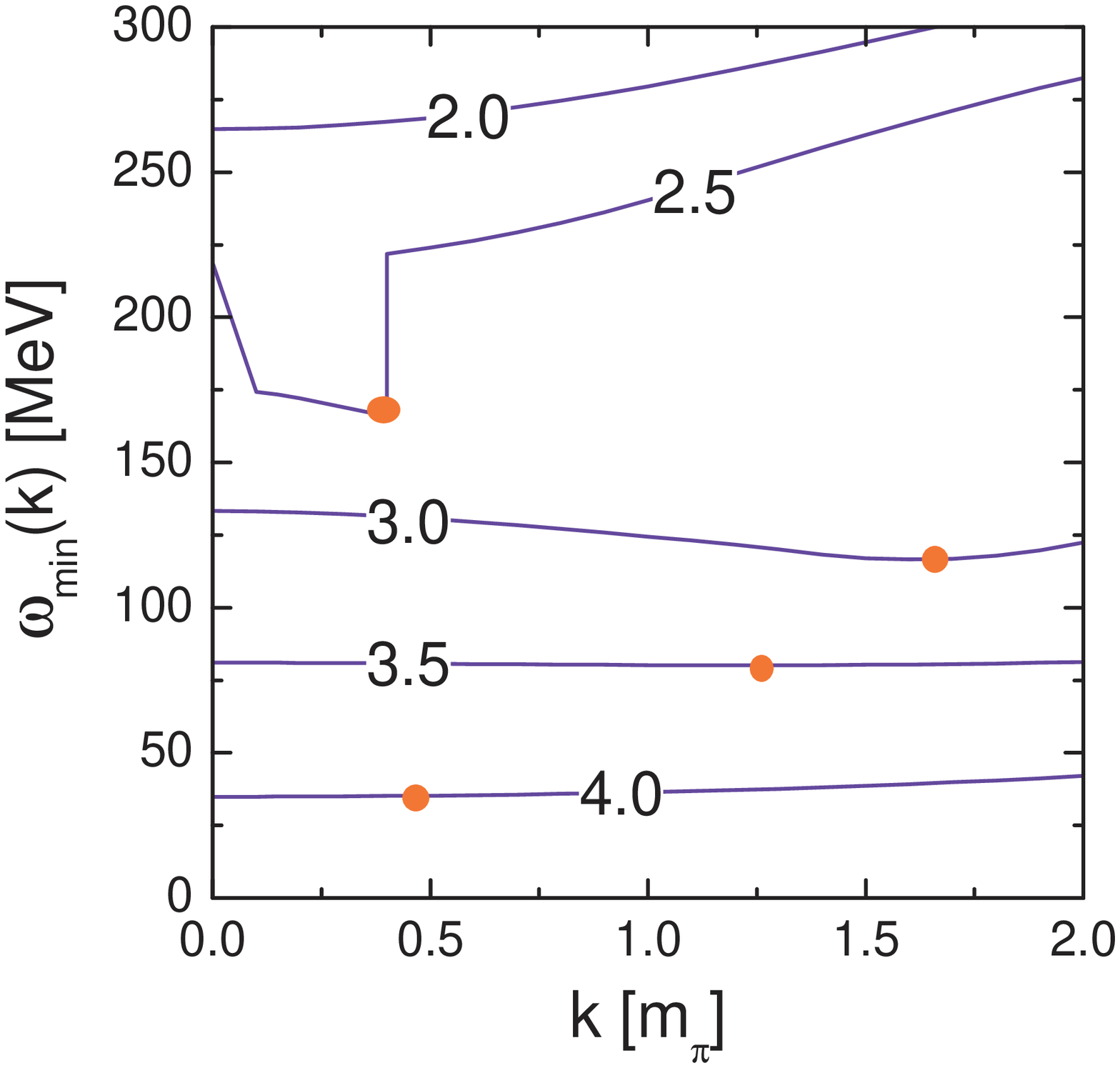}}
\qquad
\parbox{6cm}{
\includegraphics[clip=true,width=6cm]{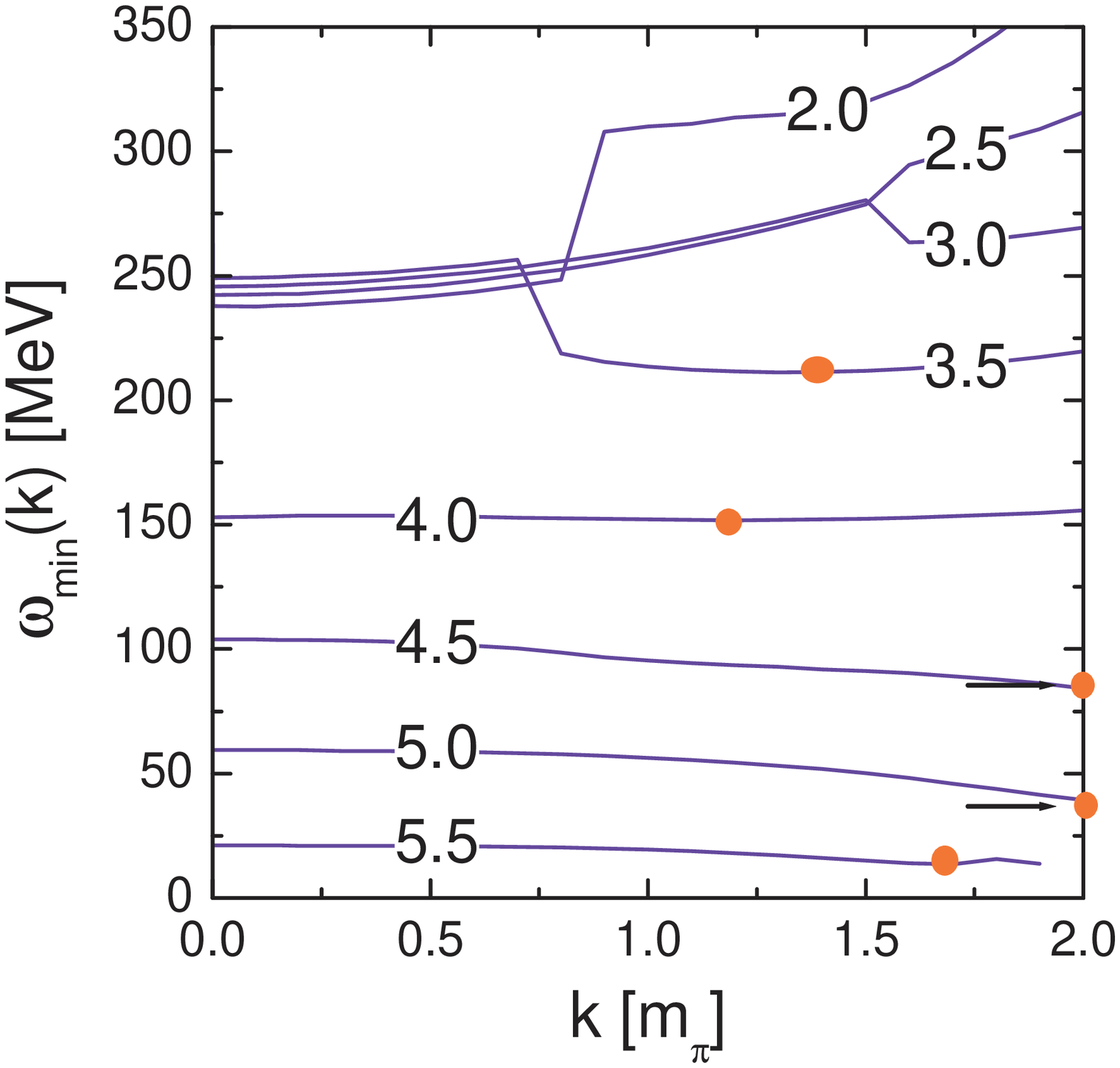}}
\caption{The energy of the lowest branch of the dispersion equation
 (\ref{dys-ful}) as function of the momentum with the
polarization operator (\ref{poltot}) for the ISM at various
densities (labels on the curves show the densities in $\rho_{0}$).
Calculations, done without and with baryon-baryon correlations,
are shown on left and right panels, respectively. Full circles
mark the position of the minimum on the spectrum.
}\label{fig:spectrumSM}
\end{figure*}

In Fig.~\ref{fig:spectrumSM} we plot the lowest branch of the
$K^-$ excitation spectrum  $\om_{\rm min} (|\vec{k}|)$ in the ISM
at various densities. In the left panel baryon-baryon correlations
are switched off and in right panel, switched on. We see that for
rather large densities ($\geq (2.5\div 3.5)\rho_0$) the spectra
have minima (dots in Fig. \ref{fig:spectrumSM}) at finite values
of the kaon momentum $\vec k_m\neq 0$. It signals that the
transition from NHM to a dense ISM would occur as the first-order
phase transition into the state with the p-wave kaon condensate.
The calculations shown in
Figs.~\ref{fig:sc-difm}-\ref{fig:spectrumSM} are done for
$g'_{\Sigma^*}=g_{\Sigma}$, $V_{\Sigma^*}=V_{\Sigma}$ and
$g_{\sigma\Sigma^*}=g_{\sigma\Sigma}$. The results obtained with
the $\Sigma^*$  detached from the mean field potentials are
checked to be of minimal difference.

\begin{figure*}
\begin{center}
\parbox{6cm}{
\includegraphics[clip=true,width=6cm]{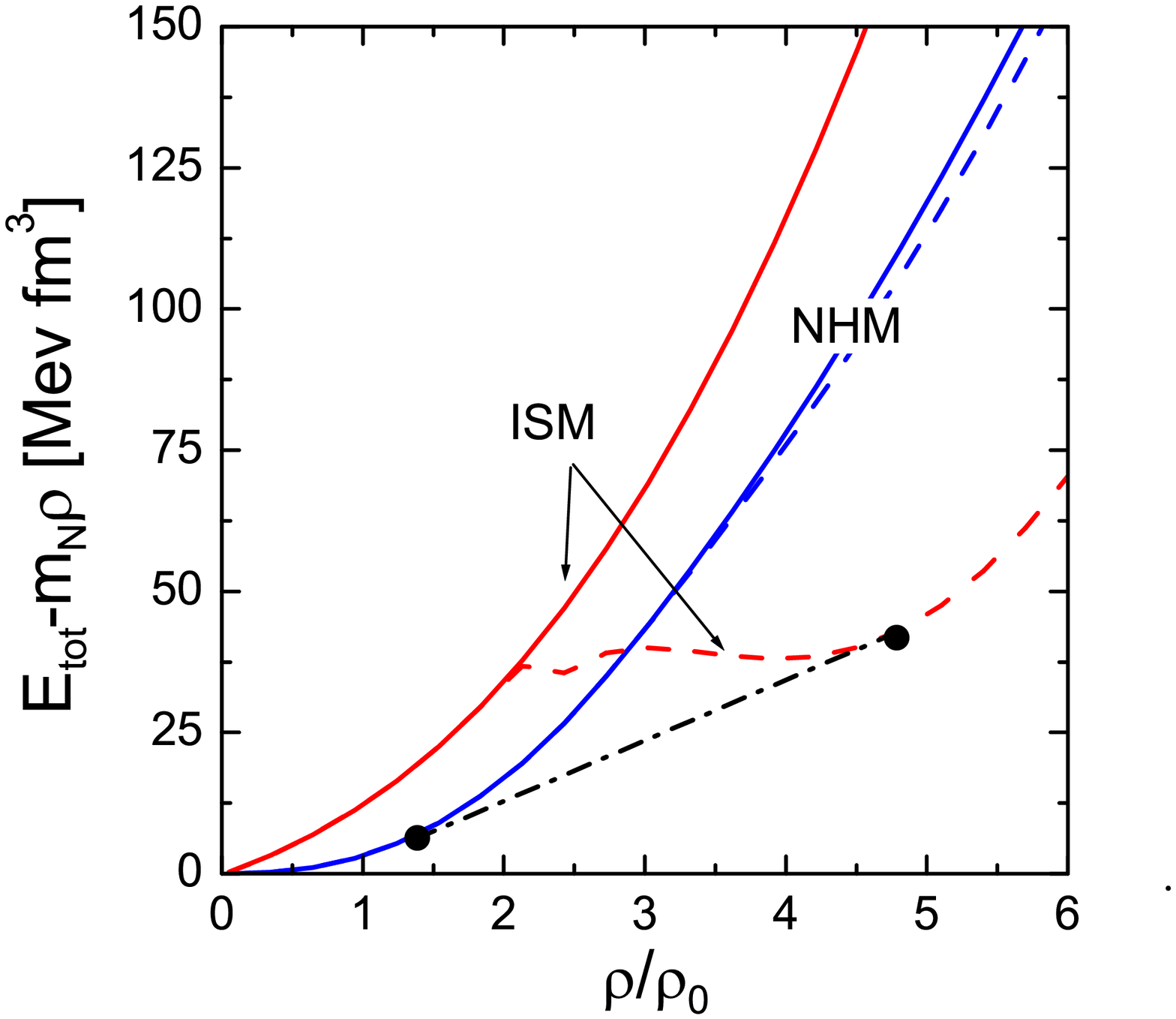}}\qquad
\parbox{6cm}{
\includegraphics[clip=true,width=6cm]{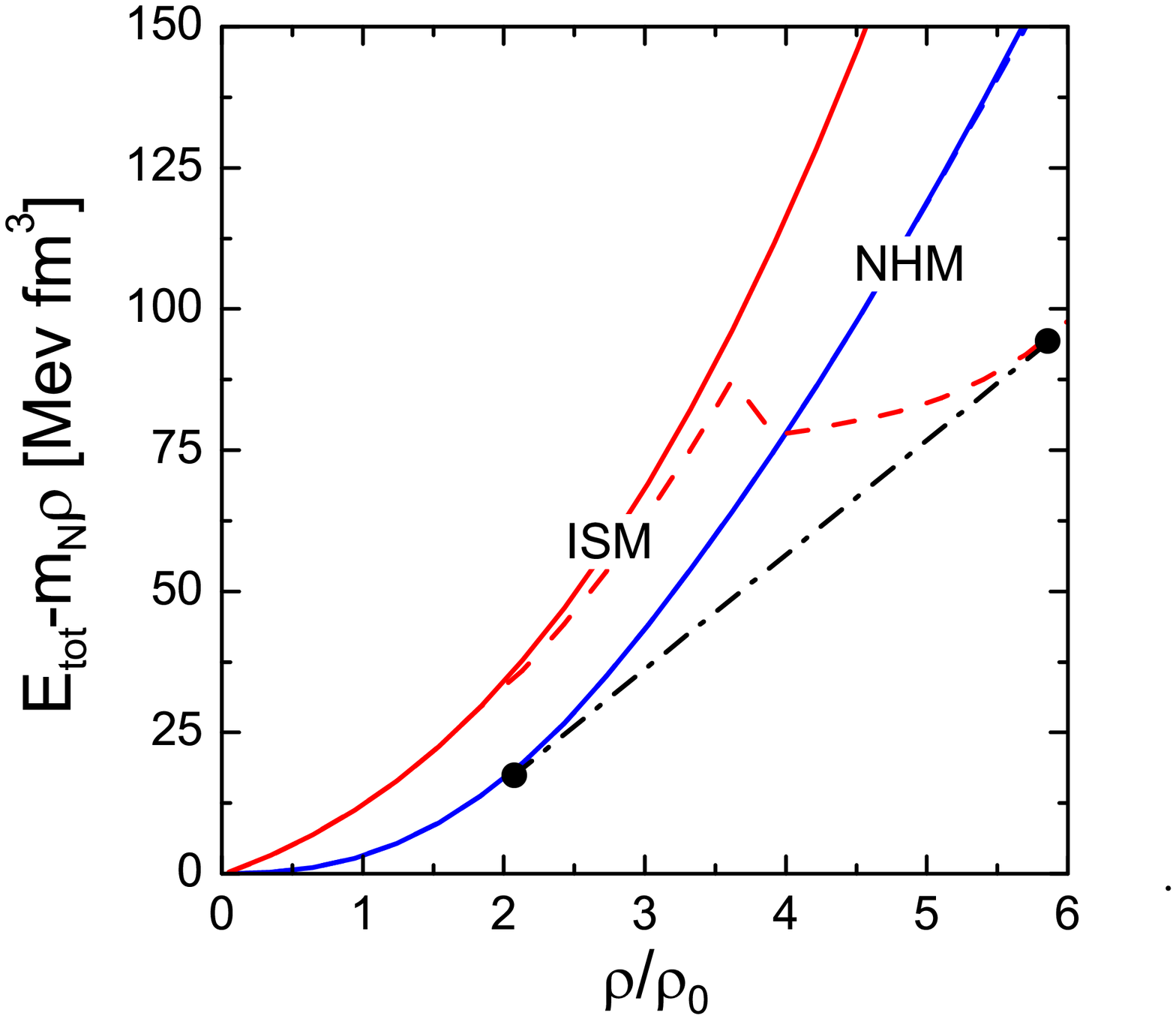}}
\end{center}
\caption{ Energy densities of NHM and ISM with and without the
$K^-$ condensate shown by dashed and solid lines, respectively.
For NHM the forces of case I are used.  Calculations with and
without baryon-baryon correlations are shown in left and right
panel respectively. The dash-dotted lines represent the
double-tangent Maxwell constructions between two phases. }
\label{fig:maxwell}
\end{figure*}

In the assumption that  the surface tension  is larger than a
critical value   the initial and final state densities  are
determined by the double-tangent Maxwell
construction~\cite{HPS93,VYT01}. In Fig.~\ref{fig:maxwell} we show
the double-tangent construction between NHM and ISM states. We see
that the critical density for the beginning  of the first-order
phase transition is equal to $\rho_{\rm c}^{\rm I}\simeq
1.4\rho_0$ without correlations and $\simeq 2.1\rho_0$ with
correlations. The critical densities of the final state are
$\rho_{\rm fin}^{\rm I}\simeq 5\rho_0$ and $\simeq 6\rho_0$,
accordingly \footnote{If the quark matter at $\rho_{\rm fin}^{\rm
I}$ is thermodynamically more favorable compared to the kaon
condensate hadron matter, then during the kaon condensate
transition the system may undergo an extra phase transition to the
quark phase. Here we disregard such a possibility, restricting
ourselves to the consideration of only the kaon condensate phase
transition.}.

 If  surface tension is smaller than the
critical value, the phase transition results in the mixed
phase~\cite{G92,HPS93,CGS00,VYT01}. Then the local
charge-neutrality condition is relaxed  being replaced to the
global charge-neutrality condition. In this case the critical
density for the appearance  of the kaon condensate droplets within
mixed phase is still smaller than that value given by the Maxwell
construction. The presence of the mixed phase may have interesting
observable consequences, see \cite{G01} and references  therein.

Thus, relying on the  analysis above, we argue that the actual
critical density of the
first-order phase transition  can be even smaller than $2\rho_0$ and
that this transition occurs into the p-wave condensate state.

\section{Conclusion}\label{sec:Conclusions}

In this work
we constructed  the $K^-$
polarization operator in dense baryonic matter of arbitrary isotopic
composition, including both the
s- and p-wave $K^-$-baryon interactions, and, using the relativistic
 mean field model to describe the baryon properties.
We applied the derived polarization operator to the issue of the
s- and p-wave $K^-$ condensations in neutron star interiors. We
considered two different models of the equation of state, cf.
Sect. \ref{sec:Baryon} and Appendix \ref{sec:OtherEos} below.
Finite temperature effects can be easily incorporated in our
general scheme.

To describe the kaon - nucleon interaction we used the
kaon-nucleon scattering amplitude obtained as a solution of the
coupled-channel Bethe-Salpeter equation with the interaction
kernel derived from the relativistic chiral SU(3) Lagrangian with
the large $N_c$ constrains of QCD. We  calculated explicitly  the
pole terms of the $K^-$ polarization operator related to $\Lambda
p^{-1}$, $\Sigma N^{-1}$, $\Sigma^* N^{-1}$ excitations with $K^-$
quantum numbers and analyzed effects of the filling of the hyperon
$H=(\Lambda,\Sigma ,\Xi )$ Fermi seas at densities above the
hyperonization point $\rho >\rho_{c,H}\simeq (2.5\div 3)\rho_0$.

In Fig.~\ref{fig:swave}  we compared the s-wave
regular part of our  polarization operator
with the simplified form, which is widely used in the literature.
The $\Sigma$-term extracted
from this comparison ($\Sigma \simeq 150$~MeV) is found to be
two-three times smaller compared to that allows for the
s-wave $K^-$ condensation in ordinary
neutron star matter composed mostly of neutrons.
However, we found the
essential attractive support
from the hyperon exchange terms of the p-wave scattering amplitude
contributing to the s-wave part of the
polarization operator, see (\ref{pols}). Inclusion of these terms,
which were omitted  in previous works,  makes a
second-order phase transition to the s-wave $K^-$
condensate state possible (at densities $\simeq 3\rho_0$ in given
model, when the correlation effects are not included yet).

We evaluated baryon-baryon correlation parameters and corrected
all the s- and p-wave terms of the polarization operator,
accordingly. Inclusion of  the correlations pushes the critical
point of the second order phase transition to the s-wave $K^-$
condensate state up to larger densities ($\gsim (4\div 5)\rho_0$,
cf. Figs.~\ref{fig:swave-col} and~\ref{fig:swave-col-new}). We
studied how much the results are sensitive to a variation of the
correlation parameters. We estimated (see Appendix
\ref{sec:Fluct}) feed-back quantum fluctuation effects arguing
that  their contribution is not too large at the low kaon energies
under consideration, and at zero temperature to the first
approximation they can be neglected.

Our next observation (see Appendix \ref{sec:Fluct}) is that at
$\vec{k}=0$ the imaginary part of the pole term of the
polarization operator is finite only in a rather narrow interval
of kaon energies. Would the electron chemical potential cross  the
$K^-$ branch within this energy interval, the s-wave condensation
would not occur. However this possibility is not realized for the
parameter choice used in our model.

We checked the possibility of a second-order phase transition to
the p-wave $K^-$ condensate state. We showed that, in the vicinity
of the critical point of the s-wave $K^-$ condensation, the p-wave
part of the polarization operator, induced mainly by
$\Lambda$-proton holes and $\Sigma^*$-nucleon holes and some
regular terms, is rather large and attractive. It may change the
sign of the momentum derivative of the energy at the lowest $K^-$
spectrum branch at origin. If occurred, it would mean that we have
p-wave condensate instead of the s-wave one at somewhat smaller
density. We demonstrated that this statement, although being
rather model dependent, holds for a wide range of varying
parameters. The results essentially depend on the unknown value of
the strength of the $\Sigma^*$ mean field potential. In most
favorable case, when $\Sigma^*$ is detached from the mean field
potentials, the second order phase transition to the p-wave
condensate state may occur even at $\rho \sim 3\rho_0$ (with
correlations included), cf. Fig.~\ref{fin}
and~\ref{fig:pwave-col-new}.  This result is also sensitive to the
details of the equation of state and the parameterization of the
hyperon-nucleon interaction. For the equation of state with
parameters (\ref{lagconst2}) the critical density is increased
compared to that one calculated with parameters (\ref{lagconst}).

We discussed the possibility of a first-order phase transition to
a $K^-$ condensate state. We calculated the energies of the
baryonic matter with different compositions with and without the
inclusion of the $K^-$ condensation. We found that with account
for a $K^-$ condensation,  the isospin-symmetrical neutron-proton
matter is more energetically favorable than the standard
nucleon-hyperon-lepton matter at densities $\gsim 3\div 4\rho_0$
(depending on the values of parameters of baryon-baryon
correlations). The hyperon Fermi seas are melting at this phase
transition. Hyperons are replaced by nucleons  and electrons are
replaced by the condensate $K^-$ mesons. We demonstrated that in
dense isospin symmetrical $np$-matter   the $K^-$ excitations are
condensed in the  p-wave state. With the help of the Maxwell
construction we found that the critical density of the beginning
of the phase transition is  about $2\rho_0$ with the baryon-baryon
correlations included, cf. Figs.~\ref{fig:maxwell} and
\ref{fig:maxwell-new}. The final state density is about $(5\div
6)\rho_0$. Appearance of such a strong first order phase
transition may have interesting observable consequences as blowing
off a part of the exterior of the  neutron star, a strong neutrino
pulls, a gravitational wave, a strong pulsar glitch, etc. These
consequences have been previously discussed in relation to the
first order phase transition to the pion condensate state
\cite{MSTV90}. Here we may expect a stronger energy release
compared to the pion condensate phase transition since typical
energy scale $m_K \gg m_\pi$.

Our derivations can be helpful  not only for description of
neutron star interiors, but also for discussion of kaonic effects
in other nuclear systems, as atomic nuclei and heavy ion
collisions. Therefore, of particular interest is the further more
detailed analysis of the p-wave effects on $K^-$ spectra in
nucleus-nucleus collisions motivated by present SIS and SPS
experiments and the future SIS200 program.

In spite of a number of new effects was incorporated in our
scheme, some other effects might be also important. Present
calculations still suffer of many uncertainties, most of which are
due to the lack of experimental information, e.g. on the coupling
constants,  the absence of unambiguous way for going off-mass
shell and the lack of study of more complicated in-medium
fluctuation effects, which we just roughly estimated in the
present work. Among them, there are the pion softening
effects~\cite{MSTV90}, which can essentially  affect the results
at finite temperatures, and the contribution of the non-linear
meson-meson interaction. The latter may partially suppress the
condensate contribution to the energy beyond the critical point.

Diminishing of the uncertainties
needs further theoretical and experimental work.

\acknowledgments
Authors acknowledge J.~Knoll, T.~Kunihiro, M.~Lutz, T.~Muto, G.~Ripka,
T.~Tatsumi, and W.~Weise  for stimulating discussions.
D.N.V. highly appreciates hospitality and support of GSI Darmstadt.
His work has been supported in part by DFG (project 436 Rus 113/558/0),
and by RFBR grant NNIO-00-02-04012.

\appendix

\section{Variation of the Parameters of the Baryon
Interaction}\label{sec:OtherEos}

In this section  we investigate how much the results of
Sec.~\ref{sec:Condensation} are sensitive to the particular choice
of the EoS. For comparison  with the results obtained within the
relativistic mean-field model with the parameters (\ref{lagconst})
we use here the EoS in parameterization~\cite{HH}, which is a good
fit to the optimal EoS of the Urbana--Argonne group~\cite{Akmal}
up to a 4 times nuclear saturation density, smoothly incorporating
the causality limit at higher densities.

The parameters of the mean-field model are adjusted to the
following bulk parameters of the nuclear matter at saturation:
$\ro=0.16$~fm$^{-3}$, binding energy $E_{\rm bind}=-15.8$~MeV,
compression modulus $K=$250~MeV, symmetry energy $a_{\rm
sym}=28$~MeV, and the effective nucleon mass  $\mn^*(\ro)=0.8\,
m_N$. The corresponding  coupling constants of the Lagrangian
(\ref{lagwal}) are then as follows:
\be \nonumber
&&\frac{\dsp g_{\om N}^2\,m_N^2}{\dsp m_\om^2}=91.2506\,,\quad
\frac{\dsp g_{\sigma N}^2\,m_N^2}{\dsp m_\sigma^2}=195.597\,,
\\ \nonumber
&&\frac{\dsp g_{\rho N}^2\,m_N^2}{\dsp m_\rho^2}=77.4993\,,\quad
b=0.0867497\,,
\\\label{lagconst2}
&& \quad c=0.0805981\,.
\ee

\begin{figure}
\begin{center}
\includegraphics[clip=true,width=7cm]{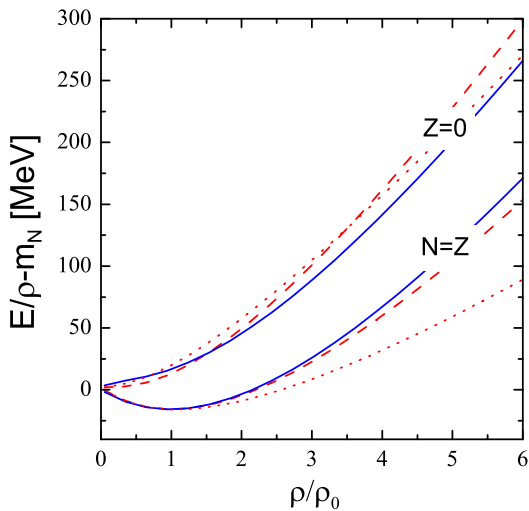}
\end{center}
\caption{EoS for ISM ($N=Z$)  and neutron ($Z=0$) matter
calculated with the parameter set (\ref{lagconst}) (dotted lines)
and with the parameter set (\ref{lagconst2}) (dashed lines). The
solid lines show the EoS from Ref.~\cite{HH}.} \label{fig:eoscomp}
\end{figure}

In  Fig.~\ref{fig:eoscomp} we show energies for the nucleon
isospin-symmetrical matter (ISM) and the pure neutron matter for
two choices of the mean-field EoS, the model (\ref{lagconst}),
simulating a softer EoS and (\ref{lagconst2}) simulating a stiffer
Urbana-Argone EoS.  In spite of the parameters (\ref{lagconst})
and (\ref{lagconst2}) essentially deviate from each other, the
energies and other thermodynamic characteristics of the neutron
star matter are rather closed to each other in  both the parameter
choices in the absence of a $K^-$ condensate. For the ISM case the
EoS with parameters (\ref{lagconst2})  is significantly stiffer
than the one calculated with parameters (\ref{lagconst}) at
$\rho\gsim 3\,\rho_0$.

\begin{figure*}
\includegraphics[clip=true,width=14cm]{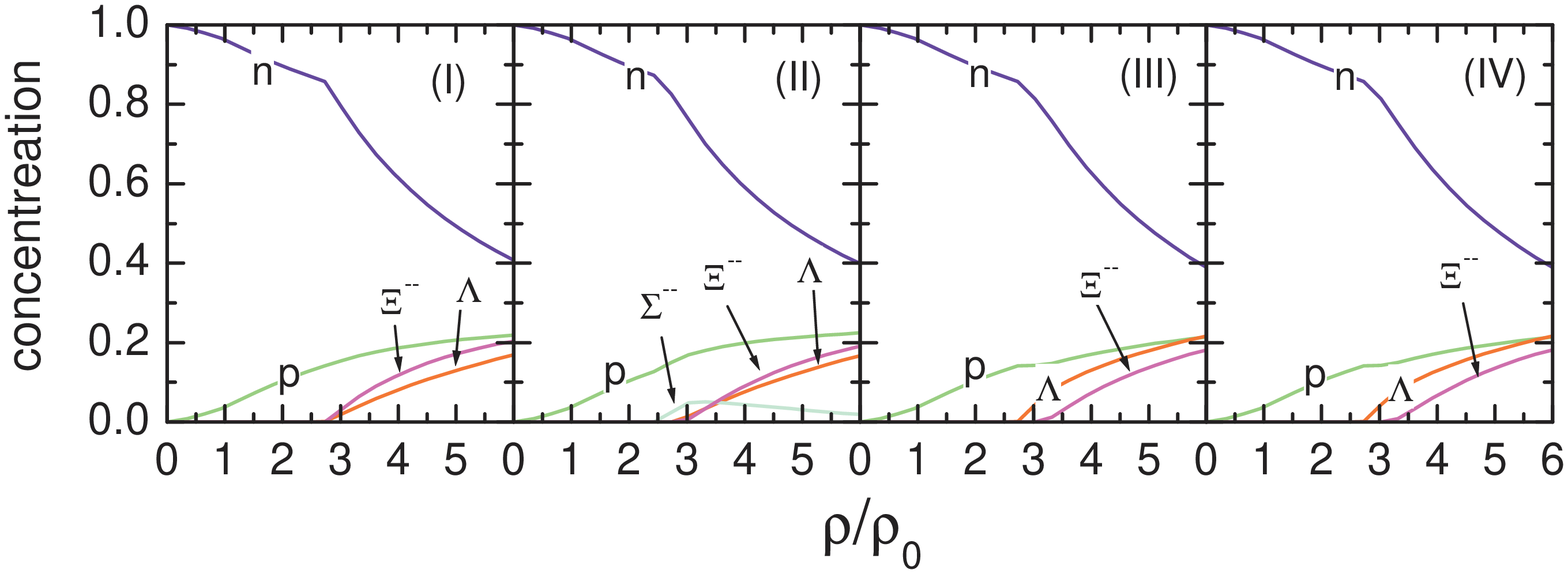}
\caption{ Concentration of baryon species in neutron star matter
for EoS (\ref{lagconst2}). Panels (I-IV) correspond to four
choices of the hyperon-nucleon interaction parameters specified in
text.} \label{fig:compos-new}
\end{figure*}

In Fig.~\ref{fig:compos-new} we show concentrations of the baryon
species in neutron star matter corresponding to EoS given by the
choice (\ref{lagconst2}) for four choices of the hyperon-nucleon
interaction (cases I - IV) which we are using in this paper. We
see that the critical density of the hyperonization is
approximately $3\rho_0$ for all the choices. General trends are
the same as the ones shown by  the corresponding curves discussed
in the main text, see Fig. \ref{fig:compos}. The most essential
difference is that the proton concentrations given by
Fig.~\ref{fig:compos-new} are  smaller than those in
Fig.~\ref{fig:compos}. This should have consequences for the
neutrino cooling of the neutron star. Indeed, when the proton
concentration enlarges $11\div 14\%$ the efficient cooling
mechanism is switched on due to allowance of the direct Urca
processes $p\rightarrow n+e+\bar{\nu}$. In principle, this
difference may allow to select a more appropriate EoS in the
future.

\begin{figure*}
\includegraphics[clip=true,width=14cm]{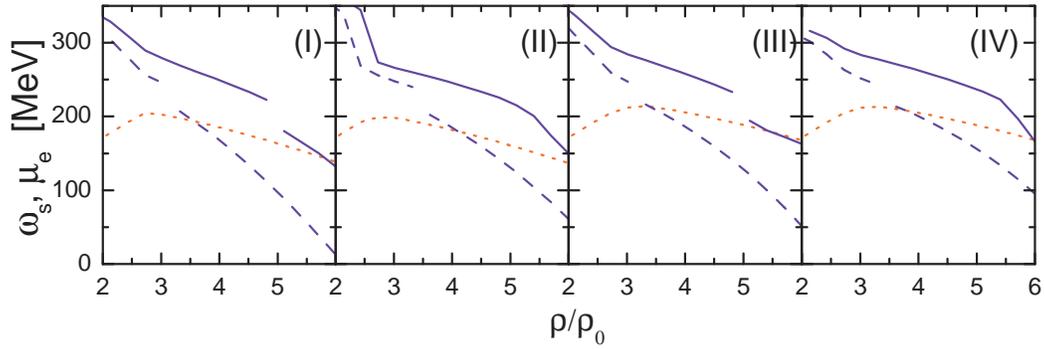}
\caption{
The same as in Fig.~\protect\ref{fig:swave-col} but for the parameters
(\protect\ref{lagconst2}).}
\label{fig:swave-col-new}
\end{figure*}

Fig.~\ref{fig:swave-col-new} demonstrates the energy of the lowest
$K^-$ branch of the dispersion equation (\ref{dys-ful}) for
$\vec{k}=0$ calculated with the mean field model with parameters
(\ref{lagconst2}). Dashed lines are computed without inclusion of
correlations and solid lines, with inclusion of correlations.
Comparing this figure with Fig.~\ref{fig:swave-col} presenting the
same calculation but with the parameter choice (\ref{lagconst}) we
see that the critical densities are increased for all the cases
I-IV for the model (\ref{lagconst2}).  The second order phase
transition to the s-wave $K^-$ condensate state occurs  within the
density interval $(3.5\div 6)\rho_0$ depending on the choice of
the parameters. For the model (\ref{lagconst}) the corresponding
interval of critical densities was $(3\div 5.5)\rho_0$.

\begin{figure*}
\includegraphics[clip=true,width=14cm]{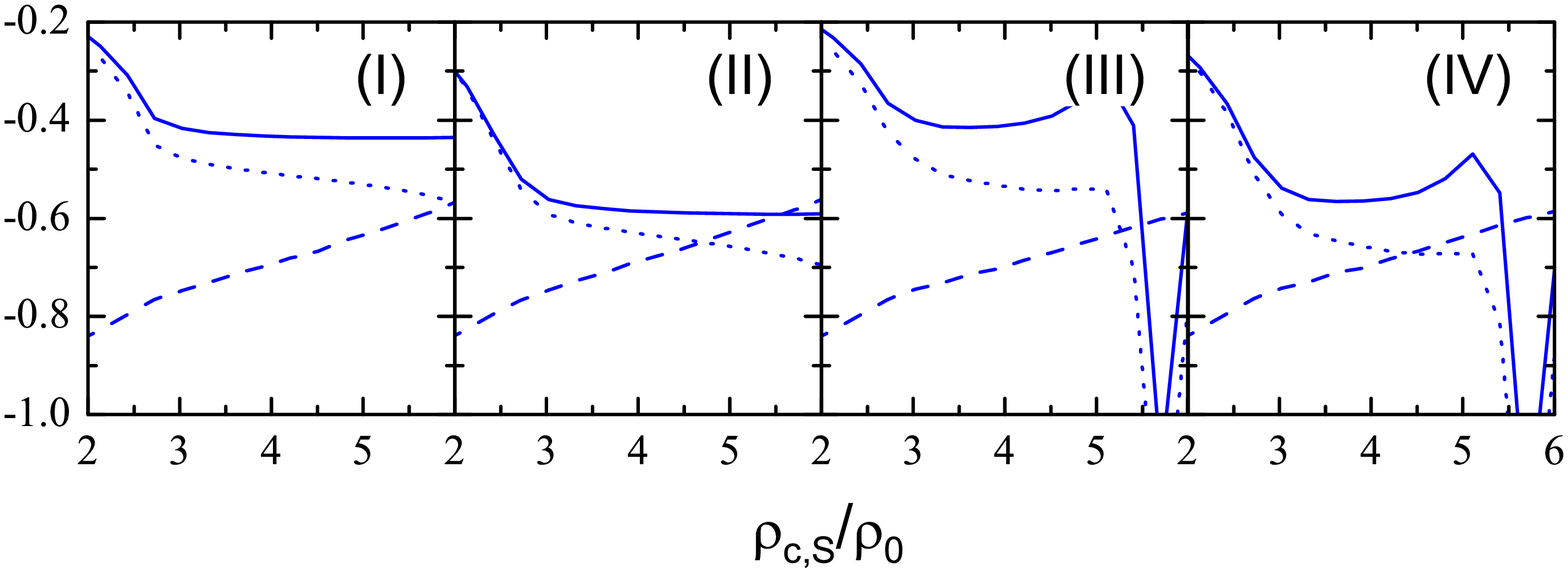}
\caption{
The same as in Fig.~\protect\ref{fin} but for the parameters
(\protect\ref{lagconst2}).}
\label{fig:pwave-col-new}
\end{figure*}

In Fig.~\ref{fig:pwave-col-new} we demonstrate the possibility of
the second order phase transition to the p-wave state within the
model with parameters (\ref{lagconst2}) at the assumption that the
system is in the vicinity of the critical point of the s-wave
condensation. Solid curves relate to the model when $\Sigma^*$ is
attached with the mean field potentials with the same strength as
$\Sigma$ and dashed curves, when $\Sigma^*$ is detached from the
mean field potentials. Comparing the curves with those curves
presented in Fig.~\ref{fin} for the parameter choice
(\ref{lagconst}) we see that the general trends are the same in
both cases. The second order phase transition to the p-wave state
may occur in the model (\ref{lagconst2})  for $\rho \geq (4.5\div
5.5)\rho_0$ in cases II - IV and it does not occur for case I up
to $6\rho_0$.

\begin{figure*}
\begin{center}
\parbox{7cm}{
\includegraphics[clip=true,width=7cm]{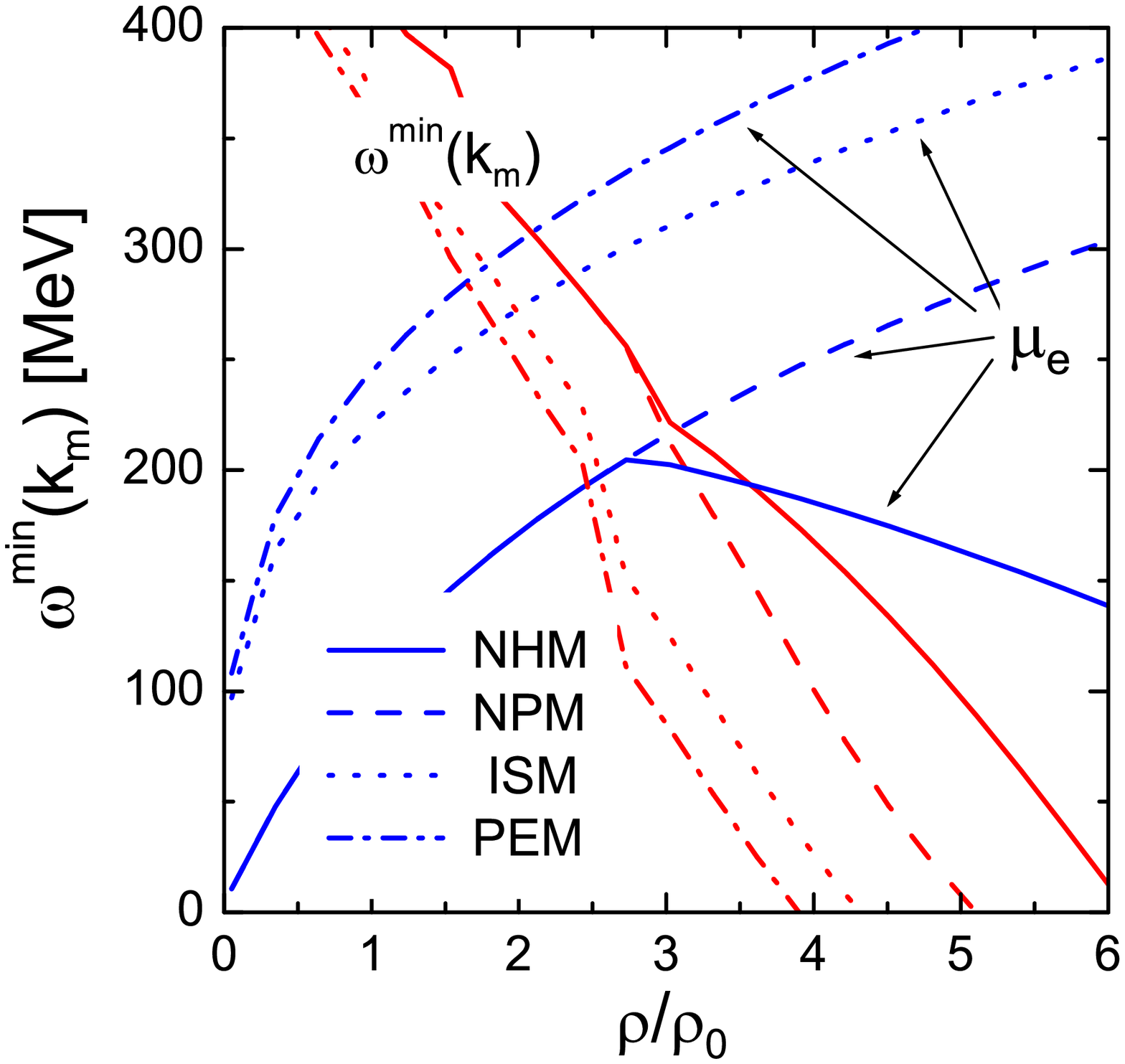}}\qquad
\parbox{7cm}{\includegraphics[clip=true,width=7cm]{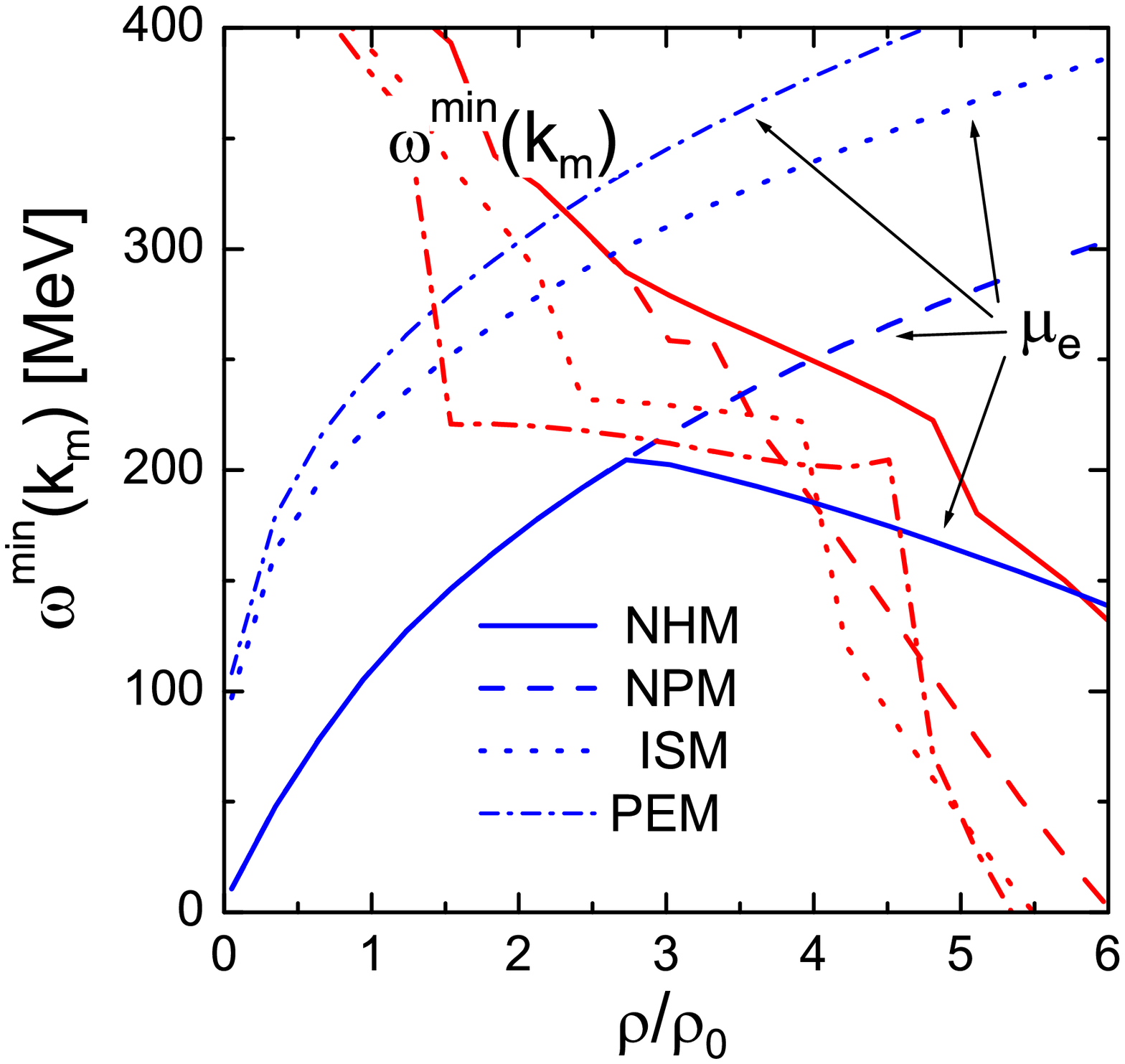}}
\end{center}
\caption{The same as in Fig.~\ref{fig:sc-difm} but for the parameter set
(\ref{lagconst2})}
\label{fig:sc-difm-new}
\end{figure*}

\begin{figure*}
\begin{center}
\parbox{7cm}{
\includegraphics[clip=true,width=7cm]{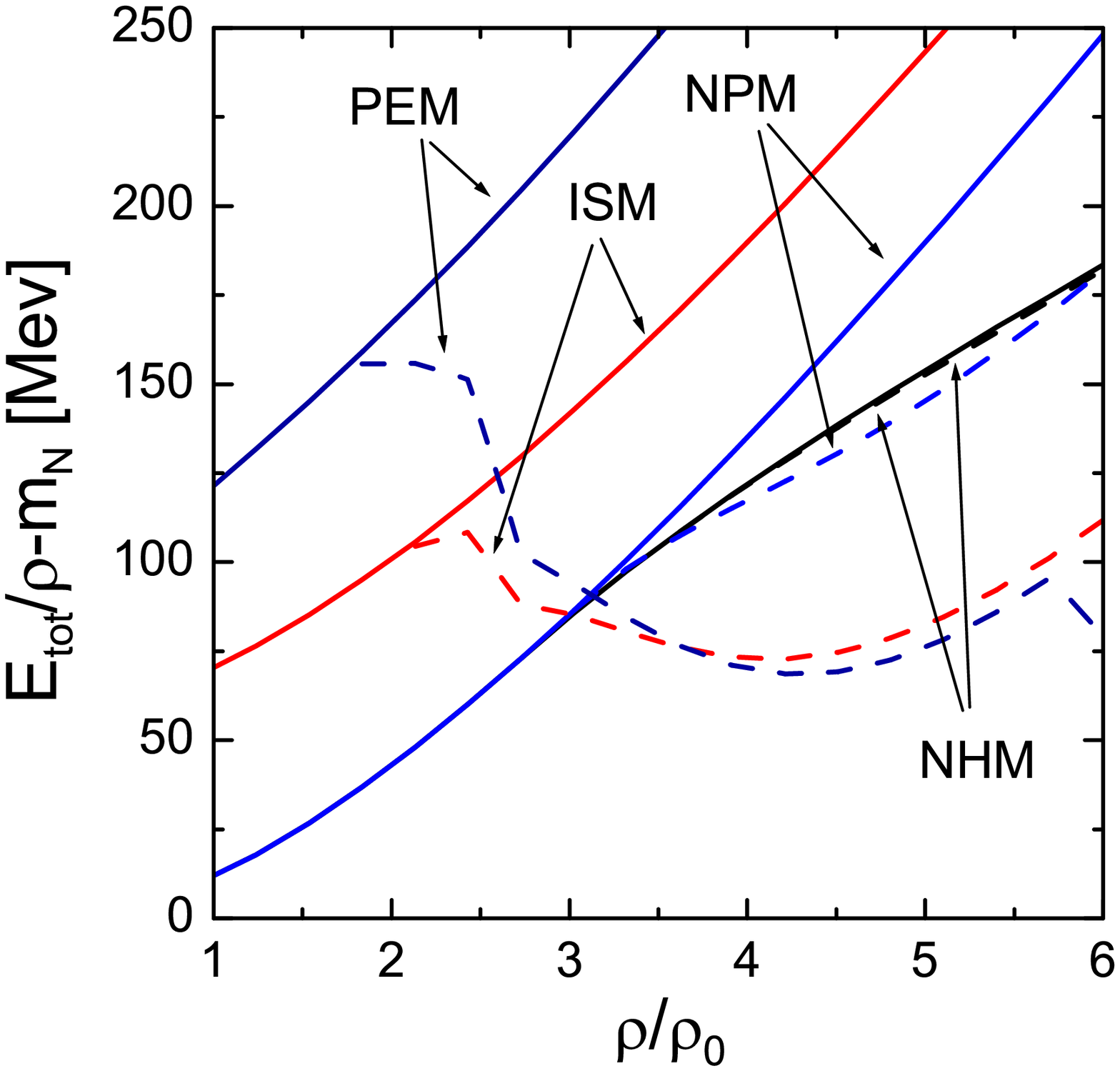}}\qquad
\parbox{7cm}{
\includegraphics[clip=true,width=7cm]{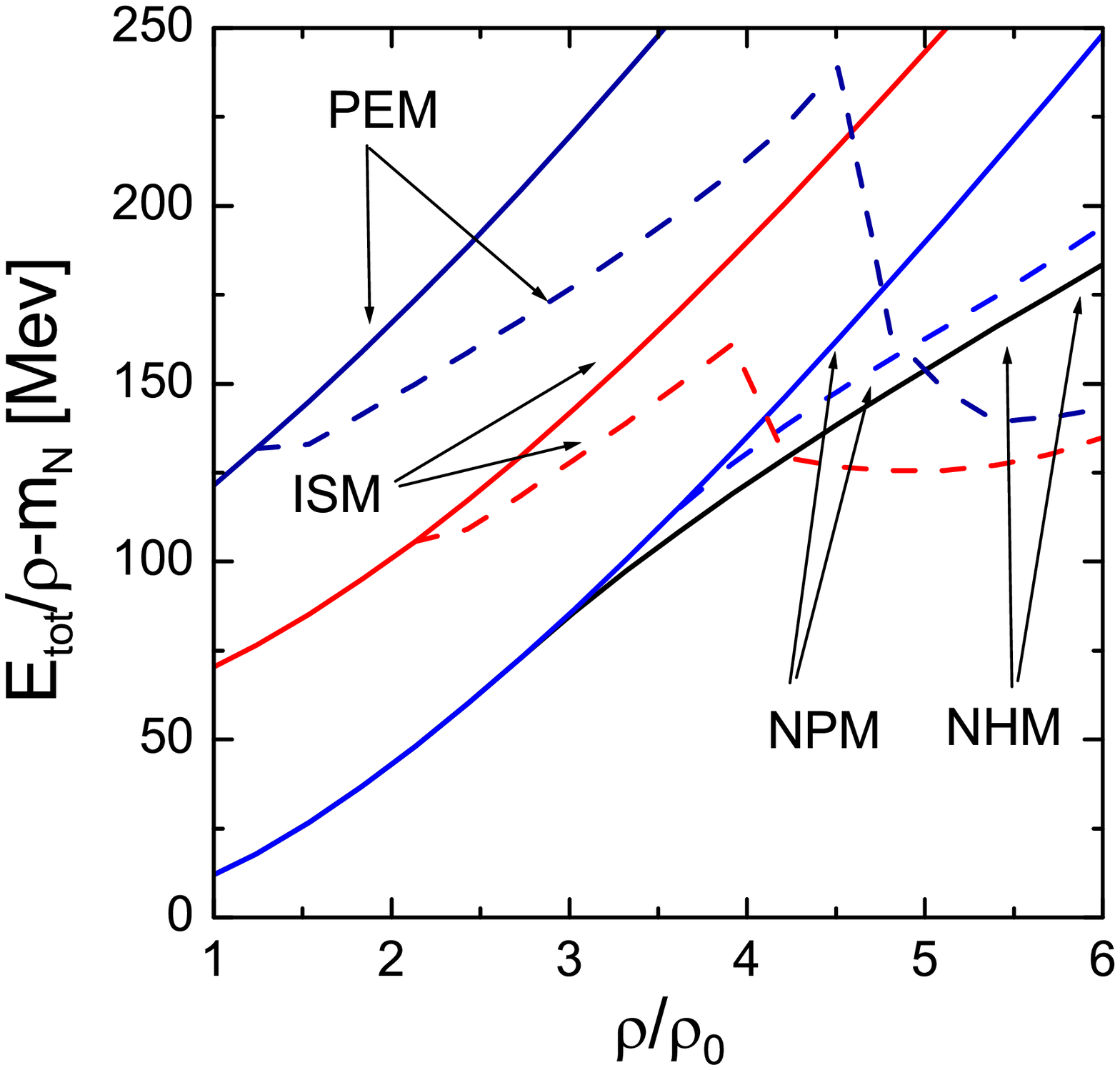}}
\end{center}
\caption{
The same as in Fig.~\protect\ref{fig:energ-s} but for the parameters
(\protect\ref{lagconst2}).}
\label{fig:energ-s-new}
\end{figure*}

\begin{figure*}
\begin{center}
\parbox{7cm}{
\includegraphics[clip=true,width=7cm]{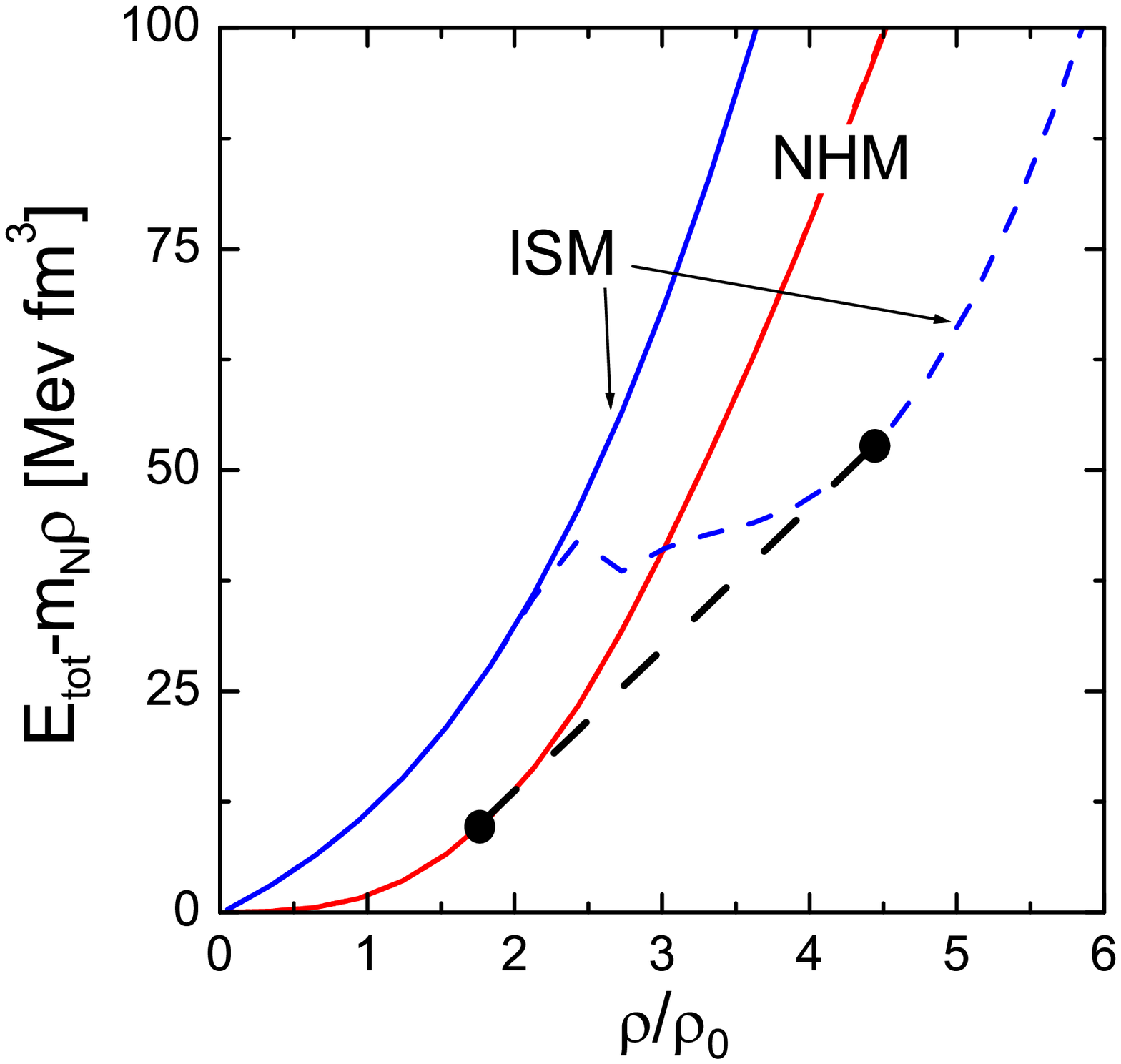}}\qquad
\parbox{7cm}{
\includegraphics[clip=true,width=7cm]{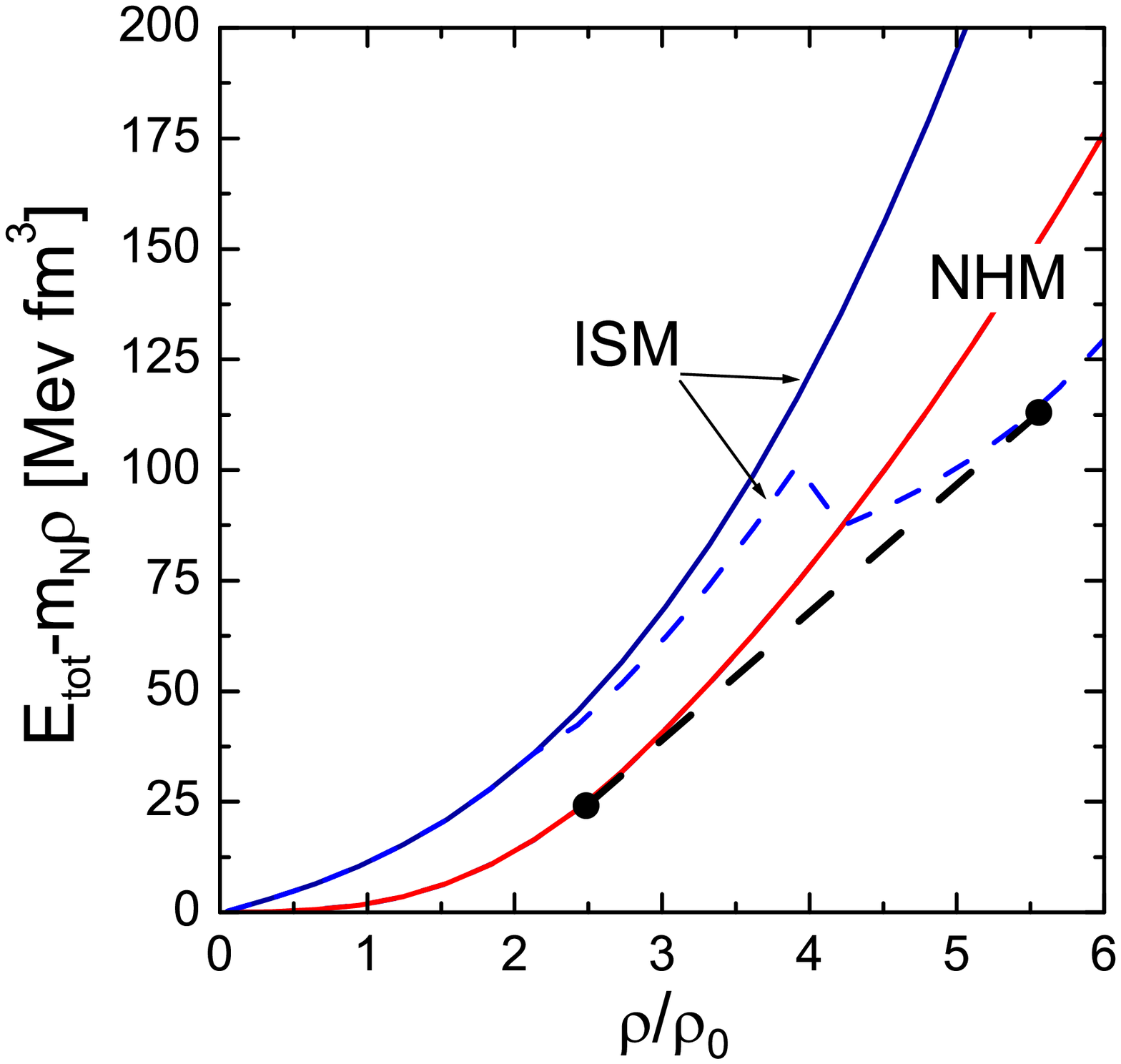}}
\end{center}
\caption{ The same as in Fig.~\ref{fig:maxwell} but for the
parameters (\ref{lagconst2}). } \label{fig:maxwell-new}
\end{figure*}

Figs.~\ref{fig:sc-difm-new}, \ref{fig:energ-s-new},
\ref{fig:maxwell-new} study the possibility of the first-order
phase transition for the EoS with parameters (\ref{lagconst2}).
Again we see the same trends as given by corresponding
Figs.~\ref{fig:sc-difm}, \ref{fig:energ-s}, \ref{fig:maxwell}.
From Fig.~\ref{fig:sc-difm-new} we see that the energy of the PEM
crosses the electron chemical potential at a smaller density
compared to ISM and NHM cases. Fig.~\ref{fig:energ-s-new}
demonstrates that the energy of the ISM with the condensate (with
correlations included)  becomes to be lower than the energy of the
NHM for $\rho \geq 4 \rho_0$. This value ($4\rho_0$) is about the
same  as that for the EoS with parameters (\ref{lagconst}).

In Fig. \ref{fig:maxwell-new} we show the double-tangent
construction for the EoS with parameters (\ref{lagconst2}). The
first order phase transition starts at the density $\rho^{\rm I}_c
\simeq 2.5 \rho_0$ (with correlations included). This value is
only  slightly larger than that ($2.1\rho_0$) given by the EoS
with parameters (\ref{lagconst}). The final state density is
$\rho^{\rm I}_{\rm fin} \simeq 5.5 \rho_0$, i.e. slightly less
than the value ($5.9\rho_0$) given by the model (\ref{lagconst}).

Thus the main conclusion, what we can do from the above comparison
of the results obtained for two EoS with parameters
(\ref{lagconst}) and (\ref{lagconst2}), is that all the general
trends in the behavior  of the kaon condensation are the same. The
critical densities of the s- and p-wave condensations are only
slightly higher in the framework of the model with parameters
(\ref{lagconst2}).

\section{Non-Equilibrium Green's functions}\label{Nonequilibrium}

Two-point (2P) functions (a Green's function or a self-energy)
are introduced within the  Schwinger-Baym-Kadanoff-Keldysh approach as, cf.
\cite{IKV00},
%
\begin{eqnarray}\nonumber
&&i F(x,y) =
\left(\begin{array}{ccc}
i F^{--}(x,y)&&i F^{-+}(x,y)\\[3mm]
i F^{+-}(x,y)&&i F^{++}(x,y)
\end{array}\right)
\\ \label{Fxy}
&&=
\left(\begin{array}{ccc}
\left<{\cal T}\widehat{A}(x)\widehat{B}(y)\right>&\hspace*{5mm}&
\mp \left<\widehat{B}(y)\widehat{A}(x)\right>\\[5mm]
\left<\widehat{A}(x)\widehat{B}(y)\right>
&&\left<{\cal T}^{-1}\widehat{A}(x)\widehat{B}(y)\right>
\end{array}\right),
\end{eqnarray}
%
where ${\cal T}$ and ${\cal T}^{-1}$ are the usual time and anti-time ordering
operators.
Note that, compared to  Green's functions, ``$\pm\,\mp$'' self-energies
 would have extra  sign $-$
since they include vertices $V^- =-iV_0$ and $V^+ =+iV_0$.

Not all four components of $F$ are independent.
The following relations between non-equilibrium and usual
retarded and advanced functions are fulfilled
%
\begin{eqnarray}\label{Fretarded}
F^R(x,y)&=&F^{--}(x,y)-F^{-+}(x,y)\nonumber\\&=&F^{+-}(x,y)-F^{++}(x,y)\nonumber\\
&:=&\theta(x_0-y_0)\left(F^{+-}(x,y)-F^{-+}(x,y)\right),\nonumber\\
F^A (x,y)&=&F^{--}(x,y)-F^{+-}(x,y)\nonumber\\&=&F^{-+}(x,y)-F^{++}(x,y)\nonumber\\
&:=&\theta(y_0-x_0)\left(F^{-+}(x,y)-F^{+-}(x,y)\right),
\end{eqnarray}
%
where $\theta(x_0-y_0)$ is the step function of the time difference.
For such 2P-functions complex conjugation implies
%
\begin{eqnarray}\label{ComplexConjugate}
\left(i F^{-+}(x,y)\right)^{\rm C}&=&i F^{-+}(y,x)
\quad\Rightarrow\quad i F^{-+}(X,p)=\mbox{real},\nonumber\\
\left(i F^{+-}(x,y)\right)^{\rm C}&=&i F^{+-}(y,x)
\quad\Rightarrow\quad i F^{+-}(X,p)=\mbox{real},\nonumber\\
\left(i F^{--}(x,y)\right)^{\rm C}&=&i F^{++}(y,x) \nonumber\\
\quad&\Rightarrow&\quad \left(i F^{--}(X,p)\right)^*=i F^{++}(X,p),\nonumber\\
\left(F^R(x,y)\right)^{\rm C}&=&F^A(y,x)\nonumber \\
\quad\hspace*{3.5mm}&\Rightarrow&\quad \left(F^R(X,p)\right)^*=F^A(X,p),
\end{eqnarray}
%
where the right parts specify the  properties of the 2P-functions in the Wigner
representation,
\be\label{wigner}
F(X,p)=\intop \frac{{\rm d}^4 \xi}{(2\pi)^4}\, e^{-i\,(p\cdot \xi)}
F(X+\xi,X-\xi)
\ee with $X=\frac12(x+y)$ and $p=(\epsilon,\vec{p}\, )$.
The conjugation operation is defined as a
$\big(\dots\big)^{\rm C}=\big(\dots\big)^*$ for bosons and
$\big(\dots\big)^{\rm C}=\gamma_0\,\big(\dots\big)^\dagger\,\gamma_0$ for fermions.

We  denote the  fermionic Green's function and self-energy as
$\widehat G^{a,b}$ and $\widehat \Sigma^{a,b}$, respectively, and
the bosonic ones as $D^{a,b}$ and $\Pi^{a,b}$.
The hats on the fermionic 2P-functions point on their spin structure, e.g.,
$\widehat{G}^{R,A}=[p\!\!\!/-m_{\rm f}
-\widehat{\Sigma}^{R,A}(p)]^{-1}$.

For  systems in equilibrium, the ``$-+$'' Green's functions,
the spectral functions and occupation numbers are related by the
Kubo-Schwinger-Martin condition:
\be
&&i \widehat G^{-+}(p)=-\widehat A_{\rm f}(p)\, n^{\rm f}(\epsilon)\, ,\,
\nonumber\\
&&i \widehat G^{+-}(p)=\widehat A_{\rm f}(p)\, [1- n^{\rm f}(\epsilon)] \,,
\nonumber \\
&&i D^{-+}(p)=A_{\rm b}(p)\, n^{\rm b}(\epsilon)
\,,\,\nonumber \\
&&i D^{+-}(p)=A_{\rm b}(p)\, [1+n^{\rm b}(\epsilon) ]\,,
\label{mp} \\
&&i\widehat \Sigma^{-+}(p)=\widehat \gamma_{\rm f}(p)\, n^{\rm f}(\epsilon)\, ,\,
\nonumber \\
&&i\widehat  \Sigma^{+-}(p)=-\widehat \gamma_{\rm f}(p)\, [1- n^{\rm f}(\epsilon) ]
\,,
\nonumber \\
&&i\Pi^{-+}(p)=-\gamma_{\rm b}(p)\, n^{\rm b}(\epsilon)\, ,\,
\nonumber \\ \label{mps}
&&i \Pi^{+-}(p)=-\gamma_{\rm b}(p) \,[1+n^{\rm b}(\epsilon) ]\,,
\ee
where $\widehat A_{\rm f}(p) =-2\Im\widehat{G}^R_{\rm f}(p)$,
$A_{\rm b}(p) =-2\Im D^R_{\rm b}(p)$ are the
fermion and the boson spectral functions,
$\widehat \gamma_{\rm f}(p) =-2\Im\widehat\Sigma^R_{\rm f}(p)$,
$\gamma_{\rm b} =-2\Im\Pi^R_{\rm b}$  are the corresponding widths,
and
\be
n^{{\rm f,b}}(\epsilon) = \{\exp[(\epsilon -
\mu_{{\rm f,b}} /T] \pm 1\}^{-1}
\ee
are fermion/boson occupation numbers, with
$\mu_{\rm f, b}$ standing for  the  fermionic and bosonic chemical potentials.

In the quasiparticle approximation
($\gamma_{\rm b} \rightarrow 0$ in the bosonic Green's functions) we have
for bosons
\be
A_{\rm b} (q )& \approx &2\pi
\delta [q_0^2-\vec{q}^2  - m_{\rm b}^2  -\Re\Pi^R (q_0,\vec q\,)]
\nonumber \\ \label{qpb}
&=&\sum_i 2\pi Z_{\vec q}^{{\rm b},i}
\delta (q_0 -\omega_{\rm b}^i ({\vec q}))\, ,
\nonumber\\
\Re D_{\rm b}^R(q) & \approx & \sum_i \frac{Z_{\vec q}^{{\rm
b},i}} {q_0 -\omega_{\rm b}^i ({\vec q}) }, \ee
where $Z_{\vec
q}^{{\rm b},i} =1/[ 2q_0 -\partial \mbox{Re}\Pi^R /\partial q_0
]_{q_0=\omega_{\rm b}^{i} (\vec{q})}$ are quasiparticle
normalization corresponding to the given spectrum branch $\om_{\rm
b}^{i}(\vec q\, )$.

As a step to a non-relativistic limit it is convenient to
approximate the spin structure of the fermionic Green's functions
as $\widehat G_{\rm f}^{R}(p)=(p\!\!\!/ +m_{\rm f})\, G_{\rm
f}^R(p)$ and $\widehat A_{\rm f}(p)= (p\!\!\!/ +m_{\rm f})\,
A_{\rm f}(p)$. In the quasiparticle approximation ($\gamma_{\rm f}
\rightarrow 0$) we have \be A_{\rm f} (p )&\approx&2\pi \delta
[\epsilon^2  - \epsilon^2_{0} (\vec{p})-\Re\Sigma^R(\epsilon  ,
\vec{p})] \nonumber \\ \label{qp} &=&  Z_{\vec p}^{\rm f}\,
\frac{\pi}{ \epsilon_{\vec p}}\,  \delta ( \epsilon
-\epsilon_{\vec p}),  \,
\nonumber\\
\Re G_{\rm f}^{R}(\epsilon,\vec p\, ) &\approx&
\frac{1}{2\,\epsilon_{\vec p}}\,\frac{Z_{\vec p}^{\rm f}}
{\epsilon-\epsilon_{\vec p}}
\nonumber \\
Z_{\vec p}^{\rm f} &=& \left[ 1 - \partial
\Re\Sigma^R/\partial\epsilon^2 \Big|_{\epsilon =\epsilon_{\vec
p}}\right]^{-1},
 \ee
where $\epsilon_{\vec p}$ obeys the dispersion equation
 $\epsilon_{\vec p}^2 =\epsilon^{2}_{0} (\vec{p})+
\Re\Sigma^R(\epsilon_{\vec p}, \vec{p})$ with $\epsilon_0
(\vec{p}) =\sqrt{m_{\rm f}^{ 2}+\vec{p}^{\,2}}$. The self-energy
$\Sigma^R$ includes averaging over the spin structure,
$\Sigma^R=\frac12\,{\rm Tr}\{\widehat \Sigma^{R}\cdot
(p\!\!\!/+m_{\rm
  f})\}$

The mean-field solutions (\ref{baren}) of the equation of motion
for a baryon $B$,  following from (\ref{lagwal}), can be
parameterized by the self-energy $\Sigma^R(\epsilon,\vec p\,)= 2\,
V_B \,\epsilon +V_B^2-2\,
 m_N\,g_\sigma\, \sigma+g_{\sigma B}^2\,
\sigma^2$.
Then we have  $\epsilon_{\vec{p}}= E_{B}(\vec p\, )$, with $E_B(\vec
 p\, )$ defined in (\ref{baren})  and
$Z_{\vec p}^{B}=[1-V_B/E_B(\vec p\,)]^{-1}=1+V_B/\epsilon_B(\vec
 p\,)\approx 1$.

The in-medium distributions of particles of given species over the
momenta are \footnote{They differ by prefactors from those at
infinity, cf. \cite{SV88}}
\be\label{distr} \frac{{\rm d}^3 N_{\rm
b}}{{\rm d}^3 X{\rm d}^3 k /(2\pi)^3}&=&
\int_{0}^{\infty}\frac{{\rm d}\epsilon}{2\pi} 2\epsilon A_{\rm b}
(n^{\rm b}_{\epsilon}
+1/2 ),\\
\frac{{\rm d}^3 N_{\rm f}}{{\rm d}^3 X{\rm d}^3 k /(2\pi)^3}&=&
\int_{0}^{\infty}\frac{{\rm d}\epsilon}{2\pi} A_{\rm f} (n^{\rm f}_{\epsilon}
-1/2 ).
\ee
The terms $\mp 1/2$ are due to quantum fluctuations and should be properly
renormalized with the corresponding subtraction of the  vacuum pieces.
The contributions corresponding to the integration over the negative
energies are to be included in the distribution of antiparticles.


\section{Imaginary part of the Lindhard's
function}\label{sec:Imag}

The imaginary part of the Lindhard's function (\ref{lind}) is
determined by the integral
\be\nonumber
\Im\Phi_{iH}(\om,\vec{k})&=& -\frac{m_N^*}{2\,\pi\,|\vec{k}|}
\intop_{m_N^*}^{\epsilon_i(p_{{\rm F} i})} {\rm d} E\,
\theta\Big[4\,\vec{k}^2\,(E^2-m_N^{*\,2})
\\ \nonumber
&-&(2\,\om\,E+\om^2-\vec{k}  ^2
+ m_N^{*\, 2}-m_H^{*\, 2})^2\Big]\,.
\ee
For brevity of notations  we omitted here the vector
potentials of baryons which can be recovered at the very end by
the gauge invariant substitution $\om\to\om+v_{iH}$.
The area where the imaginary part is finite is determined by the
inequality
$$4\,\vec{k}^2\,(E^2-m_N^{*\,2})-(2\,\om\,E+\om^2-\vec{k}  ^2
+ m_N^{*\, 2}-m_H^{*\, 2})^2>0,$$
which has solutions
$E>E^+_{iH}(\om,\vec{k})$ and $E<E^-_{iH}(\om,\vec{k})$
for $\om^2<\vec{k}^2$, and
$E^-_{iH}(\om,\vec{k})<E<E^+_{iH}(\om,\vec{k})$ for
$\om^2<(m_\Lambda^*-m_N^*)^2$ and
$\om^2>(m_\Lambda^*+m_N^*)^2+\vec{k}^2$ with
\be\nonumber
E_{iH}^\pm(\om,\vec{k})=\frac{
\,\left(-\om\,\bar{E}(k^2,m_N^{*\,2},m_H^{*\,2})\pm
|\vec{k}|\,Q(k^2,m_N^{*\,2},m_H^{*\,2})\right)}{\sqrt{\om^2-\vec{k}^2}}
\,.\ee
One can check that $E_{iH}^-(\om,\vec{k})<m_N^*$, if
$\om^2<\vec{k}^2$, and  $E_{iH}^-(\om,\vec{k})>m_N^*$, if
$\om^2>\vec{k}^2$. Therefore, in the case $\om^2<\vec{k}^2$ we
obtain
\be\nonumber
\Im \Phi_{iH}^{(1)}(\om,\vec{k})&=&-\frac{m_N^{*\,2}}{2\,\pi\,|\vec{k}|}\,
\Big(\epsilon_i(p_{{\rm F} i})-E_{iH}^+(\om,\vec{k})\Big)
\\ \nonumber
&\times& \theta\Big(\epsilon_i(p_{{\rm F} i})-E_{iH}^+(\om,\vec{k})\Big).
\ee
The condition $\epsilon_i(p_{{\rm F} i})>E_{iH}^+(\om,\vec{k})$
imposes the following restriction on the frequency:
\be\label{omrange}
\epsilon_H(p_{{\rm F} i}-|\vec{k}|)
<\om+\epsilon_i(p_{{\rm F} i})<
\epsilon_H(p_{{\rm F} i}+|\vec{k}|)\,.\ee
In the other case $\vec{k}^2<\om^2<(m_H^*-m_N^*)^2+\vec{k}^2$  we find
\be\nonumber
&&\Im \Phi_{iH}^{(2)}(\om,\vec{k})
=-\theta\Big(\epsilon_i(p_{{\rm F} i})-E_{iH}^-(\om,\vec{k})\Big)
\\ \nonumber
&&\times \frac{m_N^{*\,2}}{2\,\pi\,|\vec{k}|}\,
\Big(\min\{\epsilon_i(p_{{\rm F} i}),E_{iH}^+(\om,\vec{k})\}
- E_{iH}^-(\om,\vec{k})\Big) .
\ee
For $ \vec{k}^2<\om^2$ the condition $\epsilon_i(p_{{\rm F}
i})<E_{iH}^-(\om,\vec{k})$
puts the same restriction on frequency, as  (\ref{omrange}).
Within this interval  we always have  $\epsilon_i(p_{{\rm F}
i})>E_{iH}^+(\om,\vec{k})$ and hence
\be\nonumber
\Im \Phi_{iH}^{(2)}(\om,\vec{k})
&=&-  \frac{m_N^{*\,2}}{\pi}\,
\frac{Q(k^2,m_N^{*\,2},m_H^{*\,2})}{\sqrt{\om^2-\vec{k}^2}}
\\ \nonumber
&\times&
\theta\Big(\epsilon_i(p_{{\rm F} i})-E_{iH}^-(\om,\vec{k})\Big).
\ee

\section{Evaluation of Fluctuation Terms}\label{sec:Fluct}

Let us discuss corrections  to the
kaon polarization operator which we did not take into account yet.
We mean the  fluctuation terms, which describe changes of the
baryon propagation due to radiation and capture of virtual
mesons, analogous to the Lamb shift in QED.  Calculating
the corresponding diagrams we shall use the
Schwinger-Baym-Kadanoff-Keldysh (SBKK) technique, albeit,
 for systems at equilibrium and zero temperature the standard
Feynman technique is equally appropriate. In the SBKK technique
together with usual causal Green's functions $G^{- -}$ there
appear time-disordered objects $G^{-+}$ and $G^{+-}$ which have
meaning of Wigner's densities, cf. (\ref{Fxy}). In general
non-equilibrium case these Wigner's densities obey the generalized
kinetic equation \cite{IKV00}, being therefore very convenient for
description of the fluctuative effects. Some relations between
non-equilibrium Green's functions were introduced in
Appendix~\ref{Nonequilibrium}. We will  estimate fluctuation
effects to the baryon self-energies and  feedback from the kaon
modification in medium to the polarization operator.

\subsection{Correction of the Baryon Green's Functions Due to Kaon Fluctuations}

In Refs. \cite{LF01,Oset,smed,RO00,ske,KL01} it was argued
that the fluctuation contributions to the hyperon self-energy
are essential  at low densities and at the kaon
energies not far away from the mass-shell. This is, particularly, due to the
presence of the dynamically generated $\Lambda(1405)$ resonance close the
kaon-nucleon threshold. This resonance dominates the s-wave kaon polarization
operator at $\om\sim m_K$ and is very sensitive to the Pauli-blocking effect
and to the modification of the kaon spectral density.  At lower kaon  frequencies,
typical in our case, the influence  of the  $\Lambda(1405)$ resonance
is small. Thus, there remains  to analyze the self-energy contributions of
$\Lambda(1116)$, $\Sigma(1195)$ and $\Sigma^*(1385)$ resonances.
Here we will demonstrate that in the
low energy region and at the proton-neutron
densities of our interest these contributions are rather
suppressed.

Let us first show this on example of the diagram
\be\label{fl}
\Sigma =
\parbox{3cm}{\includegraphics[clip=true,width=3cm]{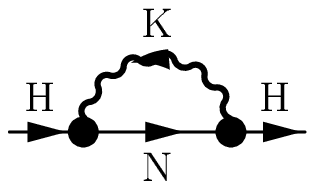}}\,\,\,.
\ee This is the self-energy insertion to the full hyperon Green's
function due to  the $K N$ intermediate states. In (\ref{fl}) we
draw full vertices and the nucleon  line represents the full
Green's function. To be specific let us concentrate on   the
kaon-proton  self-energy insertion for the $\Lambda$ hyperon. This
corresponds to $H=\Lambda$, $K=K^+$ (according to selected arrow
direction), $N=p$ in the diagram (\ref{fl}).

Within the  SBKK diagram technique,
using relations (\ref{Fretarded}) between the retarded and
``$-,+$'' Green's functions and self-energies we obtain, cf.
\cite{KV96},
\be
-i \widehat\Sigma^R (p)&=&\int \widehat  V\left[
\widehat G^{-+}(p +q )D^R (q)\right.\nonumber \\
&+& \widehat G^R (p +q )D^{-+}(q)-i \widehat G^{-+}(p +q )2\Im D^R
(q)
\nonumber \\ \label{expl}
 &+&\left. \widehat G^{R}(p +q )\, D^R (q) \right]\, \widehat V
 \frac{{\rm d}^4 q}{(2\pi )^4}.
\ee Here $-i \widehat V =-i\widehat V_0\, \Gamma$ and $-i\widehat
V_0 $ are   the full and  bare ``$-$'' vertices, and $i\widehat
V=i\widehat V_0 \,\Gamma$, $i\widehat V_0$ are the corresponding
``$+$'' vertices, $\Gamma$ is a scalar form-factor, which
includes short-range nucleon-baryon correlations and simulates,
thereby, the difference between the bare and the full vertices.

The last term in (\ref{expl}) vanishes identically, since the both
retarded Green's functions have the pole in the same complex
$q_0$-semi-plane.

To separate the contributions from particles $K^+$ and
antiparticles $K^-$, we  may use the following decompositions of
the retarded Green's functions and the Wigner's densities
\be\label{expl-1} D^R (q)&=&\theta (q_0)D^R_{K^+}(q)+\theta
(-q_0)D^A_{K^-}(-q)\,,
\\ \label{expl-2}
D^{-+}(q)&=&\theta (q_0)D^{-+}_{K^+}(q)+\theta
(-q_0)D^{+-}_{K^-}(-q)\,. \ee This allows to reduce integration
over $q_0$ in (\ref{expl}) to the positive values.
for $\Lambda$ self-energy we have
 \be
\Sigma^R (p)&=&\frac12\,{\rm Tr}\{\widehat\Sigma^R\,
(p\!\!\!/+m_\Lambda^*)\}= \int  \frac{{\rm d}^4
q\,\theta(q_0)}{(2\pi )^4}\,2\,m_\Lambda^*\,\overline{V^2}
\nonumber \\
&\times& \Big\{
iG^{-+}(p +q )\, [D^R_{K^+}(q)-2\,i\,\Im D_{K^+}^R(q)]
\nonumber\\
&&\quad+ G^R (p +q )\, i D^{-+}_{K^+}(q)
\nonumber \\
&&\quad+iG^{-+}(p -q )\, [D^R_{K^-}(q)-2\,i\,\Im D_{K^-}^R(q)]
\nonumber\\
&&\quad+ G^R (p -q )\, i D^{-+}_{K^-}(q) \Big\}\, , \ee where we
used notation
 \be\label{vert} \overline{V^2} &=&
\overline{V_0}^2\, \Gamma^2(q)\,, \quad
\nonumber \\
 \overline{V_{0}^2}& = &
C^2_{KN\Lambda}\,
\nonumber \\
&\times&\frac12 {\rm Tr}\Big\{
\frac{(p\!\!\!/+m^{*}_{\Lambda})}{2\, m^*_\Lambda}\,
q\!\!\!/\gamma_5\, \frac{(p\!\!\!/-q\!\!\!/+m^{*}_{N} )}{2
m^*_N}\, q\!\!\!/\gamma_5\Big\} . \ee
Having in mind that we are interested in the self-energy insertion to
the $\Lambda-p^{-1}$ loop, expanding  denominators of
the fermionic Green's functions near the poles, and assuming that
fermions are nonrelativistic, we may put $(p \cdot q)\simeq
m_N^{*}\om$, $p^2\simeq m_N^{* 2}$ and $\om,|\vec{q}|,
(m_\Lambda^*-m_N^*)\ll m_\Lambda^* , m_N^*$. Then we come to the
non-relativistic p-wave vertex $\overline{V^2_0} \simeq v_0^2
\vec{q}^{\,2}$,  with $v_0^2 = C_{KN\Lambda}^2\,m_N^*/m_\Lambda^*
\simeq (0.3\div 0.5)/m_{\pi}^2$. For $\Sigma$ hyperon we would
have  $v_0^2 = C_{KN\Sigma }^2\,m_N^*/m_\Sigma^* \simeq (0.07\div
0.09) /m_{\pi}^2$, and for $\Sigma^*$ the corresponding coupling
is $v_0^2 =\frac23\, C_{KN\Sigma^* }^2\,m_N^*/m_{\Sigma^*}^*
\simeq (0.3\div 0.5) /m_{\pi}^2$.

To evaluate $\mbox{Im}\Sigma^R$ and $\mbox{Re}\Sigma^R$ we use
the quasiparticle approximation for the spectral functions
(\ref{qp}). As we shall see below
non-quasiparticle corrections to the intermediate fermion Green's
functions, although exist,  produec small contributions
to the self-energies at kaon energies of interest.

We first evaluate $\Im\Sigma^R$.
Using  (\ref{mp}), (\ref{qpb}), (\ref{qp})
 we get from (\ref{expl})
\be\label{1-im}
&&\frac{\Im\Sigma^R  (\epsilon , \vec{p})}{2 m_\Lambda^*} =
-\sum_i \frac{1}{2}\int\frac{{\rm d}^3 q}{(2\pi )^2}\, v_0^2
\vec{q}^{\,2} \Gamma^2
\nonumber \\
&&\times\Big\{Z_{\vec{q}}^{K^-,i}\, [1-n^p_{\vec{p}-\vec{q}}]\,
\delta (\Delta^i_-(\epsilon,\vec p,\vec q\,) )
\nonumber\\
&&+Z_{\vec{q}}^{K^+,i} n^p_{\vec{p}+\vec{q}}\,
\delta(\Delta^i_+(\epsilon,\vec p,\vec q)  )
 \Big\},
\nonumber \\
&& \Delta^i_{\pm}(\epsilon,\vec p,\vec q\,)= \epsilon\pm
\omega_{K^\pm}^i (\vec{q}) -E_p(\vec{p}\pm\vec{q}), \ee where $i$
runs over  all the $K^-$ and $K^+$ branches. Here $\omega_K^i
(\vec{q})$ are functions of $\vec{q}^{\,2}$, and $n_{\vec
p}^p=\theta(p_{{\rm F},p}-|\vec p\,|)$ is the proton occupation
function. Following  Appendix~\ref{Nonequilibrium} we  put $Z_{\rm f}=1$.
We used also that
occupations of real $K^+$ and $K^-$ mesons are absent at $T=0$ (we do not
consider here the processes on the kaon condensate field).

As it is seen from the $\delta$ functions in (\ref{1-im}), in
the interesting for us region
$\om_{K^-}\lsim E_\Lambda (0)- E_p (0)$
there is only the contribution of the lowest $K^-$ spectrum branch.

Further evaluation is easily done taking
$\vec{p}=0$, for simplicity.
Thus, we obtain that for positive $\epsilon$
the imaginary part differs from zero only  for $\epsilon>
\om_{K^-}(0)+E_p(0)$ and is equal to
\be  \label{sigm-1}
\frac{\Im\Sigma^R (\epsilon , 0)}{2\, m_\Lambda^*} =-\frac{v_0^2}{4\pi\,
\alpha}\, Z_{\bar q}^{K^-}\, \Gamma^2 (\bar{q}) \, \bar{q}^{3}
 \theta (\bar q -p_{{\rm F},p})
\,, \ee
where $\bar {q}$ is a solution of the equation
$\Delta_-(\epsilon,0,\bar q)=0$ and  $\alpha=-
\partial \Delta_-(\epsilon,0,\vec q\,)/\partial
\vec{q}^{\,\,2}|_{\vec{q}^2=\bar{q}^2}>0$.
If $\bar q$ is small, namely, $-\Delta_-(\epsilon,0,0)\,\frac{\prt
\Delta_-(\epsilon,0,q)}{\prt \vec q^2}\Big|_{\vec q^2=0}\ll 1$, we
may approximate $\bar q\approx \bar
q_0=\sqrt{\Delta_-(\epsilon,0,0)/\alpha_0}$, where
$\alpha_0=-\frac{\prt \Delta_-(\epsilon,0,q)}{\prt \vec
q^2}\Big|_{\vec q^2=0}$\,. As it follows from the $\theta$
function  in (\ref{sigm-1}), fluctuations contribute to the
imaginary part of the given diagram only, if $\epsilon >
\omega_{K^-} (0) + E_p(0) + p_{{\rm F}, p}^2 \alpha_0$.

From Fig.~\ref{fig:spectrumSM} we see that $\omega_{K^-}
(\vec{q})$ is very flat function of $\vec{q}^2$, and we can,
therefore, neglect the momentum dependence of the $K^-$ spectrum
for $0<|\vec{q}| \lsim 2~m_\pi$. Then we can estimate the
imaginary part of the diagram  at finite $\vec p$ as follows \be
&&\frac{\Im\Sigma^R (\epsilon , \vec p)}{2\, m_\Lambda^*} =
-v_0^2\!\! \intop_{||\vec p\,|-\bar q_0|}^{|\vec p\,|+\bar
q_0}\!\! \frac{{\rm d} |\vec{q}| \,|\vec{q}|^3}{8\pi\,|\vec p|}\,
Z_{\vec q}^{K^-}\,\Gamma^2\,
\nonumber \\
&& \times 2\,[m_N^*+\Delta_-(\epsilon,0,0)]\,\theta[\Delta(\epsilon,p_{{\rm
F}, p},0)]
\nonumber \\
&& \approx  -\frac{v_0^2\,m_N^*}{2\pi}
\, Z_{\bar q_0}^{K^-}\, \Gamma^2 (\bar{q}_0) \,
\bar{q}_0\,(\bar{q}_0^2+\vec{p}^2)\,
\theta[\Delta(\epsilon,p_{{\rm
F}, p},0)]\nonumber\,.\\ \label{finp}
\ee

The energy and momentum of the $\Lambda$ within the  $\Lambda
p^{-1}$ loop contributing to the $K^-$ polarization operator at
the zero external momentum $\vec{q}$
 are $\epsilon = E_p(\vec{p}) +\om_{K^-}(0) $
with $|\vec{p}| \le p_{{\rm F},p}$. Therefore the condition of
non-zero  $\theta$-functions in (\ref{sigm-1}, \ref{finp}) is not
fulfilled within the momentum integration interval of the $\Lambda
p^{-1}$ loop .

For $\Re\Sigma^R$ using  (\ref{mp}), (\ref{qp}), (\ref{expl-1}),
(\ref{vert}) from (\ref{expl})
we get
\be\label{1-re}
\frac{\Re\Sigma^R (\epsilon ,\vec{p}) }{2\, m_\Lambda^*}
&=&-\sum_i \int\frac{{\rm d}^3 \vec{q}}{(2\pi )^3}\,  v_0^2  \vec{q}^{\,2}
\Gamma^2
\nonumber \\
&\times& \Big\{
\frac{ n^{\rm p}_{\vec{p}-\vec{q}} \, Z_{\vec q}^{K^-,i}}
{\Delta_-^i(\epsilon,\vec p,\vec q\,)}
+\frac{n^{\rm p}_{\vec{p}+\vec{q}}\, Z_{\vec q}^{K^+,i}}
{\Delta_+^i(\epsilon,\vec p,\vec q\,)} \Big\}\,. \ee For relevant
$\epsilon$ the main  contribution  is given by the first term with
the lowest $K^-$ branch. For $| \vec{p}| =0$  the integral in
(\ref{1-re}) is determined by $|\vec{q}|\sim \bar q$, and we can
expand
$\Delta_-(\epsilon,0,\vec q)\approx \alpha\,(\bar{q}^2-\vec{q}^2)$. If we have
$$\frac{\partial Z_{\vec q}^{K,i}}{\partial  {\vec q}^2 }\Big|_{q=\bar q}\,
p_{{\rm F}, p}^2 \ll Z_{\bar q}^{K,i}\,,\quad \frac{\partial
\Gamma }{\partial  {\vec q}^2}\Big|_{q=\bar q}\, p_{{\rm F},p}^2
\ll \Gamma (\bar q ),$$ the remaining integration is
straightforward, and we obtain \be\label{sigm-1-r}
&&\frac{\Re\Sigma^R (\epsilon , 0)}{2\, m_\Lambda^*}
\approx\frac{v_0^2 \Gamma^2 (\bar{q}) Z_{\bar{q}}^{K^-} }{2\pi^2
\, \alpha}\,
\nonumber \\
&&\times\left[\frac{p_{{\rm F},p}^3}{3}
 +\bar{q}^2\,p_{{\rm F},p}
-\bar{q}^2\,|\bar{q}| \,\frac12\ln\left|\frac{|\bar q|+p_{{\rm F},p }} {|\bar
q|-p_{{\rm F},p }}\right|
 \right] \,.
\ee

The extension for finite $\vec p$ can be easily  done  if we
neglect at all the $\vec q$ dependence of the kaon spectrum, that
is minor, as it is justified by our numerical analysis  for
$|\vec{q}|\lsim 2\, m_\pi$. Then we can approximate
$\Delta_-(\epsilon,\vec p-\vec q,0)
\approx (\bar{q}_0^2-(\vec p-\vec q\,)^2)/2\, m_N^*$ and the integration gives
\be\label{sigm-1-rp} \frac{\Re\Sigma^R (\epsilon , \vec p)}{2\,
m_\Lambda^*} &\approx&\frac{v_0^2\, m_N^* }{\pi^2 }\, \Gamma^2
(\bar q_0) Z_{\bar q_0}^{K^-}\,\Bigg[\frac{p_{{\rm F},p}^3}{3}
 +(\bar q_0^2+\vec p^2)\,p_{{\rm F},p}
\nonumber\\
&-&(\bar q_0^2+\vec p^2)\,|\bar q_0|\,\frac12\ln\left|\frac{|\bar q_0|+p_{{\rm F},p }}
{|\bar q_0|-p_{{\rm F},p }}\right|
 \Bigg] \,.
\ee

Let us estimate the real part of the hyperon self-energy at $\vec
p=0$ and $\epsilon=E_p(0)+\om_{K^-}(0)$. We have then
$\bar{q}_0\to 0$ and \be\label{sigm-1-re}\nonumber
&&\frac{\Re\Sigma^R}{2\, m_\Lambda^*} \approx c \equiv  v_0^2\,
m_N^*\, \Gamma^2 (0) Z_{0}^{K^-} \rho_p\sim 0.3\, m_\pi\,\Gamma^2
(0) \, \frac{\rho_p}{\rho_0}\,, \ee where we have used
$m_N^*\simeq 0.6\,m_N$ and $Z_0^{K^-}=Z_{\bar{q} =0}^{K^-}\sim
1/(2\, \om_{K^-}(0))\sim 1/(2\,\mu_e)$\,. Assuming that the
modification of the p-wave $KN\Lambda$ vertex is determined by the
graphical equation (\ref{verteq}) with the $\Lambda p^{-1}$
intermediate states, we can use the corresponding part of
(\ref{pi-pole-cor}) and write
\be\nonumber \Gamma (0)\simeq
\frac{1}{1-f'_\Lambda\, C_0\, \Phi_{p\Lambda}(\om_{K^-}(0),0)}\,.
\ee
In order to estimate the Lindhard's function here, we use
(\ref{phi-exp}), having \be \label{lines}
&&\Phi_{p\Lambda}(\om_{K^-}(0),0)\nonumber
\\
&&\approx\frac{2\,m_N^*\,
\rho_p}{\Delta_{p\Lambda}^-(\om_{K^-}(0),0,p_{{\rm F}, p})}\simeq
\frac{\rho_p}{\Delta},
\\ \nonumber
&&\Delta=\om_{K^-}(0)-E_\Lambda(0)+E_p(p_{{\rm F}, p})\,. \ee The
value of $E_\Lambda(0)-E_p(p_{{\rm F}, p})$ can be estimated from
Fig.~\ref{fig:swave} where it is shown by the dash-dotted line.
Using it, we  estimate $\Delta\sim  - m_\pi$ and therefore
$\Phi_{p\Lambda}\sim -0.5 m_\pi^2\, \rho_p/\rho_0$. Thus, we
obtain $\Gamma (0)\simeq 1/(1+0.3\,\rho_p/\rho_0)$ and
\be \label{est-re} \frac{\Re\Sigma^R}{2\,
m_\Lambda^*} \lsim 0.2 m_\pi\, \ee
for $\rho_p \lsim 3\rho_0$. We
conclude that absolute value of the fluctuation contribution is
small at $\vec p\to 0$ and can be mimic by a variation of  weakly
constrained parameters of Lagrangian (\ref{lagwal}).

From (\ref{sigm-1-r}) and (\ref{sigm-1-rp}) one can see that $\Re
\Sigma^R$ is logarithmically divergent at $|\vec p|\to p_{{\rm F},
p}$ when $\bar{q}\to p_{{\rm F}, p}$. In order to analyze, which
effects it can induce on the $\Lambda p^{-1}$ contribution to the
polarization operator, we separate the leading divergent term
\be\nonumber
\left( \frac{\Re\Sigma^R}{2\, m_\Lambda^*}\right)_{\rm div}
&\simeq &
v_0^2\, m_N^*\, \Gamma^2 (p_{{\rm F},p})
Z_{p_{{\rm F},p}}^{K^-}\, \frac{p_{{\rm F}, p}^3}{\pi^2}\,
\ln\frac{p_{{\rm F}, p}-|\vec{p}|}{p_{{\rm F}, p}}
\\ \label{est-c}
& \simeq & 3c\, \ln\frac{p_{{\rm F}, p}-|\vec{p}|}{p_{{\rm F}, p}}\,,\,\,
 \ee and estimate the variation of the Lindhard's
function (\ref{lind}):
\be\nonumber &&\delta\Phi_{p\Lambda}\simeq
\intop_0^{p_{{\rm F},p}} \frac{{\rm d}|\vec{p}|\,|\vec{p}|^2}{\pi^2}\,
\left[
\frac{1}{\Delta-3\,c\,\ln(1-p/p_{{\rm F},p})}-
\frac{1}{\Delta}\right]\, . \ee
Using estimation (\ref{est-re}) we
write \be\label{est-phi} \frac{\delta
\Phi_{p\Lambda}}{\Phi_{p\Lambda}}\simeq
F\Big(\frac{3c}{\Delta}\Big)\,,\,\, F(a)=a\, \intop^1_0{\rm d}
x\,\frac{(1-x)^2}{1-a\, \ln x}\,, \ee and find that
$$\left|\frac{\delta \Phi_{p\Lambda}}{\Phi_{p\Lambda}}\right|\lsim
0.2\,\quad \mbox{for} \quad \left|\frac{3c}{\Delta}\right|\lsim
0.6\,.$$

Thus, the above estimations prove that the self-energy corrections of the
$\Lambda$ propagator  induced by the kaon fluctuations (diagram (\ref{fl}))
do not modify properties of $K^-$ excitations with small energies and
momenta, which we consider in the  main part of this paper.
The same estimation can be done also for $\bar K^0 n$ contribution to
the $\Lambda$ self-energy and for  $\Sigma$ and $\Sigma^*$ self-energies.
Note that the results (\ref{est-re},\ref{est-c}) and (\ref{est-phi})
are the density-independent estimations from above.

\subsection{Correction of the Baryon Green's Functions Due to Pion Fluctuations}

There is  another type of diagrams
\be\label{d1}
\parbox{3cm}{\includegraphics[clip=true,width=3cm]{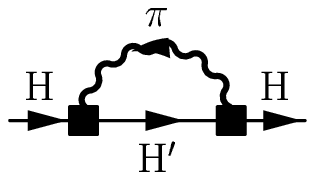}}
\ee relating to  pion fluctuations. Pions are rather soften in
nucleon matter already at densities $\sim \rho_0$. This softening
effect may result in the pion condensation at densities $\rho
>\rho_{c\pi}$, where  the critical density for the pion
condensation, $\rho_{c\pi}$, might be smaller than that for the
kaon condensation \cite{Migdal78,MSTV90}. However, concentrating
in this work on the analysis of the kaon polarization effects we
disregard such a possibility assuming $\rho < \rho_{c\pi}$.

Most essential  contribution to (\ref{d1}) comes from the width of
soft pions due to the Landau damping. Maximum of the pion spectral
function is achieved at small pion energy $q_0 < m_{\pi}$ and
finite momentum $| \vec{q}| =| \vec{q}_{\rm m}| $ for  soft pions.
In the diagram with $\pi^0 $ intermediate states typical momenta
of the hyperons are $\sim p_{{\rm F}n}$, if this self-energy
insertion enters  the $\Sigma n^{-1}$, $\Sigma^* n^{-1}$ parts of
the $K^-$ polarization operator, or they are $\sim p_{{\rm F}p}$,
if this self-energy insertion enters the $\Lambda p^{-1}$, $\Sigma
p^{-1}$, $\Sigma^* p^{-1}$ parts of the $K^-$ polarization
operator. In a simplifying assumption $| \vec{q}_{\rm m}| \ll
2p_{{\rm F}n},2p_{{\rm F}p}$ the nucleon Green's function can be
factored out from the integral. The result is then reduced to the
calculation of the pion tadpole \cite{Dyugaev} \be\label{d2}
\parbox{3cm}{\includegraphics[clip=true,width=2cm]{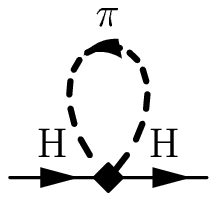}}\,\,\,\,.
\ee

At zero temperature the in-medium contribution of pion
fluctuations given by (\ref{d2}) is numerically small except for a
narrow region near the pion condensation critical point, cf.
\cite{Dyugaev,MSTV90}. For finite temperature such contributions
are substantially increased in the vicinity of the pion
condensation critical point  \cite{VM81}.

\subsection{Fluctuation Contributions to the $K^-$ Polarization Operator}

The fermion self-energy insertions  discussed above, enter
corresponding loop diagrams of the kaon polarization operator.
Since we argued that these contributions are rather small we may
still work with the quasiparticle fermion Green's functions in the
loop diagrams treating fermions on the mean-field level, as we did
it in main part of the text.

One  of important fluctuation processes in the kaon polarization
operator is given by the diagram \be\label{diag-at}
-i\delta\Pi=\parbox{3cm}{\includegraphics[clip=true,width=3cm]{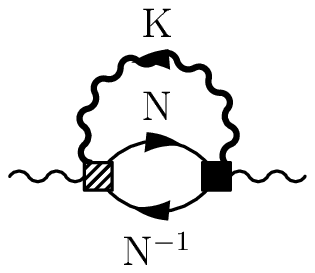}
}\,. \ee The  diagram with free vertices was under extensive
discussion \cite{LF01,Oset} in application to kaonic atoms. Both
the vertices were for simplicity taken  constants \cite{LF01},
being equal to $V_0 \simeq 4\pi (1+m_K
/m_N)\sqrt{(a_{\overline{K}N}^{\rm I=0})^2
+3(a_{\overline{K}N}^{\rm I=1})^2}$ for isospin symmetrical case,
where $a_{\overline{K}N}^{\rm I}$ is the $\overline{K} N$
scattering length for isospin  $I$. In the threshold region  one
obtains  a large value of the vertex $V_0\sim 4\pi /m_{\pi} $. We
will also use the energy independent bare vertex (hatched box) but
estimate $V_0$ using our amplitudes of Figs.~\ref{fig:ampl-s},
\ref{fig:ampl} in the far off-mass shell region. As we see in this
energy region $V_0\sim 1/m_{\pi}$, i.e. a much less value than
that in the threshold region. Already due to this we expect
significant suppression of the contribution of this diagram to the
kaon polarization operator. Another suppression comes from
short-range correlations taken into account in the full box (right
vertex).

With the help of relations (\ref{Fretarded}) from (\ref{diag-at}) we
obtain
\be
&&\delta \Pi^A (k)= \int \Big[ -i\mathcal{P}^{+-}(k+q)\, D^A (q)
\nonumber\\ && +\mathcal{P}^A (k+q)\,iD^{+-}(q) -
\mathcal{P}^{+-}(k+q)2\,\Im D^A (q) \nonumber \\\label{pi-fl} &&
+i\mathcal{P}^A(k+q)\, iD^{A}(q)\Big] \frac{{\rm d}^4 q}{(2\pi)^4}
, \ee
where $\mathcal{P}$ corresponds to the $NN^{-1}$ loop
(including vertices). The last term in (\ref{pi-fl}) is equal to
zero. For further convenience we use here the advanced rather than
the retarded quantities.

Let us consider first the real part of the diagram.
Using (\ref{mp}), (\ref{mps}), we obtain
\be\label{pi-fl-1}
&&\delta \Re\Pi^A (k) =\int
\Big[ [1+n^{K}({k_0 -q_0})]\, \gamma_{\mathcal{P}}(k-q) \Re D^A (-q)
\nonumber\\
&&
+[1+n^{K}({q_0})]\,A_{K}(q)\Re\mathcal{P}^A (k+q)\Big]
\frac{{\rm d}^4 q}{(2\pi)^4}\,,
\ee
where $\gamma_{\mathcal{P}} =2\Im\mathcal{P}^A$
and in the first term we have replaced $q\rightarrow -q$.
Using relations $1+n^{\rm b}_{-q_0}=-n^{\rm b}_{q_0}$
and (\ref{expl-1}) we  reduce integration
in (\ref{pi-fl-1}) to the positive energies:
\be
&&\delta \Re\Pi^A (k) =
\int \frac{{\rm d}^4 q}{(2\pi)^4} \theta (q_0 )
\nonumber \\
&&\qquad\Big[
[1+n^{K}({k_0 -q_0})] \gamma_{\mathcal{P}}(k-q)\, \Re D^A_{K^-}(q)
\nonumber\\
&&\qquad\,\, +[1+n^{K}({k_0 +q_0})]\gamma_{\mathcal{P}}(k+q)\,
  \Re D^A_{K^+}(q)
\nonumber\\
&& \qquad\,\,+[1+n^{K}({q_0})]\,A_{K^{+}}(q)\,\Re\mathcal{P}^A
(k+q) \nonumber \\ \label{pi-fl-10} &&\qquad\,\,
+n^{K}({q_0})\,A_{K^{-}}(q)\,\Re\mathcal{P}^A (k-q) \Big] . \ee
The $N N^{-1}$ loop is
suppressed at large frequencies
$\om > |\vec{k}| v_{{\rm  F},N}+\vec{k}^2/2\, m_N^*$.
Therefore, the
second and the third terms give a small contribution in the energy
region of our interest.  The fourth term is identically  zero
($n^K(q_0)=0$ for $q_0>0$ at $T=0$). Thus we may keep only the
first term in (\ref{pi-fl-10}) and write \be \delta \Re\Pi^A (k)
&\approx& \int \frac{{\rm d}^4 q}{(2\pi)^4} \theta (q_0 )\,V_0^2\,
\Gamma^2\,[1+n^{K}({k_0 -q_0})]
\nonumber \\
&&\times \gamma_0(k-q)\, \Re D^A_{K^-}(q)\,. \ee Here we
introduced  $\gamma_{\mathcal{P}}=V_0^2\, \Gamma^2\, \gamma_0$
using the structure of the correlation factor $\Gamma$ being
expressed via the Lindhard's function  $\Phi^A$, that leads to the
$\Gamma^2$ factor. In the space-like region the
$\gamma_0(k)=\gamma_0(k_0,\vec k\, )$ has a compact analytic
presentation, cf. \cite{V95}, which is further simplified at $T=0$
of our interest: \be\label{gam-0} \gamma_0 (k_0,\vec k\,) =
\frac{m_N^{*2}\,k_0 }{2\pi | \vec{k}|},\, 0 <k_0 < \Omega_-(\vec
k)\,, \ee and \be\label{gam-01} \gamma_0 (k_0,\vec k\,) &=&
\frac{m_N^{*3}}{4\pi | \vec{k}|^3}\, (\Omega_-(\vec{k})+k_0)\,
(\Omega_+(\vec{k})-k_0)\,,
\nonumber \\
&& 0<\Omega_-(\vec k)<k_0<\Omega_+(\vec k )\,,
\nonumber \\
&&\Omega_\pm(\vec k)=\frac{2\,|\vec k\,| p_{{\rm F}, N}\pm \vec
k^2}{2\,m_N^*}\,. \ee

Let us discuss the  low energy contribution (\ref{gam-0}) to the
polarization operator which we label below by subscript ``1''.
Integration in $q_0$ is easily done using that
$$ \max(\nu ,0)< q_0 <k_0\,,\, \nu =k_0  -\Omega_-(\vec k-\vec
q)\,.
$$ Then we get \be\label{pi-fl-11}
 \Re\delta\Pi_1^A (k) &\approx &\frac{V_0^2\, m_N^{*2}}{(2\pi )^2}
\int_{| \vec{k}-\vec{q}|<2\, p_{{\rm F},N}} \frac{{\rm d}^3 \vec{q}}{(2\pi)^3}
\frac{\Gamma^2 Z_{\vec q}^{K^-}}{| \vec{k}-\vec{q}|}
\nonumber\\
&&\times\Big[ (k_0 - \omega_{K^-} (\vec{q}))\ln\frac{(k_0
 -\omega_{K^-} (\vec{q}))}
{\max(\nu,0) -\omega_{K^-} (\vec{q})}
\nonumber\\
&&\quad -k_0+\max(\nu,0) \Big]\,. \ee This contribution can be
easily estimated  for $|k_0 -\om_{K^-}(0)|/k_0 \ll \om_{K^-}(0)$
and $\vec k=0$. We note that due to the flatness of the kaon
spectrum we have $\om_{K^-}(\vec q)\simeq \om_{K^-}(0)$ under the
integral. Hence, $k_0$ is near the branch $\om_{K^-}(\vec q\,)$ we
can neglect the first term in (\ref{pi-fl-11}). Then we have
\be\label{pi-fl-0} &&\Re\delta\Pi_1^A (k_0,0) \simeq
-\frac{V_0^2\, m_N^{*2}}{(2\pi )^2}
\nonumber\\
&&\quad\times\intop_0^{2\, p_{{\rm F},N}} \frac{{\rm d}
|\vec{q}|\, |\vec{q}|}{2\pi^2} \Gamma^2 Z_{\vec
q}^{K^-}\,\min(k_0,\Omega_-(\vec{q}))\,. \ee
For $k_0\sim\om_{K^-}(0)>p_{{\rm F},N}^2 /(2m_N^{*})$ there is  always $k_0
>\Omega_-(\vec q)$, and we  find
\be\label{est2}
\Re\delta \Pi_1^A (k_0,0) &\simeq&
-\frac{V_0^2\,m_N^*}{4\, \pi^2} \Gamma^2 (0)Z_0^{K^-} \rho_N \,
p_{{\rm F},N}\nonumber
\\ &\equiv& - \frac{C\,p_{{\rm F},N}}{4\pi^2}
\,. \ee
For $k_0\sim\om_{K^-}(0)<p_{{\rm F},N}^2 /(2m_N^{*})$,
inequalities $k_0 >\Omega_-(\vec{q})$ or $k_0
<\Omega_-(\vec{q})$ together with $0<|\vec{q}| <2p_{{\rm F},n}$
determine three regions of the momentum integration,
\be
&&\Re\delta \Pi_1^A (k_0,0)\simeq -\frac{V_0^2\,m_N^{*2}}{(2\, \pi
) ^2} \Gamma^2 (0) Z_0^{K^-}
\nonumber\\
&&\times\left[
\left(\intop_0^{q_-(k_0)}+\intop_{q_+(k_0)}^{2\, p_{{\rm F},N}}\right)\frac{{\rm
  d} |\vec{q}|\, |\vec{q}|}{2\, \pi^2}\, \Omega_-(\vec{q})+
\intop_{q_-(k_0)}^{q_+(k_0)} \frac{{\rm
  d} |\vec{q}|\, |\vec{q}|}{2\, \pi^2}\, k_0
\right] , \nonumber \ee
bordered by
$q_\pm(k_0)=p_{{\rm F},N}\pm\sqrt{p_{{\rm F},N}^2 -2m_N^{*} k_0}$ satisfying $k_0
=\Omega_-(q_\pm)$\,. Integration gives \be\label{est2-2}
&&\Re\delta \Pi_1^A (k_0,0)\simeq -\frac{C\,p_{{\rm F},N}}{4\pi^2}
\nonumber \\
&&\times\left[1-\left(1-\frac{2\,k_0\, m_N^*}{p^2_{{\rm
F},N}}\right)^{3/2} \right]\,. \ee

Now we discuss a contribution to the polarization operator from
the energy region (\ref{gam-01}) which we indicate by subscript 2.
\be\nonumber
&&\Re\delta\Pi_2^A (k_0,0)\simeq \frac{V_0^2\,
m_N^{*3}}{(2\, \pi)^4} \intop_0^{2 p_{{\rm F},N}}\frac{{\rm d}
|\vec{q}| }{|\vec{q}|}\, \Gamma^2\, Z^{K^-}_{\vec{q}}\,
\theta(k_0-\Omega_-(\vec{q}))
\\ \nonumber
&&\times \Big[ \int_{k_0-\Omega_+(\vec{q}\,)}^{k_0-\Omega_-(\vec{q}\, )}
 {\rm d} q_0
\frac{(\Omega_-(\vec{q} \,)+k_0-q_0)\,(\Omega_+(\vec{q}\, )-
k_0+q_0)}{q_0-\om_{K^-}} \ee
For $k_0 \sim \om_{K^-}(0)$ and $\vec k=0$ it contributes as
\be\nonumber &&\Re\delta\Pi_2^A (k_0,0)\simeq \frac{V_0^2\,
m_N^{*3}}{(2\, \pi)^4} \intop_0^{2 p_{{\rm F},N}}\frac{{\rm d}
|\vec{q}| }{|\vec{q}|}\, \Gamma^2\, Z^{K^-}_{\vec{q}}\,
\\ \nonumber
&&\times \theta(k_0-\Omega_-(\vec{q}))\,
\Big[ \Omega_-(\vec{q})\, \Omega_+(\vec{q})\,
\ln\frac{\Omega_-(\vec{q})}{\Omega_+(\vec{q})}
\\ \nonumber
&&\quad
-\frac32\Omega_-^2(\vec{q})+2\,\Omega_-(\vec{q})\Omega_+(\vec{q})
-\frac12\Omega_+^2(\vec{q}) \Big]\,, \ee We have neglected the
terms $\propto (k_0-\om_{K^-}(\vec{q}))$ under integral, which are
small due to the flatness of the $K^-$ spectrum. For
$k_0\sim\om_{K^-}(0)>p_{{\rm F},N}^2 /(2m_N^{*})$
 we  find
\be\label{est3}
\Re\delta\Pi_2^A (k_0,0)\simeq 3\frac{C\, p_{{\rm
F},N}}{4\,\pi^2}\, G(0)
\,, \ee
and for $k_0\sim\om_{K^-}(0)<p_{{\rm F}N}^2 /(2m_N^{*})$
we have
\be\nonumber &&\Re\delta\Pi_2^A (k_0 , 0) \simeq
3\,\frac{C\, p_{{\rm F},N}}{4\,\pi^2}\,G(x)\,,
\\ \label{est3-3}
&&x=\sqrt{1-{2 k_0 m_N^*}/{p^2_{{\rm F}, N}}}\,,\ee
where we introduce the function
\be
\nonumber\\
G(x)&=&\frac1{16}\Big(\!\intop_0^{1-x}+\intop_{1+x}^2\!\Big){\rm d} t \, t\,
\nonumber\\
&&\qquad\times\Big[(4-t^2)\ln\frac{2-t}{2+t}-4\,t^2+4\,t\Big]\,,
\nonumber
\\&=&-\frac23+\frac38\, x+ \frac{7}{24}\, x^3
\nonumber\\
&-&\frac{(1+x)(3-x)^2}{64}\, \ln\frac{1+x}{3-x}
\nonumber \\
&+& \frac{(1-x)^2 (3+x)^2}{64}\, \ln\frac{1-x}{3+x}\,,
\nonumber \\
G(0)&=&-\frac23\,,\quad G(x\to 1) \approx -\frac54\,(1-x)\,.
 \ee
The suppression
of the nucleon--nucleon-hole loop in the scalar-isoscalar channel
can be taken into account as in Ref.~\cite{MSTV90} (for recent
review see Ref.~\cite{V01}, eq.~(4 - 8)) $\Gamma=1/[1-2\,(f+f')\,
C_0\, A_{NN}(\om=0,q=p_{{\rm F}, N})]$, where $A_{NN}$ is the
nucleon-nucleon-hole loop (without spin degeneracy factor 2), and
$A_{NN}(\om=0, q=p_{{\rm F}, N})\simeq -m_N\, p_{{\rm F},
  N}/(2\,\pi^2)$.
The Landau-Migdal parameters of $NN$ interaction are $f\simeq 0$
and $f'\simeq 0.5\div 0.6$~\cite{lmpar}. Thus we have $\Gamma\sim
1/(1+0.3\,(\rho_N/\rho_0)^{1/3})$.  For $\rho_N\lsim (3\div 5)
\rho_0$ we have $C\sim m_{\pi}$. When  $\om_{K^-}(0)>p_{{\rm
F},N}^2/(2 m_N^*)$, we estimate the attractive contribution of the
diagram (\ref{diag-at}) as \be\nonumber -\Re\delta\Pi^A \lsim 0.3
m_\pi^2 . \ee For larger densities, when  $\om_{K^-}(0)<p_{{\rm
F},N}^2/(2 m_N^*)$, this contribution is attractive. Its absolute
value is additionally suppressed by the ratio $\om_{K^-}(0)
m_N^*/p_{{\rm F},N}^2 .$

The imaginary part of the diagram under consideration describes
the processes when a kaon excitation dissolves into multi-particle
nucleon--nucleon-hole modes. If the $K^-$ energy reached the
electron chemical potential in the region, where there is  the
width contribution given by the imaginary part of this diagram,
the $K^-$ condensation would not occur.

Using (\ref{pi-fl}) we obtain
\be\label{pi-fl-im}
\Im\delta \Pi^A &=&
\int\frac{{\rm d}^4 q}{(2\pi)^4}
\left[
i\mathcal{P}^{+-}(k+q)\Im D^A (q)\right.
\nonumber \\
&&\quad\quad +\left.\Im\mathcal{P}^A (k+q)\,iD^{+-}(q) \right] .
\ee
With the help of  relations (\ref{mp}), (\ref{mps}) we find
\be\label{pi-fl-im1}
\Im\delta \Pi^A &=&
\int\frac{{\rm d}^4 q}{(2\pi)^4}\,\frac{1}{2}\gamma_{\mathcal{P}} (k+q) A_{K} (q)
\nonumber \\
&&\qquad\times  \left[n^{\rm b}({q_0}) - n^{\rm b}({k_0 +q_0})
\right] . \ee
Using relations $1+n^{\rm b}({-q_0})=-n^{\rm
b}({q_0})$ and (\ref{expl-1}) we  reduce integration in
(\ref{pi-fl-im1}) to the positive energies: \be\label{pi-fl-im111}
&&\Im\delta \Pi^A = \int\!\!\theta (q_0 )\frac{{\rm d}^4
q}{(2\pi)^4}
\nonumber\\
&&\times\frac{1}{2} \Big\{
\gamma_{\mathcal{P}} (k+q) A_{K^+} (q) [n^{\rm b}({q_0}) -
n^{\rm b}({k_0 +q_0})]\nonumber \\
&& +\gamma_{\mathcal{P}} (k-q)A_{K^-}(q)\,[1+n({q_0}) + n({k_0
-q_0})]\Big\}. \ee The second term gives the main contribution.
Taking into account that at $T=0$ there is $(1+n_{q_0}^K +n_{k_0
-q_0}^K )=\theta(k_0-q_0) $, we  have \be\label{pi-fl-im2}
&&\Im\delta \Pi^A =\frac{\gamma }{2} \nonumber\\&&=
\frac{1}{2}\intop_{0}^{k_0}\frac{{\rm d} q_0}{2\pi}\, \frac{{\rm
d}^3 q}{(2\pi)^3} V_0^2\, \Gamma^2\,\gamma_0 (k -q ) A_{K^-} (q)
\,. \ee The $K^-$ spectral function has the form \be\label{sp-f}
&&A_{K^-} (q)=2\pi Z_q^{K^-}\, \delta (q_0 -\om_{K^-}
(\vec{q}))\nonumber \\
&&\quad+ \frac{\gamma (q)}{[q_0^2-m_K^2 -\Re \Pi^A (q)]^2 +
\frac14\gamma^2(q)}, \ee where we separated two different
contributions, the ordinary $\delta$-function term from the branch
and a possible term, which can be obtained only by self-consistent
solution of (\ref{pi-fl-im2}).

Let us now consider the quasiparticle contribution to the spectral
function given by the first term in  (\ref{sp-f}). \be
\gamma(k_0,\vec k\, )&=&\intop_{|\vec k-\vec q|<2\, p_{{\rm F},N}}
\frac{{\rm d}^3 q}{(2\pi)^3}\, \frac{m_N^{*2}\, V_0^2}{2\,
\pi\,|\vec
  k-\vec q\,|}\, \Bigg\{
\nonumber\\
&&[k_0-\om_{K^-}(\vec q)]\, \theta(k_0-\om_{K^-}(\vec q))\,
\nonumber\\
&&\times \theta (\om_{K^-}(\vec q)+\Omega_-(\vec k-\vec q )-k_0)
\nonumber\\
&+&\frac{m_N^*}{2\, |\vec k-\vec q\, |^2}\, [\Omega_-(\vec k-\vec
q )+k_0-\om_{K^-}(\vec{q})]
\nonumber \\
&&\times[\Omega_+(\vec k-\vec q)-k_0+\om_{K^-}(\vec{q})]
\nonumber\\
&&\times\theta(k_0-\Omega_-(\vec k-\vec q)-\om_{K^-}(\vec{q}))
\nonumber \\
&&\times\theta( \Omega_+(\vec k-\vec q
)+\om_{K^-}(\vec{q})-k_0)\Bigg\} \nonumber \ee

Let us, as before, assume that $\om_{K^-} (\vec q)$ is a very flat
function of ${\vec q}^2$.
Then we obtain
\be\label{gaminter} &&\gamma =\theta(z)\, \theta\Big(\frac{p_{{\rm
F},N}^2}{2 m_N^*}-z\Big)
\frac{m_N^{*2}\,V_0^2}{4\pi^3}\, z \intop_{q_-(z)}^{q_+(z)}
Z_{\vec q}^K \Gamma^2 |\vec{q}| {\rm d} |\vec{q}|
\nonumber \\
&&+\frac{m_N^{*3}\,V_0^2}{8\pi^3}\, \intop^{2 p_{{\rm
F},N}}_{-q_-(-z)}\frac{{\rm d} |\vec{q}|}{|\vec{q}|} Z_{\vec q}^K
\Gamma^2 \,[\Omega_-(\vec{q})+z]\,[\Omega_+(\vec{q})-z]
\nonumber\\
&&\simeq \theta(z)\, \theta\Big(\frac{p_{{\rm F},N}^2}{2 m_N^*}-z\Big)
\,
\nonumber \\ &&
\quad\times\frac{m_N^{*2}\, V_0^2 }{2\pi^3} \,z
 \, Z_{0}^{K^-}\,  \Gamma^2 \,
 p_{{\rm F},N}
\sqrt{p_{{\rm F},N}^2 -2 m_N^{*}z}
\nonumber\\
&&+ \theta(z)\, \theta\Big(4\,\frac{p_{{\rm F},N}^2}{m_N^*}-z\Big)
\,\nonumber\\
&&\quad\times\frac{m_N^{*}\, V_0^2 }{32\pi^3}
 \, Z_{0}^{K^-}\,  \Gamma^2 \,p_{{\rm F},N}^4\,H(2\,m_N^*\,z/p_{{\rm F},N}^2),
 \nonumber\\
 && H(x)=\intop_{\sqrt{1+x}-1}^2\frac{{\rm d} t}{t}\,[4\,t^2-(t^2-x)^2]
 \nonumber\\
&&= (1+\frac{x}{2}+\sqrt{1+x})^2-x^2+ x^2\,\ln\frac{\sqrt{1+x}-1}{2}
\nonumber \\
&&H(x\to 0)\approx 4\, (1+x)\,,\, H(x\to 8)\approx \frac13 (x-8)^2
\,, \ee
here $z=k_0 -\om_{K^-} (0 )$\,. As follows from the
$\theta$-functions there is no width for  energies $k_0
<\min_q\{\om_{K^-} (\vec q)\}=\om_{K^-} (\vec q_m)$ and for $k_0 <
\om_{K^-}(0)$. The width exists only above the branch. Exactly at
the branch, i.e. for $k_0 =\om_{K^-} (\vec{k})$, the
$\delta$-function contributes only for $\vec k \neq 0$. Thus  the
quasiparticle part of the $K^-$ spectral function generates no
width at $\vec{k}=0$ at the critical point of the second-order
phase transition to the s-wave $K^-$ condensation: $k_0 =\mu_e
=\om_{K^-} (0))$.

In the region of low energies, $k_0 < \om_{K^-}(\vec{k})$, where
the $\delta$-function term does not contribute, there might appear
another contribution from the self-consistent solution given by
the second term of (\ref{sp-f}).

Let us find this
self-consistent solution and demonstrate that the width exists
even for $\vec{k}=0$ in this case, thus affecting critical
condition for the s-wave condensation. In order to avoid rather
cumbersome expressions, in our estimate we will do several
simplifying assumptions. We will assume that $k_0 \sim \om_{K^-}
(\vec k ) \ll \epsilon_{{\rm F}N}$, and
$\left(\frac{\partial\om_{K^-}}{\partial\vec{q}^2}\right)_0 $ is
very small in the interval $0<| \vec{q}| < k_0 /v_{{\rm F}N}$.
Then the spectral function simplifies as
\be\label{sp-simple}
A_K(q_0)\simeq
\frac{\big(Z_{0}^{K^-}\big)^2\,\gamma (q_0)}{[q_0-\om_{K^-}(0)]^2 +
\frac14\big(Z_{0}^{K^-}\,\gamma(q_0)\big)^2}
\ee
and the
self-consistent solution of equation (\ref{pi-fl-im2}) for the
width $\gamma$ is determined by
\bwt\be
&&\gamma(k_0)\approx\frac{m_N^{*2}\, V_0^2}{8\,
\pi^4}\,\big(Z_{\vec
  {0}}^{K^-}\big)^2\,\Gamma^2
\intop_0^{2\, p_{{\rm F},N}} {\rm d} |\vec{q}| \,
|\vec{q}|\intop_0^{\min\{k_0,\Omega_-(\vec{q})\}}{\rm d} z\,
\frac{z\, \gamma (k_0-z)}{(k_0-z-\om_{K^-}(0))^2 +
\frac14\,\big(Z_{0}^{K^-}\,\gamma(k_0-z)\big)^2}
\nonumber\\
&&+\frac{m_N^{*3}\, V_0^2}{16\,
\pi^4}\,\big(Z_{\vec{0}}^{K^-}\big)^2\,\Gamma^2 \intop_0^{2\,
p_{{\rm F},N}} \frac{{\rm d} |\vec{q}|
}{|\vec{q}|}\intop_{\Omega_-(\vec{q})}^{\min\{k_0,\Omega_+(\vec{q})\}}{\rm
d} z\,
\frac{(\Omega_-(\vec{q})+z)\,(\Omega_+(\vec{q})-z)\,\theta(k_0-
\Omega_-(\vec{q}))\,\gamma (k_0-z)} {(k_0-z-\om_{K^-}(0))^2 +
\frac14\,\big(Z_{0}^{K^-}\,\gamma(k_0-z)\big)^2}
\nonumber\\
&&=\frac{m_N^{*2}\, V_0^2}{8\, \pi^4}\,\big(Z_{\vec
{0}}^{K^-}\big)^2\,\Gamma^2 \left\{
\left[\intop_0^{q_-(k_0)}+\intop_{q_+(k_0)}^{2p_{{\rm F}, N}}
\right]{\rm d} |\vec{q}| |\vec{q}|\intop_0^{\Omega_-(\vec{q})}
{\rm d} z+ \intop_{q_-(k_0)}^{q_+(k_0)} {\rm d}
|\vec{q}||\vec{q}|\intop_0^{k_0} {\rm d} z \right\}
\nonumber\\
&&\qquad\qquad\qquad\times\frac{z\,\gamma (k_0-z)}{(k_0-z-\om_{K^-}(0))^2 +
\frac14\,\big(Z_{0}^{K^-}\,\gamma(k_0-z)\big)^2}
\nonumber\\
&&+\frac{m_N^{*3}\, V_0^2}{16\, \pi^4}\,\big(Z_{0}^{K^-}\big)^2\,\Gamma^2
\left[\intop_0^{q_-(k_0)}+\intop_{q_+(k_0)}^{2p_{{\rm F}, N}}
\right] \frac{{\rm d}|\vec{q}|}{|\vec{q}|}
\intop_{\Omega_-(\vec{q})}^{\min\{k_0,\Omega_+(\vec{q})\}}{\rm
d} z\, \frac{(\Omega_-(\vec{q})+z)\,(\Omega_+(\vec{q})-z)\,\gamma
(k_0-z)} {(k_0-z-\om_{K^-}(0))^2 +
\frac14\,\big(Z_{0}^{K^-}\,\gamma(k_0-z)\big)^2}
\label{gamm1}
\ee
where we did the replacement $q_0=k_0-z$\,.
For $k_0\sim\om_{K^-}\ll p_{{\rm F},N}^2/2 m_N^*$ we have
$q_-\approx k_0\, m_N^*/p_{{\rm F}, N}$, and $q_+\approx 2 p_{{\rm
F},N}-k_0\, m_N^*/p_{{\rm F}, N}$\,. Then (\ref{gamm1}) reduces
to \be &&\gamma(k_0)=\frac{m_N^{*2}\, V_0^2}{8\, \pi^4}\,
\big(Z_{\vec {0}}^{K^-}\big)^2\,\Gamma^2\, \left[ \left(
\intop_0^{\frac{k_0 m_N^*}{p_{{\rm F}, N}}}|\vec{q}|{\rm d}
|\vec{q}|+ \intop_{2p_{{\rm F}, N}-\frac{k_0 m_N^*}{p_{{\rm F},
N}}} ^{2p_{{\rm F}, N}} |\vec{q}|{\rm d} |\vec{q}| \right)
\intop_0^{\frac{|\vec{q}|\, p_{{\rm F},
N}}{m_N^*}-\frac{\vec{q}^2}{2m_N^*} } {\rm d} z +
\intop_0^{2p_{{\rm F}, N}} {\rm d}|\vec{q}|
|\vec{q}|\intop_0^{k_0}{\rm d} z\right] \nonumber \\
&\times&\frac{z\,\gamma (k_0 -z )}{(k_0-z-\om_{K^-}(0))^2 +
\frac14\,\big(Z_{0}^{K^-}\,\gamma(k_0 -z)\big)^2}
\nonumber\\
&&+ \frac{m_N^{*3}\, V_0^2}{16\,
\pi^4}\,\big(Z_{0}^{K^-}\big)^2\,\Gamma^2 \,\left[
\intop_0^{\frac{k_0 m_N^*}{p_{{\rm F}, N}}} +\intop_{2p_{{\rm F},
N}-\frac{k_0 m_N^*}{p_{{\rm F}, N}}}^{2p_{{\rm F}, N}}\right]
\frac{{\rm d} |\vec{q}|}{|\vec{q}|}
\intop_{\Omega_-(\vec{q})}^{\min\{k_0,\Omega_+(\vec{q})\}}{\rm d}
z\, \frac{(\Omega_-(\vec{q})+z)\,(\Omega_+(\vec{q})-z)\,\gamma
(k_0-z)} {(k_0-z-\om_{K^-}(0))^2+
\frac14\,\big(Z_{0}^{K^-}\,\gamma(k_0-z)\big)^2}\,. \nonumber
\ee\ewt To solve this equation we assume $\gamma (k_0)\simeq
\alpha k_0$\, for small values of $k_0$ which we consider here.
The main contribution comes from the third integral of the first
line. It is  $\propto k_0$. In the first integral of the first
line typical values $|\vec{q}|\sim k_0$ and $z \sim k_0$. In the
second integral of the first line  $2p_{{\rm F}, N}-|\vec{q}|\sim
k_0$ and again $z\propto k_0$. Then we estimate the first two
integrals of the first line as $\propto k_0^2$. In the first
integral of the second line with the help of the replacement
$z=|\vec{q}|p_{{\rm F}, N}/m_N^{*} -\vec{q}^2/(2m_N^{*})+\xi$ we
easily see that $\xi \sim \vec{q}^2 \sim k_0^2$ and the integral
is $\propto k_0^4$. In the second integral of the second line
besides this replacement we introduce the replacement
$|\vec{q}|=2p_{{\rm F}, N}-k_0 m_N^{*}/p_{{\rm F}, N} +y$ and
observe that $\xi \propto y\propto k_0$. Then we come to an
estimation that the integral is $\propto k_0^2$. Thus, remaining
only the third integral of the first line we obtain \be\nonumber
&&\alpha k_0 \simeq \frac{m_N^{*2}\, V_0^2}{4\, \pi^4}\,
\big(Z_{0}^{K^-}\big)^2\,\Gamma^2\, p_{{\rm F}, N}^2
\\ \nonumber
&&\times \intop_0^{k_0} \frac{z\alpha (k_0 -z){\rm d} z}
{(k_0-z-\om_{K^-}(0))^2 +\frac14 \big(Z_{0}^{K^-}\big)^2\alpha^2
(k_0-z)^2}, \ee which has a non-trivial solution: \be \gamma
\simeq \alpha k_0 \simeq m_N^{*}V_0 \Gamma p_{{\rm F},N}k_0
/\pi^{2} \,, \ee  for $\alpha > 2/Z_{0}^{K^-}$.  As one may
expect, this inequality indeed holds for rather large densities
since $Z_{0}^{K^-}\sim 1/(2\om_{K^-}(0)) \gg 1/m_{\pi}$ for small
values  $k_0 \sim \om_{K^-}(0)$.

The possible presence of a width at low energies is a principal
question. The second-order phase transition to the s-wave $K^-$
condensation cannot occur, if $\om_{K^-} (0)$ crosses $\mu_e$ in
the energy region, where exists imaginary part of the polarization
operator. The possibility of the condensation is however
recovered, when the electron chemical potential exceeds $\om_{K^-}
(0)$ reaching the upper boarder of the region of the width. At
least for $\om_{K^-} (0) > p_{{\rm F}N}^2 /(2m_N^{*})$ we did not
find self-consistent solution for the width. Thereby we would like
to stress that the second order phase transition to the s-wave
$K^-$ condensation may not occur only if $\om_K (0)=\mu_e <
p_{{\rm F}n}^2 /(2m_N^{*})$. For realistic values of parameters we
have $\om_K (0)=\mu_e > p_{{\rm F}n}^2 /(2m_N^{*})$ without any
problem for the s- wave $K^-$ condensation.

Thus, by the above examples we demonstrated that in spite of many
new peculiarities associated with the fluctuation processes for a
rough analysis we may  drop the fluctuation contributions treating
baryons on the mean field level.

Similar diagram  to (\ref{diag-at}) (but with the pion instead of
the kaon and a  different vertex) also describes the $K^-\pi - \pi
K^-$ interaction. In a simplifying assumption that the pion
momentum $|\vec{q}_m|\ll 2p_{{\rm F},n} $ the diagram is reduced
to
\be\label{d3}
\parbox{3cm}{\includegraphics[clip=true,width=2cm]{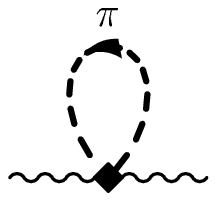}}\,\,\,\,,
\ee
the same as (\ref{d2}) but with a different vertex and the
kaon external Green's functions. As we mentioned such a
contribution can be easily estimated. At zero temperature the
in-medium contribution of pion fluctuations given by (\ref{d2}) is
numerically small except for a narrow region near the pion
condensation critical point, cf. \cite{Dyugaev,MSTV90}. For finite
temperature such contributions are substantially increased in the
vicinity of the pion condensation critical point \cite{VM81}.



\begin{thebibliography}{99}
\bibitem{hic} R.~Stock, Phys.\ Rep.\ {\bf 135}, 259 (1986);
P.~Senger and H.~Str\"obele, J.\ Phys.\ G {\bf 25}, R59 (1999);
H.R.~Schmidt and J.~Schukraft,
J.\ Phys.\ G {\bf 19}, 1705 (1993);
W.~Reisdorf, H.G.~Ritter, An. Rev. Nucl. Part. Sci {\bf 47}, 663 (1997).
\bibitem{advent}
D.~R\"ohrich, Proc. Int. Conf. "Quark Matter 2002", July 18-24, Nantes,
France (2002);
P.~Senger, Nucl.\ Phys.\ A {\bf 685}, 312 (2000).
\bibitem{krive}
V.P.~Berezovoy, I.V.~Krive, E.M.~Chudnovsky,
Sov.\ J. Nucl.\ Phys.\ {\bf 30}, 581 (1979).
\bibitem{swave}
D.~Kaplan, A.~Nelson, Phys.\  Lett.\ B {\bf 175}, 57 (1986);
\bibitem{bb} G.E.~Brown, H.A.~Bethe, Astrophys. J. {\bf 423}, 659
(1994).
\bibitem{kvk95}
E.E.~Kolomeitsev, D.N.~Voskresensky, B.~K\"ampfer,
Nucl.\ Phys.\ A {\bf 588}, 889 (1995).
\bibitem{nscool}
G.E.~Brown, K.~Kubodera, D.~Page, P.~Pizzochero,
Phys.\ Rev.\ D {\bf 37}, 2042 (1988);
H.~Fujii, T.~Muto, T.~Tatsumi,  R.~Tamagaki, Nucl.\ Phys.\ A {\bf 571}, 758
(1994).
\bibitem{swave-1}
T.~Muto, T.~Tatsumi, Phys.\  Lett.\ B {\bf 283}, 165 (1992);
G.E.~Brown, V.~Thorsson, K.~Kubodera, M.~Rho, Phys.\ Lett.\ B {\bf
291}, 355 (1992).
\bibitem{swave-2}
G.E.~Brown, C.H.~Lee, M.~Rho, V.~Thorsson, Nucl.\ Phys.\ A {\bf 567}, 937
(1994);
C.H.~Lee, G.E.~Brown, D.P.~Min, M.~Rho, Nucl.\ Phys.\ A {\bf 585}, 401 (1995).

\bibitem{tpl94}
V.~Thorsson, M.~Prakash, J.M.~Lattimer, Nucl.\
phys.\ A {\bf 572}, 693 (1994); ibid. A {\bf 574}, 851 (E) (1994).
\bibitem{gs}
N.K.~Glendenning, J.~Schaffner--Bielich, Phys.\ Rev.\ C {\bf 60}, 025803 (1999).
\bibitem{pw}
T.~Muto, Prog.\ Theor.\ Phys.\ {\bf 89}, 415 (1993);
H.~Yabu, S.~Nakamura, F.~Myhrer, K.~Kubodera, Phys.\ Lett.\ B {\bf 315}, 17
(1993).
\bibitem{muto}
T.~Muto, Nucl.\ Phys.\ A {\bf 697}, 225 (2002).
\bibitem{kv99}
E.E.~Kolomeitsev, D.N. Voskresensky, eprint: nucl-th/0001062
\bibitem{kv98}
E.E.~Kolomeitsev, D.N. Voskresensky, Phys.\ Rev.\ C {\bf 60}, 034610 (1999).
\bibitem{lk01}
M.F.M.~Lutz, E.E.~Kolomeitsev, Nucl.\ Phys.\ A {\bf 700}, 193 (2002).
\bibitem{kvk96}
E.E.~Kolomeitsev, D.N.~Voskresensky, B.~K\"ampfer,
Int. J. Mod. Phys. E {\bf 5}, 313 (1996).
\bibitem{kaos}
F.~Laue et al., Phys.\ Rev.\ Lett.\  {\bf 82}, 1640 (1999)
C.~Sturm et al., J.\ Phys.\ G {\bf 28}, 1895 (2002).
\bibitem{LF01}
M.F.M.~Lutz, W.~Florkowski,
Acta Phys.\ Polon.\ B {\bf 31}, 2567 (2000); eprint: nucl-th/0004020.
\bibitem{Oset}
G. Garcia-Recio, J.~Nieves, E.~Oset, A.~Ramos, Nucl.\ Phys.\ A {\bf 703}, 271
(2002).
\bibitem{walecka} N.K.~Glendenning, Astrophys.\ J.\ {\bf 293}, 470 (1985).
\bibitem{softeos}
B.~Friedman, V.R.~Pandharipande,
Nucl.\ Phys.\ A {\bf 361}, 502 (1981).
\bibitem{Lombardo} U.~Lombardo, W. Zuo, in "Isospin Physics in
Heavy-Ion Collisions at Intermediate Energies", Eds. B.A.~Li,
W.U.~Schr\"oder, Nova Science Publisher (2001, N.Y.).
\bibitem{MSTV90}
A.B.~Migdal, E.E.~Saperstein, M.A.~Troitsky, D.N.~Voskresensky,
 Phys.\ Rep.\ {\bf 192}, 179 (1990).
\bibitem{Piek} J.~Piekarewicz, eprint: nucl-th/0205007.
\bibitem{Akmal}
A.~Akmal, V.R.~Pandharipande, D.G.~Ravenhall,
Phys.\ Rev.\ C {\bf 58}, 1804 (1998).
\bibitem{sbg} J.~Schaffner-Bielich, A.~Gal, Phys.\ Rev.\ C {\bf 62 },
034311 (2000).
\bibitem{glenmos}
N.K.~Glendenning, S.A.~Moszkowski, Phys.\ Rev.\ Lett.\ {\bf 67}, 2414 (1991).
\bibitem{lnucl} H. Band\={o}, T.~Motoba, J. \v{Z}ofka,
Int.\ J.\ Mod.\ Phys.\ A {\bf 5}, 4021 (1990)\\
B.F.~Gibson, E.V.~Hungerford~III, Phys.\ Rep.\  {\bf 257}, 349 (1995).
D.J.~Millener, C.B.~Dover, A.~Gal, Phys. Rev. C {\bf  38}, 2700  (1988).
\bibitem{sbind} C.B.~Dover, D.J.~Millener, A.~Gal, Phys.\ Rep.\ {\bf 184}, 1
(1989);
C.J.~Batty, E.~Friedman, A.~Gal, Phys.\ Lett.\ B {\bf 335}, 273
(1994);
J.~Mare\v{s}, E.~Friedman, A.~Gal, B.K.~Jennings, Nucl.\ Phys.\ A {\bf 594},311
(1995).
\bibitem{xibind}
C.J.~Batty, E.~Friedman, A.~Gal, Phys.\ Rev.\ C {\bf 59}, 295 (1999).
\bibitem{gcoupl}
N.K.~Glendenning, Phys.\ Rev.\ C  {\bf 64}, 025801 (2001).
\bibitem{kaiser}
N.~Kaiser, P.~Siegel, W.~Weise, Nucl.\ Phys.\ A {\bf 612},
297 (1997).
\bibitem{krippa}B.~Krippa,
Phys.\ Rev.\ C {\bf 58}, 1333 (1998);
B.~Krippa, J.T.~Londergan,
Phys.\ Rev.\ C {\bf 58}, 1634 (1998).
\bibitem{or98}
E.~Oset,  A.~Ramos, Nucl.\ Phys.\ A {\bf 635}, 99 (1998).
\bibitem{om01}
J.A.~Oller, U.-G.~Mei\ss{}ner, Phys.\ Lett.\ B {\bf 500}, 263 (2001).
\bibitem{kmatrix}
A.D.~Martin, Nucl.\ Phys.\  B{\bf 179}, 33 (1981).
\bibitem{keil}
M.Th.~Keil, G.~Penner, U.~Mosel, Phys.\ Rev.\ C {\bf
63} (2001) 045202.
\bibitem{smed}
T.~Waas, N.~Kaiser, W.~Weise, Phys.\ Lett.\  B {\bf 379}, 34
(1996);\\
M.~Lutz, Phys.\ Lett.\  B {\bf 426}, 12 (1998).
\bibitem{RO00}
A.~Ramos, E.~Oset, Nucl.\ Phys.\ A {\bf 671}, 481 (2000).
\bibitem{ske} J.~Schaffner--Bielich, V.~Koch, M.~Effenberger,
Nucl.\ Phys.\ A {\bf 669}, 153 (2000).
\bibitem{tolos} 
L.~Tol\'os, A.~Ramos, A.~Polls,  T.T.S.~Kuo,
Nucl.\ Phys.\ A {\bf 690}, 547 (2001);
L.~Tol\'os, A.~Ramos, A.~Polls, Phys.\ Rev.\  C {\bf 65} 054907  (2002).
\bibitem{KL01}
M.F.M.~Lutz, C.L.~Korpa, Nucl.\ Phys.\ A {\bf 700}, 309 (2002)
\bibitem{Landoldt} G.~H\"ohler, in Landolt-B\"orstein, Vol. I/9b2,
ed. H.~Schopper (Springer, Berlin, 1983).
\bibitem{Julich}
B. Holzenkamp, K.~Holinde, J.~Speth, Nucl.\ Phys.\ A {\bf 500},
485 (1989).
\bibitem{EW88} T.O. Ericson, W. Weise, ``Pions in
Nuclei'', Clarendon, Oxford (1988).
\bibitem{Bruk} L.L.~Foldy, J.D.~Walecka, Ann.\ Phys.\ (N.Y.) {\bf
36}, 447
(1969);
V.R.~Pandharipande, H.A. Bethe, Phys.\ Rev.\ C {\bf 7}, 1312 (1973).
\bibitem{jb78}
M.B.~Johnson, H.A.~Bethe, Nucl.\ Phys.\ A {\bf 305}, 418 (1978);
M.B.~Johnson, B.D. Keister, Nucl.\ Phys.\ A {\bf 305},
461 (1978)
\bibitem{bbow}
G.E.~Brown, S.O.~B\"ackman, E.~Oset, W.~Weise,
Nucl.\ Phys.\ A {\bf 286}, 191 (1977);
E. Oset, H.~Toki, W.~Weise, Phys.\ Rep.\ {\bf 83}, 281 (1982).
\bibitem{FW}A.L.~Fetter, J.D.~Walecka, Quantum Thoery of
Many-Particle Systems, McGraw-Hill, 1971.
\bibitem{ee66}
M.~Ericson, T.E.O.~Ericson, Annals\ Phys.\ {\bf 36}, 323 (1966).
\bibitem{f73}
G.~F\"aldt, Nucl.\ Phys.\ A {\bf 206}, 176 (1973).
\bibitem{corr-pand}
V.R.~Pandharipande, C.J.~Pethick, V.~Thorsson,
Phys.\ Rev.\ Lett.\   {\bf 75}, 4567 (1995).
\bibitem{wrw97}
T.~Waas, M.~Rho, W.~Weise, Nucl.\ Phys.\ A {\bf 617}, 449 (1997).
\bibitem{w77}
W.~Weise, Nucl.\ Phys.\ A {\bf 278}, 402 (1977);
\bibitem{tkfs}
A.B.~Migdal, Theory of Finite  Fermi Systems and
application to atomic nuclei, (Wiley, New York, 1967).
\bibitem{correst}
G.~Baym, G.E. Brown, Nucl.\ Phys.\ A {\bf 247}, 395 (1975);
\bibitem{HPS93}
H.~Heiselberg, C.J.~Pethick, E.F.~Staubo, Phys.\ Rev.\ Lett.\ {\bf
70} (1993) 1355.
\bibitem{VYT01}
D.N.~Voskresensky, M.~Yasuhira, T.~Tatsumi, Phys.\ Lett.\ B {\bf
541}, 93 (2002); nucl-th/0208067.
\bibitem{G92}
N.K. Glendenning, Phys.\ Rev.\ D {\bf 46} (1992) 1274.
\bibitem{CGS00}
M.~Christiansen, N.K.~Glendenning, J.~Schaffner-Bielich, Phys.\
Rev.\ C {\bf 62} (2000) 025804.
\bibitem{G01}
N.~Glendenning, Phys.\ Rep.\ {\bf 342}, 393 (2001).
\bibitem{HH} H.~Heiselberg, M.~Hjorth-Jensen, astro-ph/9904214.
\bibitem{IKV00}
Yu.B.~Ivanov, J.~Knoll, D.N.~Voskresensky, Nucl.\ Phys.\ {\bf 672}, 313, (2000).
\bibitem{KV96}
J.~Knoll, D.~N.~Voskresensky, Ann.\ Phys.\ (N.Y.) {\bf 249}, 532 (1996).
\bibitem{Migdal78}
A.B.~Migdal, Rev.\ Mod.\ Phys.\ {\bf 50}, 107 (1978).
\bibitem{Dyugaev}
A.M.~Dyugaev, ZhETF {\bf 83}, 1005 (1982); Sov.\ J.\ Nucl.\ Phys.\ {\bf 38}, 680
(1983).
\bibitem{VM81}
D.N. Voskresensky, I.N. Mishustin, JETP Lett.\ {\bf 34} (1981), 303;
Sov.\ J.\ Nucl.\ Phys.\ {\bf 35} (1982) 667.
\bibitem{V95}
D.N.~Voskresensky, Phys.\ Lett.\ B {\bf 358}, 1 (1995).
\bibitem{V01}
D.N.~Voskresensky,
Lect.\ Notes Phys.\  {\bf 578}, 467 (2001); astro-ph/0101514.
\bibitem{lmpar}
E.E~Saperstein, S.V. Tolokonnikov, JETP Lett.\ {\bf 68}, 553 (1998);
S.A.~Fayans, D.~Zawischa, Phys.\ Lett.\ B {\bf 363}, 12  (1995);
I.N.~Borzov, S.V.~Tolokonnikov, S.A.~Fayans, Sov.\ J.\ Nucl.\ Phys.\
{\bf 40}, 732 (1984).
\bibitem{vkzc}
D.N.~Voskresensky, V.A.~Khodel, M.V.~Zverev, J.W.~Clark, Astrophys.\
J.\ {\bf 533}, L127 (2000).
\bibitem{SV88}
A.V.~Senatorov, D.N.~Voskresensky, Phys.\ Lett.\  B {\bf 219}, 31 (1989);
D.N. Voskresensky, O.V. Oreshkov, Sov.\ J.\ Nucl.\ Phys.\ {\bf 50}, 820 (1989).
\end{thebibliography}
\end{document}